\pdfoutput=1
\documentclass[11pt,a4paper]{article}

\usepackage{ifthen} 
\newboolean{pdflatex}
\setboolean{pdflatex}{true} 

\newboolean{articletitles}
\setboolean{articletitles}{true} 

\newboolean{uprightparticles}
\setboolean{uprightparticles}{false} 

\def\paperauthors{Anja Beck, Thomas Blake, Michal Kreps} 
\def\paperasciititle{Angular distribution of Lb->pKellell decays comprising Lambda resonances with spin up to 5/2} 
\def\papertitle{Angular distribution of \Lbpkll \\ decays comprising \Lstar resonances \\ with spin $\leq5/2$} 

\usepackage[top=1in, bottom=1.25in, left=1in, right=1in]{geometry}

\usepackage{microtype}
\usepackage{lineno}  
\usepackage{xspace} 
\usepackage{caption} 

\usepackage{graphicx}  
\usepackage{color}
\usepackage{colortbl}
\graphicspath{{./figs/}} 

\usepackage{amsmath} 
\usepackage{amssymb}
\usepackage{amsfonts}
\usepackage{upgreek} 

\newcommand*\patchAmsMathEnvironmentForLineno[1]{%
\expandafter\let\csname old#1\expandafter\endcsname\csname #1\endcsname
\expandafter\let\csname oldend#1\expandafter\endcsname\csname
end#1\endcsname
 \renewenvironment{#1}%
   {\linenomath\csname old#1\endcsname}%
   {\csname oldend#1\endcsname\endlinenomath}%
}
\newcommand*\patchBothAmsMathEnvironmentsForLineno[1]{%
  \patchAmsMathEnvironmentForLineno{#1}%
  \patchAmsMathEnvironmentForLineno{#1*}%
}
\AtBeginDocument{%
\patchBothAmsMathEnvironmentsForLineno{equation}%
\patchBothAmsMathEnvironmentsForLineno{align}%
\patchBothAmsMathEnvironmentsForLineno{flalign}%
\patchBothAmsMathEnvironmentsForLineno{alignat}%
\patchBothAmsMathEnvironmentsForLineno{gather}%
\patchBothAmsMathEnvironmentsForLineno{multline}%
\patchBothAmsMathEnvironmentsForLineno{eqnarray}%
}

\usepackage{hyperxmp}

\usepackage[bottom,flushmargin,hang,multiple]{footmisc}

\usepackage[pdftex,
            pdfauthor={\paperauthors},
            pdftitle={\paperasciititle}
            ]{hyperref}

\usepackage[all]{hypcap} 

\usepackage{xspace} 
\usepackage{upgreek}

\def\lhcb   {\mbox{LHCb}\xspace}
\def\atlas  {\mbox{ATLAS}\xspace}
\def\cms    {\mbox{CMS}\xspace}

\def\cdf    {\mbox{CDF}\xspace}

\def\lhc    {\mbox{LHC}\xspace}

\def\MagUp {\mbox{\em Mag\kern -0.05em Up}\xspace}

\ifthenelse{\boolean{uprightparticles}}%
{

 \def\Pmu         {\ensuremath{\upmu}\xspace}

 \def\Ppi         {\ensuremath{\uppi}\xspace}

 \def\PDelta      {\ensuremath{\Delta}\xspace}                 
 \def\PXi         {\ensuremath{\Xi}\xspace}                 
 \def\PLambda     {\ensuremath{\Lambda}\xspace}                 
 \def\PSigma      {\ensuremath{\Sigma}\xspace}                 
 \def\POmega      {\ensuremath{\Omega}\xspace}                 
 \def\PUpsilon    {\ensuremath{\Upsilon}\xspace}

 \def\PB      {\ensuremath{\mathrm{B}}\xspace}                 
                  
 \def\PD      {\ensuremath{\mathrm{D}}\xspace}

 \def\PK      {\ensuremath{\mathrm{K}}\xspace}

 \def\PZ      {\ensuremath{\mathrm{Z}}\xspace}                 
                  
 \def\Pb      {\ensuremath{\mathrm{b}}\xspace}

 \def\Pe      {\ensuremath{\mathrm{e}}\xspace}

 \def\Pi      {\ensuremath{\mathrm{i}}\xspace}

 \def\Pp      {\ensuremath{\mathrm{p}}\xspace}                 
 \def\Pq      {\ensuremath{\mathrm{q}}\xspace}                 
                  
 \def\Ps      {\ensuremath{\mathrm{s}}\xspace}

 \def\thebaroffset{0.0em}
}
{

 \def\Pmu         {\ensuremath{\mu}\xspace}

 \def\Ppi         {\ensuremath{\pi}\xspace}

 \mathchardef\PDelta="7101
 \mathchardef\PXi="7104
 \mathchardef\PLambda="7103
 \mathchardef\PSigma="7106
 \mathchardef\POmega="710A
 \mathchardef\PUpsilon="7107
                  
 \def\PB      {\ensuremath{B}\xspace}                 
                  
 \def\PD      {\ensuremath{D}\xspace}

 \def\PK      {\ensuremath{K}\xspace}

 \def\PZ      {\ensuremath{Z}\xspace}                 
                  
 \def\Pb      {\ensuremath{b}\xspace}

 \def\Pe      {\ensuremath{e}\xspace}

 \def\Pi      {\ensuremath{i}\xspace}

 \def\Pp      {\ensuremath{p}\xspace}                 
 \def\Pq      {\ensuremath{q}\xspace}                 
                  
 \def\Ps      {\ensuremath{s}\xspace}

 \def\thebaroffset{0.18em}
}
\newcommand{\offsetoverline}[2][\thebaroffset]{\kern #1\overline{\kern -#1 #2}}%

\makeatletter
\ifcase \@ptsize \relax
  \newcommand{\miniscule}{\@setfontsize\miniscule{4}{5}}
\or
  \newcommand{\miniscule}{\@setfontsize\miniscule{5}{6}}
\or
  \newcommand{\miniscule}{\@setfontsize\miniscule{5}{6}}
\fi
\makeatother

\DeclareRobustCommand{\optbar}[1]{\shortstack{{\miniscule (\rule[.5ex]{1.25em}{.18mm})}
  \\ [-.7ex] $#1$}}

\def\en         {{\ensuremath{\Pe^-}}\xspace}   
\def\ep         {{\ensuremath{\Pe^+}}\xspace}

\def\mup        {{\ensuremath{\Pmu^+}}\xspace}
\def\mun        {{\ensuremath{\Pmu^-}}\xspace} 

\def\mumu       {{\ensuremath{\Pmu^+\Pmu^-}}\xspace}

\def\ellm       {{\ensuremath{\ell^-}}\xspace}
\def\ellp       {{\ensuremath{\ell^+}}\xspace}
\def\ellell     {\ensuremath{\ell^+ \ell^-}\xspace}

\def\Z      {{\ensuremath{\PZ}}\xspace}

\def\quark     {{\ensuremath{\Pq}}\xspace}
\def\quarkbar  {{\ensuremath{\overline \quark}}\xspace}
\def\qqbar     {{\ensuremath{\quark\quarkbar}}\xspace}

\def\squark    {{\ensuremath{\Ps}}\xspace}
\def\squarkbar {{\ensuremath{\overline \squark}}\xspace}

\def\bquark    {{\ensuremath{\Pb}}\xspace}
\def\bquarkbar {{\ensuremath{\overline \bquark}}\xspace}
\def\bbbar     {{\ensuremath{\bquark\bquarkbar}}\xspace}

\def\pion   {{\ensuremath{\Ppi}}\xspace}

\def\pim    {{\ensuremath{\pion^-}}\xspace}

\def\kaon    {{\ensuremath{\PK}}\xspace}

\def\KorKbar {\kern \thebaroffset\optbar{\kern -\thebaroffset \PK}{}\xspace}

\def\Km      {{\ensuremath{\kaon^-}}\xspace}

\def\Kstarz  {{\ensuremath{\kaon^{*0}}}\xspace}

\def\D       {{\ensuremath{\PD}}\xspace}

\def\DorDbar {\kern \thebaroffset\optbar{\kern -\thebaroffset \PD}\xspace}

\def\Dp      {{\ensuremath{\D^+}}\xspace}
\def\Dm      {{\ensuremath{\D^-}}\xspace}

\def\DpDm    {\ensuremath{\Dp {\kern -0.16em \Dm}}\xspace}

\def\B       {{\ensuremath{\PB}}\xspace}

\def\BorBbar {\kern \thebaroffset\optbar{\kern -\thebaroffset \PB}\xspace}
\def\Bz      {{\ensuremath{\B^0}}\xspace}

\def\Bd      {{\ensuremath{\B^0}}\xspace}

\def\BdorBdbar {\kern \thebaroffset\optbar{\kern -\thebaroffset \Bd}\xspace}

\def\Bs      {{\ensuremath{\B^0_\squark}}\xspace}

\def\BsorBsbar {\kern \thebaroffset\optbar{\kern -\thebaroffset \Bs}\xspace}

\def\Y#1S{\ensuremath{\PUpsilon{(#1S)}}\xspace}

\def\proton      {{\ensuremath{\Pp}}\xspace}

\def\Lz          {{\ensuremath{\PLambda}}\xspace}

\def\LorLbar     {\kern \thebaroffset\optbar{\kern -\thebaroffset \PLambda}\xspace}

\def\Lb           {{\ensuremath{\Lz^0_\bquark}}\xspace}

\newcommand{\decay}[2]{\ensuremath{#1\!\to #2}\xspace} 

\def\to                 {\ensuremath{\rightarrow}\xspace}

\newcommand{\mLb}{{\ensuremath{m_{\Lb}}}\xspace}

\def\qsq       {{\ensuremath{q^2}}\xspace}

\def\CP                {{\ensuremath{C\!P}}\xspace}

\def\bsll     {\decay{\bquark}{\squark \ell^+ \ell^-}}

\def\AT#1     {\ensuremath{A_{\mathrm{T}}^{#1}}\xspace}           

\def\C#1      {\ensuremath{\mathcal{C}_{#1}}\xspace}                       
\def\Cp#1     {\ensuremath{\mathcal{C}_{#1}^{'}}\xspace}                    
\def\Ceff#1   {\ensuremath{\mathcal{C}_{#1}^{\mathrm{(eff)}}}\xspace}        
\def\Cpeff#1  {\ensuremath{\mathcal{C}_{#1}^{'\mathrm{(eff)}}}\xspace}       
\def\Ope#1    {\ensuremath{\mathcal{O}_{#1}}\xspace}                       
\def\Opep#1   {\ensuremath{\mathcal{O}_{#1}^{'}}\xspace}                    

\newcommand{\bra}[1]{\ensuremath{\langle #1|}}             
\newcommand{\ket}[1]{\ensuremath{|#1\rangle}}              

\newcommand{\aunit}[1]{\ensuremath{\text{\,#1}}}       

\newcommand{\tev}{\aunit{Te\kern -0.1em V}\xspace}
\newcommand{\gev}{\aunit{Ge\kern -0.1em V}\xspace}
\newcommand{\mev}{\aunit{Me\kern -0.1em V}\xspace}
\newcommand{\kev}{\aunit{ke\kern -0.1em V}\xspace}
\newcommand{\ev}{\aunit{e\kern -0.1em V}\xspace}
 
\newcommand{\mevc}{\ensuremath{\aunit{Me\kern -0.1em V\!/}c}\xspace}
\newcommand{\gevc}{\ensuremath{\aunit{Ge\kern -0.1em V\!/}c}\xspace}
\newcommand{\mevcc}{\ensuremath{\aunit{Me\kern -0.1em V\!/}c^2}\xspace}
\newcommand{\gevcc}{\ensuremath{\aunit{Ge\kern -0.1em V\!/}c^2}\xspace}

\def\deriv {\ensuremath{\mathrm{d}}}

\def\gsim{{~\raise.15em\hbox{$>$}\kern-.85em
          \lower.35em\hbox{$\sim$}~}\xspace}
\def\lsim{{~\raise.15em\hbox{$<$}\kern-.85em
          \lower.35em\hbox{$\sim$}~}\xspace}

\def\tell1  {TELL1\xspace}
\def\ukl1   {UKL1\xspace}

\newcommand{\ie}{\mbox{\itshape i.e.}\xspace}

\usepackage{longtable} 
\usepackage{lscape}

\usepackage{booktabs}
\usepackage{cite} 
\usepackage{mciteplus}

\usepackage{comment}

\usepackage{tikz}
\usetikzlibrary{decorations.markings}

\usepackage{multirow}

\usepackage{multicol} 
\usepackage[T1]{fontenc} 

\DeclareMathAlphabet{\calTom}{U}{rsfso}{m}{n}

\def\Lzero{\ensuremath{\Lz(1115)}\xspace}
\def\Lstar{\ensuremath{\Lz}\xspace}
\newcommand{\Lnum}[1]{\ensuremath{\Lz(#1)}\xspace}

\def\Lbpkmm{\decay{\Lb}{p\Km\mumu}}
\def\Lbpkll{\decay{\Lb}{p\Km\ellell}}
\def\pk{\ensuremath{\proton\Km}\xspace}
\def\Lstpk{\decay{\Lstar}{\pk}}
\def\LbLzmm{\decay{\Lb}{\Lzero\mumu}}
\def\LbLll{\decay{\Lb}{\Lstar\ellell}}
\def\LbLV{\decay{\Lb}{\Lstar V}}
\def\Vll{\decay{V}{\ellell}}

\def\jOne{\ensuremath{\tfrac{1}{2}}\xspace}
\def\jThree{\ensuremath{\tfrac{3}{2}}\xspace}
\def\jFive{\ensuremath{\tfrac{5}{2}}\xspace}

\def\pLb{\ensuremath{\mathcal{P}_{\!\!\scriptscriptstyle\Lb}}\xspace}    
\def\mLb{\ensuremath{m_{\PLambda_b}}\xspace}              
\def\mLambda{\ensuremath{m_\Lambda}\xspace}                   
\def\gLambda{\ensuremath{\Gamma_\Lambda}\xspace}              
\def\pLambda{\ensuremath{P_\Lambda}\xspace}                   
\def\phase{\ensuremath{\delta_\Lambda}\xspace}                
\def\jLambda{\ensuremath{J_\Lambda}\xspace}                   
\def\jV{\ensuremath{J_{V}}\xspace}         
\newcommand{\Kobs}[1]{\ensuremath{\overline{K}_{#1}}\xspace}  

\def\mpk{\ensuremath{m_{pK}}\xspace}                          
\def\ksq{\ensuremath{k^2}\xspace}                             

\def\angles{\ensuremath{\vec{\Omega}}\xspace}
\def\phip{\ensuremath{\phi_p}\xspace}
\def\phil{\ensuremath{\phi_\ell}\xspace}
\def\phib{\ensuremath{\phi_b}\xspace}
\def\tp{\ensuremath{\theta_p}\xspace}
\def\tl{\ensuremath{\theta_\ell}\xspace}
\def\thetab{\ensuremath{\theta_b}\xspace}

\def\lV{\ensuremath{\lambda_{V}}\xspace}            
\def\lLb{\ensuremath{\lambda_b}\xspace}           
\def\lLambda{\ensuremath{\lambda_\Lambda}\xspace} 
\def\lProton{\ensuremath{\lambda_{p}}\xspace}     
\def\lLepp{\ensuremath{\lambda_{1}}\xspace}       
\def\lLepn{\ensuremath{\lambda_{2}}\xspace}       

\newcommand{\hHad}[2]{\ensuremath{h^{\Lambda}_{{#1,#2}}}}         
\newcommand{\hLep}[4]{\ifthenelse{\equal{#1}{}}{\ensuremath{\tilde{h}^{#2}_{#3,#4}}}{\ensuremath{\tilde{h}^{#1,#2}_{#3,#4}}}} 
\newcommand{\HVL}[3]{\ensuremath{{\cal H}_{#1,#2}^{\Lambda, #3}}} 
\newcommand{\Amp}[3]{\ensuremath{{\cal A}^{Q,#1}_{#2,#3}}}  
\newcommand{\AmpJP}[5]{\ensuremath{{\cal A}^{#4^#5,#1}_{#2,#3}}}  
\newcommand{\HVLva}[3]{\ensuremath{H_{#1,#2}^{\Lambda, #3}}}      

\newcommand{\WC}[1]{\ensuremath{\mathcal{C}_{#1}}\xspace}
\newcommand{\Op}[1]{\ensuremath{\mathcal{O}_{#1}}\xspace}

\begin{document}

\renewcommand{\thefootnote}{\fnsymbol{footnote}}
\setcounter{footnote}{1}

\begin{titlepage}

\vspace*{-1.5cm}

\noindent

\vspace*{4.0cm}

{\normalfont\bfseries\boldmath\huge
\begin{center}
  \papertitle
\end{center}
}

\vspace*{2.0cm}

\begin{center}
A.~Beck\footnote{\href{mailto:anja.beck@cern.ch}{anja.beck@cern.ch}},
T.~Blake\footnote{\href{mailto:thomas.blake@cern.ch}{thomas.blake@cern.ch}},
M.~Kreps\footnote{\href{mailto:michal.kreps@cern.ch}{michal.kreps@cern.ch}}.
\bigskip\\
{\normalfont\itshape\footnotesize
Department of Physics, University of Warwick, Coventry CV4\,7AL, UK \\
}
\end{center}

\vspace{\fill}

\begin{abstract}
  \noindent
  This paper describes the angular distribution of $\Lb\to\Lstar(\to \proton\Km)\ellell$ decays. 
  A full expression is given for the case of multiple interfering spin-states with spin $\leq \jFive$.
  This distribution is relevant for future measurements of \Lbpkll decays, where different states cannot easily be separated based on their mass alone. 
  New observables arise when considering spin-\jFive states as well as interference between states.
  An exploration of their behaviour for a variety of beyond the Standard Model scenarios shows that some of these observables exhibit interesting sensitivity to the Wilson coefficients involved in \bsll transitions.
  Others are insensitive to the Wilson coefficients and can be used to verify the description of $\Lb \to \Lstar$ form-factors. 
  A basis of weighting functions that can be used to determine all of the angular observables described in this paper in a moment analysis of the experimental data is also provided. 
\end{abstract}

\vspace*{2.0cm}
\vspace{\fill}

\end{titlepage}

\pagestyle{empty}  


\newpage
\setcounter{page}{2}
\mbox{~}

\renewcommand{\thefootnote}{\arabic{footnote}}
\setcounter{tocdepth}{2}
\setcounter{footnote}{0}
\tableofcontents
\cleardoublepage

\pagestyle{plain} 
\setcounter{page}{1}
\pagenumbering{arabic}


\clearpage

\section{Introduction}
\label{sec:introduction}

In recent years, the \lhcb collaboration has published several measurements of the rates and angular distributions of \bquark- to \squark-quark flavour-changing neutral-current processes~\cite{LHCb:2014cxe,LHCb:2016ykl,LHCb:2020lmf,LHCb:2021awg,LHCb:2021zwz}.
The experimental results reveal a pattern of discrepancies with predictions based on the Standard Model of particle physics (SM). 
The measurements by the \lhcb collaboration are reinforced by compatible observations performed by the BaBar and Belle experiments and the \atlas, \cms and \cdf collaborations~\cite{BaBar:2006tnv,BaBar:2012mrf,Belle:2009zue,Belle:2016fev,Belle-II:2022fky,ATLAS:2018gqc,ATLAS:2018cur,CMS:2017rzx, CMS:2018qih,CMS:2019bbr, CMS:2020oqb,CDF:2011tds,CDF:2011buy}. 
Global analyses of \bquark- to \squark-quark transitions indicate that the measurements form a coherent picture that could be explained by several proposed extensions of the SM, see for example Refs.~\cite{Altmannshofer:2021qrr,Alguero:2021anc,Ciuchini:2020gvn,Hurth:2021nsi}. 
Measurements comparing the rates of processes involving $\bquark\to\squark\mup\mun$ and $\bquark\to\squark\ep\en$ transitions also show differences \cite{LHCb:2021trn,LHCb:2017avl}, which suggest that the underlying theory may have non-universal lepton couplings. 

Thus far, measurements have mainly focused on analyses of \B meson decays.
It is important to confirm the discrepancies in other systems. The most convenient choice for this is through the decay of the \Lb baryon, which is the lightest \bquark-baryon and is produced abundantly at the \lhc~\cite{LHCb:2019fns}. 
As of today, the \lhcb collaboration has measured the branching fraction and angular distribution of the \LbLzmm decay~\cite{Aaij:2003794,Aaij:2633197}, where the label \Lzero is used to refer to the weakly decaying ground-state baryon.
Measurements of the \LbLzmm transition have already been considered in global analyses~\cite{Blake:2019guk, Altmannshofer:2021qrr} but larger experimental data sets are needed to understand the compatibility of the measurements with those in \B meson systems.   
The \lhcb experiment also observes large signals of \Lbpkmm decays, which it has used to search for \CP violation in the decay~\cite{LHCb:2017slr} and to test lepton flavour universality by comparing $\Lb\to p\Km \ep\en$ and \Lbpkmm decays~\cite{LHCb:2019efc}. 
A unique feature of the \pk spectrum in these decays is the rich contribution from different \Lstar resonances, whose states cannot easily be separated.\footnote{In order to avoid confusion, the weakly-decaying ground-state will be labelled \Lzero and the strongly decaying resonance states will be collectively labelled \Lstar resonances when referring to the resonances in general and a mass in parentheses will be used to refer to a specific state.}

From the theoretical point of view, the semi-leptonic \Lb decay to the ground-state \Lzero baryon has been studied in detail. 
There are predictions for the form factors for the decay from light-cone sum-rule techniques~\cite{Gan:2012tt,Wang:2008sm} and lattice QCD~\cite{Detmold:2016pkz,Detmold:2012vy,Detmold:2012ug}.  
Dispersive bounds on the form factors have also been discussed in Ref.~\cite{Blake:2022vfl}. 
The angular distribution for the decay is known~\cite{Boer:2014kda}, even for the case of polarised \Lb baryons~\cite{Blake:2017une} and the full basis of new physics operators~\cite{Sahoo:2016nvx, Das:2019omf}. 
Much less is known about the decay via other \Lstar resonances. 
The form factors for \Lb to \Lnum{1520} transitions have been determined in lattice QCD~\cite{Meinel:2020owd,Meinel:2021mdj}, in the quark model~\cite{Li:2022nim}, and studied in HQET~\cite{Bordone:2021bop}. 
Dispersive bounds have also been considered in Ref.~\cite{Amhis:2022vcd}.
For other resonances, form-factor predictions are only available in the context of the quark model~\cite{Mott:2011cx,Mott:2015zma}. 
The full angular distribution of single spin-\jOne~\cite{Dey:2019myq,Das:2020cpv} and the spin-\jThree~\cite{Descotes-Genon:2019dbw,Amhis:2020phx} resonances is known but the distribution of higher-spin resonances and the more general case of overlapping, interfering, resonances has not been studied. 
The aim of this paper is to provide a description of the angular distribution, including up-to spin-\jFive resonances, and to present a method that can be used by experiments to perform a model-independent analysis of the \Lbpkll decay.

The following sections begin with a decomposition of the full \Lbpkll decay rate into subsequent two-body decays. 
The approach used holds for any decay of a spin-\jOne baryon to a final state involving a spin-\jOne baryon, a spin-0 meson, and two fermions.
The amplitudes for the two-body decays are calculated in the helicity formalism, as described in  Section~\ref{sec:helformalism}. 
Section~\ref{sec:angular:distribution} provides an expansion for the full angular distribution in terms of a set of basis functions. 
Section~\ref{sec:method-of-moments} introduces the method of moments and explains its application to the decay rate developed in the first sections. 
Section~\ref{sec:observables} provides explicit expressions for some of the observables appearing in the angular distribution.
Section~\ref{sec:individual} explores the angular distributions of individual \Lstar resonances. 
The angular distribution of the spin-\jFive \Lnum{1820} resonance, and a realistic ensemble of \Lstar states, is explored in Section~\ref{sec:np} together with the possibility of observing modifications of the angular distribution of the decay in extensions of the SM.

\section{Decomposition of the decay rate}
\label{sec:decomp}

The differential decay rate for the full decay chain can be expressed as
\begin{align}\label{eq:decrate_general}
    \deriv\Gamma = \frac{\overline{|\mathcal{M}|}^2}{2 \mLb} (2\pi)^4 \deriv\Phi_4 \,,
\end{align}
where $\mathcal{M}$ is the invariant amplitude for the decay, \mLb is the mass of the \Lb baryon and $\deriv\Phi_4$ the 4-body differential phase space. 
The \Lbpkll decay is modelled as three subsequent two-body decays, where the \Lb baryon first decays into a \Lstar resonance and a virtual vector boson, labelled below by $V$. 
The \Lstar resonance then decays strongly into a proton and a kaon, while the virtual vector boson produces the two leptons. 
In what follows, the four momenta of the \Lb baryon and the \Lstar resonance are denoted $p$ and $k$, respectively. The four momentum of the \ellell system is $q = p -k$. The four momenta of the proton and kaon are $k_1$ and $k_2$ and the four momenta of the \ellp and \ellm are $q_1$ and $q_2$. We use the notation $p^{\mu}=(p^0,\vec{p}\,)$ when referring to the energy and three momentum of the particles.

\subsection{Invariant amplitude}
\label{sec:decomp_matrixelem}

The spin-averaged invariant amplitude-squared, $\overline{|\mathcal{M}|}^2$, is obtained by summing over the possible helicites of the \Lb baryon, $\lLb=\pm\frac{1}{2}$, the proton, $\lProton = \pm\tfrac{1}{2}$, and the  two leptons,  $\lLepp = \pm\tfrac{1}{2}$ and $\lLepn= \pm\tfrac{1}{2}$,
\begin{align}
    \overline{|\mathcal{M}|}^2 = \sum_{{\lLb}}\mathcal{P}_{\lLb}\sum_{\lLepp,\lLepn,\lProton}|\mathcal{M}_{{\lLb},\lProton,\lLepp,\lLepn}|^2 \ .
\end{align}
The factor $\mathcal{P}_{\lLb}$ corresponds to the relative amount of the $\Lb$ spin state ${\lLb}$.
The sum of $\mathcal{P}_{\lLb}$ over the two spin-states is one, \ie $\mathcal{P}_{+1/2} + \mathcal{P}_{-1/2} = 1$. 

The amplitude for a given set of initial and final states corresponds to the sum over all intermediate resonances, \Lstar, and their corresponding helicity, $\lLambda$,
\begin{align}
    \mathcal{M}_{{\lLb},\lProton,\lLepp,\lLepn} = \sum_{\Lstar}\sum_{\lLambda}\mathcal{M}^{\Lstar}_{\lLambda} \ .
\label{eq:resonancesum} 
\end{align}
The helicity indices \lLb, \lProton, \lLepp, and \lLepn have been suppressed on the right-hand side of the expression for readability. 
Equation~\ref{eq:resonancesum} can be split into two pieces, representing an amplitude for the decay to $\Lstar\ellell$ and for the subsequent decay of the \Lstar resonance to \pk, \ie 
\begin{align}
    \mathcal{M}^{\Lstar}_{\lLambda} =  \mathcal{M}^{\LbLll}_{\lLambda}\mathcal{M}^{\Lstpk}_{\lLambda} \ .
\end{align}

Using a naive factorisation approach, after integrating out heavy degrees of freedom, the effective Lagrangian is
\begin{align}
    \mathcal{L}_\text{eff} = \mathcal{L}_\text{QCD} + \mathcal{L}_\text{QED} + \frac{4G_F}{\sqrt{2}}V_{tb}V_{ts}^*\sum_{i}\WC{i}\Op{i} \ ,
\end{align}
where $V_{ij}$ are elements of the Cabibbo-Kobayashi-Maskawa quark-mixing matrix and $G_F$ is the Fermi constant. 
The \WC{i} and \Op{i} represent the Wilson coefficients and the corresponding local operators of the effective theory.
The relevant dimension-four operators for \bsll transitions are 
\begin{align}
\label{eq:operators}
\begin{split}
    \Op{7^{(\prime)}} &= \frac{e}{16\pi^2}m_b(\Bar{s}\sigma_{\mu\nu}P_{R(L)}b)F^{\mu\nu} 
    \,,\\
    \Op{9^{(\prime)}} &= \frac{e^2}{16\pi^2}(\Bar{s}\gamma_\mu P_{L(R)}b)(\Bar{\ell}\gamma^\mu \ell) \,, \\
    \Op{10^{(\prime)}} &= \frac{e^2}{16\pi^2}(\Bar{s}\gamma_\mu P_{L(R)}b)(\Bar{\ell}\gamma^\mu\gamma_5 \ell)\,,
\end{split}
\end{align}
with the left- and right-handed chiral projection operators $P_{L,R} = \frac{1}{2}\left(1\mp\gamma_5\right)$. 
Four-quark current-current and QCD penguin operators, usually denoted \Op{1-6}, also contribute to \bsll transitions.
Their impact is discussed in Appendix~\ref{app:effC9} and included by using effective Wilson coefficients $\WC{i}^\text{eff}$.
Non-factorisable corrections that can not be expressed in terms of contributions to $\WC{7,9}^\text{eff}$ are neglected in this paper.

Splitting the operators into $\Lb\to\Lstar$ and dilepton current results in the expression
\begin{align}
\begin{split}
    \mathcal{M}^{\LbLll}_{\lLambda}
    = N_1\sum_i\WC{i}\bra{\ell\ell}\Op{\text{lep},i}^\mu\ket{0}\bra{\Lstar}\Op{\text{had},i}^\nu\ket{\Lb}g_{\mu\nu}
\end{split}
\end{align}
for the \LbLll decay amplitude, where the constants have been absorbed into the normalisation factor
\begin{align}
    N_1 = \frac{4G_F}{\sqrt{2}}V_{tb}V_{ts}^* \ .
\end{align}

To simplify the calculations, a projection onto an intermediate vector boson is introduced by expressing the Minkowski metric as
\begin{align}
    g_{\mu\nu} = \sum_{\lV}\varepsilon_\nu^*(\lV)\varepsilon_\mu(\lV)g_{\lV\lV} \ .
    \label{eq:minkowski}
\end{align}
The virtual vector boson has four polarization states: time-like ($\jV=\lV=0$), longitudinal ($\jV=1,\lV=0$), and transverse ($\jV=1,\lV=\pm1$).
In the following, these states will be labelled by $\lV=t,0,\pm$. 
An explicit form of the polarisation vectors for the different polarisation states is given in Appendix~\ref{app:spinors}. 

Using Equation~\ref{eq:minkowski}, the hadronic and leptonic currents can be separated
\begin{align}
    \mathcal{M}^{\LbLll}_{\lLambda}
    &= N_1\sum_{\lV}g_{\lV\lV}\sum_{i}
    \underbrace{\bra{\ell\ell}\Op{\text{lep},i}^\mu\ket{0}\varepsilon_\mu(\lV)}_{\mathcal{M}_{\lV,\Op{i}}^{\Vll}}
    \underbrace{\WC{i}\bra{\Lstar}\Op{\text{had},i}^\nu\ket{\Lb} \varepsilon_\nu^*(\lV)}_{\mathcal{M}^{\LbLV}_{\lLambda,\lV,\Op{i}}} \,,
\end{align}
and evaluated in independent reference frames.
The amplitudes $\smash{\mathcal{M}_{\lV,\Op{i}}^{\Vll}}$, $\smash{\mathcal{M}^{\LbLV}_{\lLambda,\lV,\Op{i}}}$, and $\smash{\mathcal{M}^{\Lstpk}_{\lLambda}}$ are further discussed in Section~\ref{sec:helformalism}.

\subsection{Four-body phase-space}
\label{sec:decomp_phsp}

The four-body phase-space for the \Lbpkll decay can be decomposed into three two-body phase-space elements as
\begin{align}
\begin{split}
    \deriv\Phi_4 &=
    \frac{1}{4(2\pi)^6}\frac{|\vec{k}|}{\mLb}\deriv\Omega_{\Lb}
    \times \frac{1}{4(2\pi)^6}\frac{|\vec{k}_1|}{\sqrt{\ksq}}\deriv\Omega_{pK}\cdot(2\pi)^3\deriv\ksq
    \times \frac{1}{4(2\pi)^6}\frac{|\vec{q}_1|}{\sqrt{\qsq}}\deriv\Omega_{\ell\ell}\cdot(2\pi)^3\deriv\qsq \\
    &= \frac{1}{2^6(2\pi)^{12}}\frac{|\vec{k}|}{\mLb}\frac{|\vec{k}_1|}{\sqrt{\ksq}}\frac{|\vec{q}_1|}{\sqrt{\qsq}}\deriv\Omega_{\Lb}\deriv\Omega_{pK}\deriv\Omega_{\ell\ell}\deriv\ksq \deriv\qsq \ .
\end{split}
\end{align}
Here and throughout this paper, the three-momenta $\vec{k}$, $\vec{k}_1$, and $\vec{q}_1$ are evaluated in the rest frame of their respective parent particle.

In the following, we give explicit expressions for the angular phase-space elements beginning with the \Lb decay. 
We assume that the \Lb baryons are produced at a hadron collider and refer to the centre-of-mass frame of the two colliding beams as the laboratory frame. 
The standard convention is to align the beam directions with the $\hat{z}$-axis. 

In $pp$ collisions at the LHC, \bbbar pairs are produced primarily by QCD processes (gluon-gluon fusion, gluon splitting and flavour excitation~\cite{Cacciari:1998it}). 
Consequently, \Lb baryons are expected to have no longitudinal polarisation and only a small transverse polarisation, measured against the axis $\hat{n}_\perp=\hat{p}\times\hat{p}_\text{beam}$. 
Here, $\hat{p}$ is the \Lb direction and $\hat{p}_{\text{beam}}$ is the beam direction in the laboratory frame. 
This is consistent with the measurement by the LHCb collaboration in Ref.~\cite{LHCb:2020iux} that finds no transverse polarisation in \Lb production. 
Longitudinally polarised \Lb baryons are produced in \Z decays~\cite{DELPHI:1999hkl} such that a large sample of polarized \Lb baryons could be obtained at the proposed FCC-$ee$.
In this case, the formalism outlined in this paper remains unchanged but the angular definition must be updated to reflect the different choice of polarisation axis.
The polarisation fraction measured against the corresponding axis is
\begin{align}\label{eq:Lbpol}
    \pLb=\mathcal{P}_{+1/2}-\mathcal{P}_{-1/2} \ .
\end{align}

\begin{figure}[!tb]
\centering
\begin{tikzpicture}

\node[] at (2.0,5.0) {\Lb rest frame};
\draw[] (1.9,3.4) to[bend right] (1.6,3.1);
\node[] at (1.5,3.45) {\textcolor{red}{\thetab}};
\draw[thick,blue,->] (2,3) -- (1.8,3.9);
\node[left] at (1.8,3.9){\textcolor{blue}{$\hat{n}$}};
\draw[thick,->](2,3) -- (0.4,3.4);
\draw[thick,->](2,3) -- (3.6,2.6);
\node[] at (0.5,3.0) {\Lstar};
\node[] at (3.3,2.3) {$V$};
\node[] at (3.5,4.4) {$p_{\Lb}^{\text{lab}}$};

\draw[thick,dashed] (1,1.5) -- (2,3);
\draw[thick,dashed,->] (2,3) -- (3,4.5);
\draw[thick,dashed](-2,4.0) -- (-4,4.5);
\draw[thick,dashed](6,2.0) -- (8,1.5);

\node[] at (-0.6,4.6) {{\small \textcolor{blue}{$x$}}};
\node[] at (0.2,4.3) {{\small \textcolor{magenta}{$y$}}};
\node[] at (-1.5,3.85) {{\small \textcolor{olive}{$z$}}};
\draw[thick,blue,->] (-0.4,3.6) -- (-0.6,4.5);
\draw[thick,magenta,->] (-0.4,3.6) -- (0.1,4.2);
\draw[thick,olive,->] (-0.4,3.6) -- (-1.4,3.85);

\node[] at (4.2,3.4) {{\small \textcolor{blue}{$x$}}};
\node[] at (3.8,1.75) {{\small \textcolor{magenta}{$y$}}};
\node[] at (5.5,2.15) {{\small \textcolor{olive}{$z$}}};
\draw[thick,blue,->](4.4,2.4)--(4.2,3.3);
\draw[thick,olive,->] (4.4,2.4) -- (5.4,2.15);
\draw[thick,magenta,->] (4.4,2.4) -- (3.9,1.8);


\node[] at (-2.7,1.6) {\Lstar rest frame};

\draw[-](-3.6,2.4)--(-3.2,6.4);
\draw[-](-3.2,6.4)--(-0.2,5.65);
\draw[-](-3.6,2.4)--(-0.6,1.65);
\draw[-](-0.6,1.65)--(-0.2,5.65);

\filldraw[black] (-2,4) circle (2pt);
\draw[thick,->] (-2,4) -- (-2.6,5.2);
\draw[thick,->] (-2,4) -- (-1.4,2.8);
\node[] at (-2.2,5.1) {\proton};
\node[] at (-1.85,2.9) {\Km};

\draw[orange,->] (-0.55,4.15) to[bend left] (-0.35,4.20);
\node[] at (-0.9,4.2) {\textcolor{orange}{\phip}};

\draw[] (-2.51,4.14) to[bend left] (-2.25,4.45);
\node[] at (-2.7,4.45) {\textcolor{red}{\tp}};


\node[] at (6.8,4.4) {\ellell rest frame};
\draw[-](6.1,3.35) -- (8.7,-0.05);
\draw[-](6.1,3.35) -- (3.1,4.10);
\draw[-](8.7,-0.05) -- (5.7,0.70);
\draw[-](5.7,0.70) -- (3.1,4.10);

\draw[thick,->] (6,2) -- (6.15,3.0);
\draw[thick,->] (6,2) -- (5.85,1.0);
\node[] at (5.8,2.75) {\ellp};
\node[] at (6.35,1.1) {\ellm};
\filldraw[black] (6,2) circle (2pt);

\draw[] (6.05,2.45) to[bend left] (6.44,1.90);
\node[] at (6.6,2.2) {\textcolor{red}{\tl}};

\draw[orange,->] (4.25,3.0) to[bend right] (4.03,2.88);
\node[] at (4.7,2.9) {\textcolor{orange}{\phil}};

\end{tikzpicture}
\caption{
Illustration of the different angles appearing in the expression for the differential decay rate when considering transverse polarisation. 
Three different rest-frames are used, that of the \Lb, the \Lstar and the \ellell system. 
A common axis (in blue), given by the normal to the plane containing the \Lb direction and the beam direction, is used to define the coordinate systems. 
}
\label{fig:angular:def}
\end{figure}
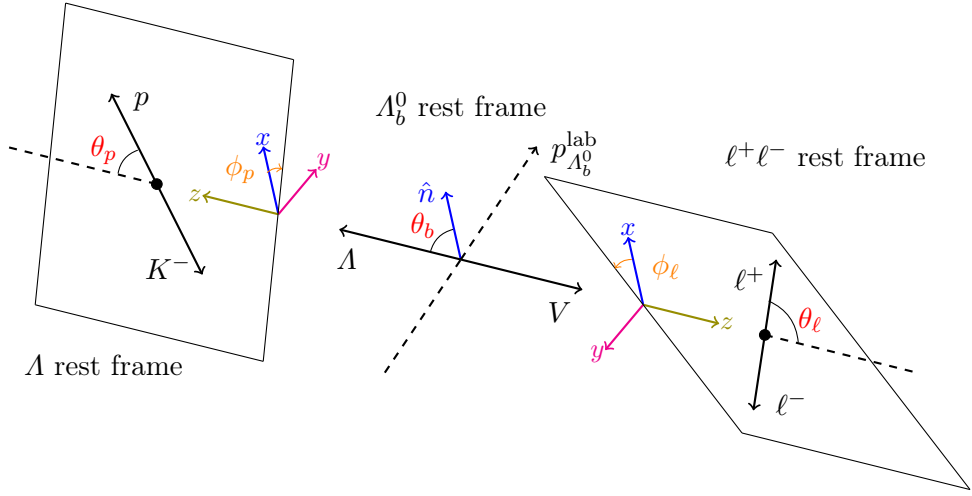

The angular phase space element $\smash{\deriv\Omega_{\Lb}}$ is $\deriv\!\cos\thetab\deriv\phib$, where \thetab and \phib are the polar and azimuthal angles of the momentum vector of the \Lstar resonance in the \Lb-baryon rest-frame measured against the polarisation axis. 
Due to rotational invariance, the amplitude for the decay is independent of \phib.  
Analogously, the angular phase space elements of the \Lstar resonance and dilepton systems are $\deriv\Omega_{pK} = \deriv\!\cos\tp\deriv\phip$ and $\deriv\Omega_{\ell\ell} = \deriv\!\cos\tl\deriv\phil$. 
Here, \tp and \phip correspond to polar and azimuthal angles of the proton in the \Lstar-resonance rest-frame and \tl and \phil the polar and azimuthal angles of the $\ell^+$ in the dilepton rest-frame. 
Figure~\ref{fig:angular:def} illustrates the angle definitions and Appendix~\ref{app:angles} provides explicit expressions for their calculation.
Using this angular convention, and integrating over \phib, leads to  
\begin{align}
    \deriv\Phi_4 &= \frac{1}{2^6(2\pi)^{11}}\frac{|\vec{k}|}{\mLb}\frac{|\vec{k}_1|}{\sqrt{\ksq}}\frac{|\vec{q}_1|}{\sqrt{\qsq}}\deriv\!\cos\thetab\deriv\!\cos\tp\deriv\phip\deriv\!\cos\tl\deriv\phil\deriv\ksq \deriv\qsq \ .
\end{align}

\section{Invariant amplitudes in the helicity formalism}
\label{sec:helformalism}

To simplify the calculation of the differential decay rate when multiple, interfering, \Lstar resonances contribute, we use the helicity formalism. 
The \LbLV decay amplitude, $\mathcal{H}$, is evaluated in the \Lstar helicity frame. 
The angular structure due to rotating the \Lb polarization axis into the \Lstar helicity frame is described by the Wigner D-matrix element $D^{1/2}_{\lLb,\lLambda-\lV}(\phib,\thetab,-\phib)^*$.
Because this is the only \phib-dependent component of the amplitude, the complex terms in the Wigner D-matrix element cancel in the invariant amplitude squared, $|\mathcal{M}_{\lLb,\lProton,\lLepp,\lLepn}|^2$.
This was anticipated in Section~\ref{sec:decomp_phsp} when integrating over \phib.
As a result, the invariant amplitude for the \LbLV transition in terms of helicity amplitude and rotation reads
\begin{align}
\label{eq:M_Lb}
\begin{split} 
    \mathcal{M}^{\LbLV}_{\lLambda,\lV,\Op{i}} = \HVL{\lLambda}{\lV}{\Op{i}}(\mpk,\qsq)d^{1/2}_{\lLb,\lLambda-\lV}(\thetab) \, ,
\end{split}
\end{align}
with the real-valued Wigner d-matrix element.
The helicity amplitude depends on the \pk invariant mass $\mpk=\sqrt{\ksq}$ and the dilepton invariant mass squared \qsq.

The amplitude for the decay of a \Lstar-resonance  with spin \jLambda to a proton and a kaon, $h$, is calculated in the proton helicity frame. 
Combined with a rotation from the \Lstar to the proton helicity frame, the amplitude for the \Lstar decay is
\begin{align}
\label{eq:M_Lstar}
    \mathcal{M}^{\Lstpk}_{\lLambda} = \sqrt{\jLambda+\frac{1}{2}}\,\hHad{\lLambda}{\lProton}\!(\mpk) D^{\jLambda}_{\lLambda,\lProton}(\phip,\tp,-\phip)^* \ .
\end{align}
A spin-dependent factor is introduced to compensate for the normalisation of the Wigner D-matrix elements.

The lepton system amplitude, $\Tilde{h}$, is calculated in the helicity frame of the positively charged lepton. 
After the rotation of the quantization axis from the lepton system helicity frame to the $\ell^+$ helicity frame, the amplitude is
\begin{align}\label{eq:M_ll}
    \mathcal{M}^{\Vll}_{\lV,\Op{i}} &= \hLep{\Op{i}}{\lV}{\lLepp}{\lLepn}(\qsq)D^{\jV}_{\lV,\lLepp-\lLepn}(\phil,\tl,-\phil)^* \ .
\end{align}
Note, that the spin of the virtual vector boson, \jV, is implicitly contained in its polarization \mbox{$\lV=t,0,\pm$}.

Combining equations~\ref{eq:M_Lb}--\ref{eq:M_ll}, with the results of Section~\ref{sec:decomp}, 
\begin{equation}
\begin{split}
\frac{\deriv^{7}\Gamma}{\deriv\qsq\,\deriv \mpk\deriv\angles}
&= \frac{1}{\mLb^2}\frac{N_1^2}{2^6(2\pi)^7} \frac{|\vec{k}| |\vec{k}_1||\vec{q}_1|}{\sqrt{\qsq}}
\sum_{\lLb} \mathcal{P}_{\lLb} \sum_{\lLepp,\lLepn,\lProton}\Big|
\sum_{\Op{i}}\sum_{\Lambda}\sqrt{\jLambda+\frac{1}{2}}\sum_{\lLambda} g_{\lV\lV} \\
&\qquad\qquad\times\HVL{\lLambda}{\lV}{\Op{i}}(\qsq,\mpk)\,d^{1/2}_{\lLb,\lLambda - \lV}(\thetab) \\ 
&\qquad\qquad\times \hLep{\Op{i}}{\lV}{\lLepp}{\lLepn}(\qsq) D^{\jV}_{\lV,\lLepp-\lLepn}(\phil,\tl,-\phil)^* \\
&\qquad\qquad\times \hHad{\lLambda}{\lProton}(\mpk) D^{\jLambda}_{\lLambda,\lProton}(\phip,\tp,-\phip)^*
\Big|^2 \,,
\end{split}
\label{eq:base}
\end{equation}
where $\angles = (\cos\thetab,\,\cos\tp,\,\phip,\,\cos\tl,\,\phil)$. 
It is convenient to replace the sum over the \Lb helicities, \lLb, by the spin-density matrix
\begin{align}
\begin{split}\label{eq:spindensity}
\rho_{\lLambda-\lV,\lLambda'-\lV'} &= \sum_{\lLb}\mathcal{P}_{\lLb} d^{1/2}_{\lLb,\lLambda-\lV}(\thetab)d^{1/2}_{\lLb,\lLambda'-\lV'}(\thetab) \\
&=\frac{1}{2}\left(
    \begin{array}{cc}
        1+ \pLb \cos\thetab & \pLb \sin\thetab \\
        \pLb \sin\thetab & 1-\pLb \cos\thetab
    \end{array}\right)\,,
\end{split}
\end{align}
where \pLb is the \Lb polarisation defined in Equation~\eqref{eq:Lbpol} and the upper-left (lower-right) element corresponds to $\rho_{+1/2,+1/2}(\rho_{-1/2,-1/2})$ and the off-diagonal elements correspond to $\rho_{\pm1/2,\mp1/2}$.

\subsection[Helicity amplitudes for the \texorpdfstring{\LbLV}{Lb -> Lstar V} decay]{Helicity amplitudes for the \texorpdfstring{\boldmath\LbLV}{Lb -> Lstar V} decay}

Separate amplitudes need to be considered for hadronic operators with different Lorentz structures, $\Op{{\rm had.},i}^{\mu} =\Bar{s}\Gamma^\mu_i P_{L,R}b$. 
The relevant amplitudes for this paper are 
\begin{alignat}{4}
    \HVL{\lLambda}{\lV}{7^{(\prime)}}(\qsq,\mpk)
    &=- \frac{2m_b}{\qsq} &&\frac{\WC{7^{(\prime)}}^\text{eff}}{2} &&e^{i\phase}&&\left(\HVLva{\lLambda}{\lV}{T}\mp \HVLva{\lLambda}{\lV}{T5}\right)\,, \nonumber \\
    \HVL{\lLambda}{\lV}{9^{(\prime)}}(\qsq,\mpk)
    &= &&\frac{\WC{9^{(\prime)}}^\text{eff}}{2} &&e^{i\phase}&&\left(\HVLva{\lLambda}{\lV}{V}\mp \HVLva{\lLambda}{\lV}{A}\right)\,, \\
    \HVL{\lLambda}{\lV}{10^{(\prime)}}(\qsq,\mpk)
    &= &&\frac{\WC{10^{(\prime)}}}{2} &&e^{i\phase}&&\left(\HVLva{\lLambda}{\lV}{V}\mp \HVLva{\lLambda}{\lV}{A}\right)~. \nonumber
\end{alignat}
The labels $V$, $A$, $T$ and $T5$ refer to vector, axialvector, tensor and axialtensor currents with the Lorentz structures $\Gamma^\mu=\gamma^\mu$, $\gamma^\mu\gamma_5$, $i\sigma^{\mu\nu}q_\nu $ and $i\sigma^{\mu\nu}\gamma_5q_\nu$, respectively. 
A common complex phase, \phase, arises from QCD separately for each \Lstar resonance. 
The amplitudes, $\HVLva{\lLambda}{\lV}{\Gamma^\mu}$, are
\begin{align}\label{eq:helamps}
\begin{split}
    \HVLva{\lLambda}{\lV}{\Gamma^\mu} &= \varepsilon^*_\mu(\lV)\bra{\Lambda}\bar{s}\Gamma^\mu b\ket{\Lb} \ .
\end{split}
\end{align}
Note, the polarization vectors for the vector-boson used in this paper are defined in the opposite direction to those of Ref.~\cite{Descotes-Genon:2019dbw}.

The current $\bra{\Lambda}\bar{s}\Gamma^\mu b\ket{\Lb}$ can be decomposed in terms of its underlying Lorentz structure.
One choice is to expand the currents in terms of $\gamma^\mu$ and the \Lb and \Lstar 4-velocities, $v_p$ and $v_k$. 
This approach is taken in Ref.~\cite{Mott:2011cx}, where the currents for $\jLambda = \jOne$ are
\begin{align}
\label{eq:matrixelements:1over2}
\bra{\Lambda}\bar{s}\Gamma^\mu b\ket{\Lb} = \bar{u}(k,\lLambda)\left[
     X_{\Gamma1}(\qsq)\gamma^\mu
    +X_{\Gamma2}(\qsq)v_p^\mu
    +X_{\Gamma3}(\qsq)v_k^\mu
    \right]u(p,\lLb)\,,
\end{align} 
for $\jLambda=\jThree$ are 
\begin{align} 
\label{eq:matrixelements:3over2}
    \bra{\Lambda}\bar{s}\Gamma^\mu b\ket{\Lb} = \bar{u}_\alpha(k,\lLambda)\left[v_p^\alpha\left(
     X_{\Gamma1}(\qsq)\gamma^\mu
    +X_{\Gamma2}(\qsq)v_p^\mu
    +X_{\Gamma3}(\qsq)v_k^\mu\right)
    +X_{\Gamma4}(\qsq)g^{\alpha\mu}
    \right]u(p,\lLb)\,,
\end{align} 
and for $\jLambda=\jFive$ are 
\begin{align} 
\begin{split} 
\label{eq:matrixelements:5over2}
   \bra{\Lambda}\bar{s}\Gamma^\mu b\ket{\Lb} = \bar{u}_{\alpha\beta}(k,\lLambda)v_p^\alpha\left[v_p^\beta\left(
     X_{\Gamma1}(\qsq)\gamma^\mu  \right. \right. &  \left. 
    +X_{\Gamma2}(\qsq)v_p^\mu
    +X_{\Gamma3}(\qsq)v_k^\mu\right) \\ &
    \left. 
    +X_{\Gamma4}(\qsq)g^{\beta\mu}
    \right]u(p,\lLb)\,.
\end{split} 
\end{align}
In the notation of Ref.~\cite{Mott:2011cx,Mott:2015zma}, the form-factors $X_{\Gamma i}$ are $X_{V i} = F_i$, $X_{A i} = G_i$, $X_{T i} = F_i^T$ and $X_{T5 i} = G_i^{T}$. 
Reference~\cite{Mott:2011cx} provides predictions for these form factors for most $\Lb\to\Lstar$ transitions in a quark model. Predictions for some of the states are available from lattice QCD~\cite{Detmold:2016pkz,Meinel:2020owd}. The lattice predictions use an alternative expansion of the currents. 
A translation between the two expansions is provided in Appendix~\ref{app:FFtranslation}.
For \Lstar states with $\jLambda=\jOne$, the \Lstar spinors $u(k,\lLambda)$ are the standard Dirac spinors. For \Lstar states with higher spin, the \Lstar spinors $u_\alpha(k,\lLambda)$ and $u_{\alpha\beta}(k,\lLambda)$ are Rarita-Schwinger objects constructed from coupling an integer-spin tensor-object of order $\jLambda-\tfrac{1}{2}$ and a standard Dirac spinor~\cite{Chung:1971ri}. 
This is described further in Appendix~\ref{app:spinors}.

The amplitudes in Equation~\ref{eq:helamps} are calculated by evaluating the spinor products in Equations~\ref{eq:matrixelements:1over2}-\ref{eq:matrixelements:5over2}. 
The time-like ($\jV=0$ and $\lV=t$) helicity amplitudes for natural parity states, with $\pLambda=(-1)^{\jLambda-\frac{1}{2}}$, are
\begin{align}
\begin{split} 
    H^{V(T)}_{\pm1/2,t} = \hphantom{\pm} N_{\jLambda} \sqrt{\frac{s_+}{\qsq}} \left[
     F_1^{(T)} \left( |\vec{q}|\sqrt{\frac{s_-}{s_+}} + q^0 \right) \right. & 
    +F_2^{(T)} q^0 
    +F_3^{(T)} \left( \frac{\mLb q^0-\qsq}{\mpk} \right) \\ & \left. 
    +F_4^{(T)} \left( q^0 + |\vec{q}|\frac{s_-+s_+}{2\sqrt{s_-s_+}} \right)
    \right]\,,
\end{split} 
\end{align} 
\begin{align}
\begin{split}
H^{A(T5)}_{\pm1/2,t} = \pm N_{\jLambda} \sqrt{\frac{s_-}{\qsq}} \left[
     G_1^{(T)} \left( |\vec{q}|\sqrt{\frac{s_+}{s_-}} + q^0 \right) \right. & 
    -G_2^{(T)} q^0 
    -G_3^{(T)} \left( \frac{\mLb q^0-\qsq}{\mpk} \right) \\ & \left.
    -G_4^{(T)} \left( q^0 + |\vec{q}|\frac{s_-+s_+}{2 \sqrt{s_-s_+}} \right)
    \right]\,,
\end{split} 
\end{align}
where \mbox{$s_\pm = (\mLb \pm \mpk)^2 -\qsq$}. 
The longitudinally polarised ($\jV=1,\lV=0$) helicity amplitudes for natural parity states are
\begin{align}
\begin{split}
    H^{V(T)}_{\pm1/2,0} = \hphantom{\pm} N_{\jLambda} \sqrt{\frac{s_+}{\qsq}} \left[
     F_1^{(T)} \left( |\vec{q}| + q^0\sqrt{\frac{s_-}{s_+}} \right)  \right. &
    +F_2^{(T)} |\vec{q}|
    +F_3^{(T)} \left( \frac{\mLb}{\mpk}|\vec{q}| \right)  \\ & \left. 
    +F_4^{(T)} \left( |\vec{q}| + q^0\frac{s_-+s_+}{2 \sqrt{s_-s_+}} \right)
    \right]\,,
\end{split} 
\end{align} 
\begin{align}
\begin{split} 
    H^{A(T5)}_{\pm1/2,0} = \pm N_{\jLambda} \sqrt{\frac{s_-}{\qsq}} \left[
     G_1^{(T)} \left( |\vec{q}| + q^0\sqrt{\frac{s_+}{s_-}} \right) \right. & 
    -G_2^{(T)} |\vec{q}|
    -G_3^{(T)} \left( \frac{\mLb}{\mpk}|\vec{q}| \right)  \\ & \left. 
    -G_4^{(T)} \left( |\vec{q}| + q^0\frac{s_-+s_+}{2 \sqrt{s_-s_+}} \right)
    \right]\,.
\end{split} 
\end{align}
The transverse ($\jV=1,\lV=\pm1$) polarised helicity amplitudes for natural parity states are
\begin{align}
    H^{V(T)}_{\pm1/2,\pm1} &= - N_{\jLambda} \sqrt{2s_-}
    \left[ F_1^{(T)} - F_4^{(T)} \frac{\mLb\mpk}{s_-} \right] \,,\\
    H^{A(T5)}_{\pm1/2,\pm1} &= \mp N_{\jLambda} \sqrt{2s_+}
    \left[ G_1^{(T)} - G_4^{(T)} \frac{\mLb \mpk }{s_+} \right] \,,\\
    H^{V(T)}_{\pm3/2,\pm1} &= \hphantom{\mp} N_{\jLambda} N_{\jLambda}^{3/2} \frac{\mLb\mpk}{\sqrt{s_-}} F_4^{(T)}\,, \\
    H^{A(T5)}_{\pm3/2,\pm1} &= \mp N_{\jLambda} N_{\jLambda}^{3/2} \frac{\mLb\mpk}{\sqrt{s_+}} G_4^{(T)}\,.
\end{align}
These share the same structure for all \Lstar-resonance spins and only differ by spin-dependent normalisation factors $N_{3/2}^{3/2} = \sqrt{6}$, $N_{5/2}^{3/2} = 2$, and 
\begin{align}
    N_{1/2} = 1 \quad,\quad 
    N_{3/2} = \left(\frac{\sqrt{s_-s_+}}{\mLb\mpk}\right)\frac{1}{\sqrt{6}} \quad,\quad 
    N_{5/2} = \left(\frac{\sqrt{s_-s_+}}{\mLb\mpk}\right)^2 \frac{1}{2\sqrt{10}} \,.
\end{align}

The amplitudes for unnatural parity states are obtained by swapping vector and axialvector (or tensor and axialtensor) amplitudes and swapping \mbox{$F_i^{(T)}\longleftrightarrow G_i^{(T)}$}. For example, the time-like amplitudes for unnatural parity are
\begin{align}
\begin{split} 
    H^{A(T5)}_{\pm1/2,t} = \hphantom{\pm} N_J \sqrt{\frac{s_+}{\qsq}} \left[
     G_1^{(T)} \left( |\vec{q}|\sqrt{\frac{s_-}{s_+}} + q^0 \right) \right. & 
    +G_2^{(T)} q^0
    +G_3^{(T)} \frac{\mLb q^0-\qsq}{\mpk}  \\ & \left. 
    +G_4^{(T)} \left( q^0 + |\vec{q}|\frac{s_-+s_+}{2\sqrt{s_-s_+}} \right)
    \right] \,,\\
\end{split} 
\end{align} 
\begin{align}
\begin{split}
    H^{V(T)}_{\pm1/2,t} = \pm N_J \sqrt{\frac{s_-}{\qsq}} \left[
     F_1^{(T)} \left( |\vec{q}|\sqrt{\frac{s_+}{s_-}} + q^0 \right) \right. & 
    -F_2^{(T)} q^0
    -F_3^{(T)} \frac{\mLb q^0-\qsq}{\mpk}  \\ & \left. 
    -F_4^{(T)} \left( q^0 + |\vec{q}|\frac{s_-+s_+}{2 \sqrt{s_-s_+}} \right)
    \right]\,.
\end{split} 
\end{align}

\subsection[Helicity amplitudes for the \texorpdfstring{\Lstar}{Lstar}-resonance decay]{Helicity amplitudes for the \texorpdfstring{\boldmath\Lstar}{Lstar}-resonance decay}
\label{sec:helamp:Lstar}

The helicity amplitudes for a natural-partity \Lstar resonance decay are given by
\begin{align}
    \hHad{\lLambda}{\lProton}(\mpk) = \frac{g}{(\mpk^2-\mLambda^2) - i\mpk\Gamma(\mpk;\vec{k}_1^\Lstar,\vec{k}_1)}
    \bar{u}(k_1,\lProton)\gamma_5 U(k,\lLambda)~,
\end{align}
and for unnatural parity states by
\begin{align}
    \hHad{\lLambda}{\lProton}(\mpk) = -\frac{g}{(\mpk^2-\mLambda^2) - i\mpk\Gamma(\mpk;\vec{k}_1^\Lstar,\vec{k}_1)} \bar{u}(k_1,\lProton) U(k,\lLambda)~, 
\end{align}
where a Breit-Wigner line shape is used to model the \mpk dependence, \mLambda is the pole-mass of the resonance, and $g$ is the strong coupling constant.  
The spinor objects respresenting the \Lstar resonance are 
\begin{align}
    U(k,\lLambda) = \begin{cases}
    u(k,\lLambda) &,\jLambda=\jOne \\
    k_1^\mu u_\mu(k,\lLambda) &,\jLambda=\jThree \\
    k_1^\mu k_1^\nu u_{\mu\nu}(k,\lLambda) &,\jLambda=\jFive
    \end{cases}\,,
\end{align}
and $\Gamma$ is the mass dependent width
\begin{align}\label{eq:massdependent-width}
    \Gamma(\mpk;\vec{k}_1^\Lstar,\vec{k}_1) = \gLambda \frac{\mLambda}{\mpk} \left(B_{l}(\vec{k}_1,\vec{k}_1^\Lstar)\right)^2\left(\frac{|\vec{k}_1|}{|\vec{k}_1^\Lstar|}\right)^{2l + 1} \,.
\end{align}
Here, \gLambda is the total width of the \Lstar resonance, $l$ is the orbital angular-momentum in the proton-kaon system, $B$ is a Blatt-Weisskopf form factor~\cite{Blatt:1952ije} and $\vec{k}_1^\Lstar$ is the momentum of the proton evaluated at the \Lstar-resonance pole-mass.
An explicit form of the Blatt-Weisskopf form-factors is given in Appendix~\ref{app:BW_FF}.
Our definition of the width is consistent with Ref.~\cite{Descotes-Genon:2019bud} and fulfils the narrow-width approximation
\begin{align}
    \int_0^\infty\frac{1}{(\mLambda^2-\mpk^2) - i\mpk\Gamma(\mpk;\vec{k}_1^\Lstar,\vec{k}_1)}\deriv\mpk = \frac{\pi}{\mLambda\gLambda} \ .
\end{align}
The strong coupling constant can be expressed in terms of the partial width as~\cite{Carbone:1969fx}
\begin{align}\label{eq:strongcoupling}
    \frac{g^2}{4\pi} &= \frac{(2\jLambda)!}{2^{\jLambda-\frac{1}{2}}[(\jLambda-\frac{1}{2})!]^2}\frac{\mpk}{(k_1^0\mp m_p)|\vec{k}_1|^{2\jLambda}}\Gamma_\text{partial} \ ,
\end{align}
where the $-$($+$) in the denominator appears for a natural (unnatural) parity state and the partial width to $\proton\Km$ can be expressed in terms of the mass-dependent decay width, $\Gamma(\mpk;\vec{k}_1^\Lstar,\vec{k}_1)$, and the branching fraction, $\mathcal{B}_\Lambda$, of the \Lstar resonance decay to \pk
\begin{align}
\Gamma_{\text{partial}} =  \Gamma(\mpk;\vec{k}_1^\Lstar,\vec{k}_1)\,\mathcal{B}_{\Lstar}~.
\end{align} 
Inserting the strong coupling constant and the spinor product into the amplitude yields
\begin{align}
    \hHad{\lLambda}{+1/2}(\mpk) = (-1)^{\jLambda-\frac{1}{2}} 
   \sqrt{8\pi}\sqrt{\frac{\mpk\mLambda}{|\vec{k}_1|}}\frac{\sqrt{\Gamma_\text{partial}}}{(\mpk^2-\mLambda^2) - i\mpk\Gamma(\mpk;\vec{k}_1^\Lstar,\vec{k}_1)} \,.
      \label{eq:amp:replacing:g}
\end{align}

The momentum $\vec{k}_1^\Lstar$ is not defined for resonances with a pole mass below the \pk threshold.
A common solution is to replace the Breit-Wigner shape by a Flatt\'e model (see for example the description of the \Lnum{1405} state by the LHCb collaboration in Refs.~\cite{LHCb:2015yax,LHCb:2022ouv}).
In the Flatt\'e model, the total width is expressed as the sum of partial widths for the decays $\Lstar\to\Sigma^+\pi^-$ and $\Lstar\to\pk$.
Identical widths are assumed for both decays, up-to phase-space factors. 
When evaluating the partial widths, $\vec{k}_1^\Lstar$ is replaced by the momentum of the $\pim$ at the pole-mass of the resonance.
Equation~\ref{eq:amp:replacing:g} only includes the contribution from the decay to $\proton\Km$ in $\Gamma_\text{partial}$.
This approach is also used for the \Lnum{1405} state in this work. 

The amplitude with the opposite proton helicity, $\hHad{\lLambda}{-1/2}(\mpk)$, is given by parity conservation
\begin{equation}\label{eq:hadparity}
    \hHad{\lLambda}{-1/2}(\mpk) = \pLambda(-1)^{l+1}\hHad{\lLambda}{1/2}(\mpk) \,,
\end{equation}
where \pLambda is the parity of the \Lstar resonance.
\subsection{Helicity amplitudes for the leptonic current}

The lepton amplitudes are the projections of the lepton currents onto polarization vectors $\varepsilon_\mu$ and have the general form 
\begin{equation}
    \hLep{}{\jV}{\lLepp}{\lLepn} = \varepsilon_\mu(\lLepp-\lLepn)\Bar{u}(q_2,\lLepn)\Gamma^\mu v(q_1,\lLepp) \ ,
\end{equation}
where $\lLepp-\lLepn=0$ with $\jV=0$ ($\jV=1$) corresponds to the time-like (longitudinal) polarization.
Explicit expressions for the polarisation vectors are given in Appendix~\ref{app:spinors}. 
The lepton amplitudes are calculated in the positively-charged lepton helicity-frame.
There are two relevant Lorentz structures, corresponding to vector ($\Gamma^\mu = \gamma^{\mu}$) and axialvector ($\Gamma^\mu = \gamma^{\mu}\gamma_5$) currents. 
The vector current appears with Wilson coefficients \WC{7^{(\prime)}} and \WC{9^{(\prime)}} and the axialvector current with \WC{{10}^{(\prime)}}. 
Inserting the Lorentz structures into the amplitudes yields
\begin{equation}\label{eq:lepton_amp}
\begin{aligned}
    & \hLep{V}{0}{+1/2}{+1/2} && = 0\,,\qquad &&  \hLep{A}{0}{+1/2}{+1/2} = 2 m_\ell\,, \\
    & \hLep{V}{1}{+1/2}{+1/2}  && = 2 m_\ell\,,\qquad && \hLep{A}{1}{+1/2}{+1/2}  =  0\,, \\
    & \hLep{V}{1}{+1/2}{-1/2}  && = -\sqrt{2 \qsq}\,,\qquad && \hLep{A}{1}{+1/2}{-1/2}  = \sqrt{2 \qsq} \beta_{\ell}~, \\
    & \hLep{V}{\jV}{-\lLepp}{-\lLepn}  && = -\hLep{V}{\jV}{+\lLepp}{+\lLepn}\,,\qquad && \hLep{A}{\jV}{-\lLepp}{-\lLepn} \ \ = \hLep{A}{\jV}{+\lLepp}{+\lLepn}\,,  
\end{aligned}
\end{equation}
where $V$ and $A$ refer to (axial)vector and $\beta_\ell$ is the lepton velocity in the dilepton rest frame, \ie
\begin{equation}
\beta_{\ell} = \frac{|\vec{q}_1|}{q_1^0} =  \sqrt{ 1 - \frac{4 m_{\ell}^{2}}{\qsq} }\ .
\end{equation}

\section{Angular distribution}
\label{sec:angular:distribution}

Expanding the expression for the differential decay rate and performing sums over all of the relevant helicities and \Lstar resonances up-to $\jLambda = \jFive$ yields
\begin{align}
\begin{split}
   \frac{32\pi^2}{3}\frac{\deriv^{7}\Gamma}{\deriv\qsq\,\deriv \mpk\,\deriv\angles} = &\sum_{i=1}^{178}K_i(\qsq,\mpk)f_i(\angles) \,.
   \end{split}
   \label{eq:angularStructure}
\end{align}
The $K_{i}$ are bilinear combinations of products of the amplitudes for the \Lb and \Lstar decays and will be examined in more detail in Sec.~\ref{sec:observables}.
We simplify the expansion of the differential decay rate by noting that the helicity of the \Lstar resonance can take any value within $|\lLambda|\leq\jLambda$ and that the helicity combinations are constrained by $|\lLambda-\lV|=J_{\Lb}=\jOne$.
As a result, only helicities of $\pm\tfrac{1}{2}$ and $\pm\tfrac{3}{2}$ are allowed for the \Lstar resonance, regardless of its spin.

There is no unique choice of basis for the functions $f_i(\angles)$. 
In this paper, we choose to group terms using orthogonal functions. 
The expansion of the differential decay rate involves products of Wigner-D matrices, 
\begin{align} 
D^{\jV}_{\lV,\lLepp-\lLepn}(\phil,\tl,-\phil)^* 
D^{\jV'}_{\lV',\lLepp-\lLepn}(\phil,\tl,-\phil) 
D^{\jLambda}_{\lLambda,\lProton}(\phip,\tp,-\phip)^*
D^{\jLambda'}_{\lLambda',\lProton}(\phip,\tp,-\phip)\,,
\end{align}
which can be written in terms of products of associated Legendre polynomials~\cite{Beaujean:2015xea,Gratrex:2015hna}.

In the unpolarised case our angular basis functions are
\begin{align}\label{eq:basisfunc_nopol}
\begin{split} 
    f(\angles;l_\text{lep},l_\text{had},|m|) &= \sqrt{\frac{8}{3}}n^m_{l_\text{lep}}n^m_{l_\text{had}}P_{l_\text{lep}}^{|m|}(\cos\tl) P_{l_\text{had}}^{|m|}(\cos\tp) \\
    &\times\begin{cases} 
\sin ( |m|(\phil + \phip) ) & m < 0 \\
\frac{1}{\sqrt{2}} & m=0 \\ 
\cos ( |m|( \phil + \phip ) ) & m > 0\end{cases} \,,
\end{split}
\end{align}
where $P_l^m(\cos\theta)$ are the associated Legendre polynomials with the normalisation
\begin{align}
    n^m_l = \sqrt{\frac{(2l+1)(l-m)!}{2(l+m)!}} \ ,
\end{align}
$l_\text{lep}$ is in the range 0 to 2\jV (\ie 0 to 2), $l_\text{had}$ is in the range $0$ to $2\jLambda$, $|m| \leq l_\text{had}$ and $|m| \leq l_\text{lep}$.
This results in 46 different angular basis functions that are independent of the angle \thetab and are either independent of \phil and \phip or depend only on the angle between the \Lstar-resonance and dilepton-system decay-planes, $\phi = \phil + \phip$.
The angular functions arising in the unpolarised case are given in Table~\ref{tab:legendre:basis}.
In order to reduce the number of arguments, the basis functions are labelled $f_i(\angles)$ with an index ranging from $i=$1--46.

In the polarised case, there are 46 additional terms that are proportional to $\cos\thetab$ but otherwise have the same dependence on the remaining angles as the 46 terms appearing in the unpolarised case.
The $\cos\thetab$ dependent basis functions can be obtained by multiplying the unpolarised ones in Eq.~\eqref{eq:basisfunc_nopol} with $\sqrt{3}\cos\thetab$ and the corresponding observables by $\tfrac{1}{\sqrt{3}}\pLb$.
An additional 86 terms also arise proportional to $\sin\thetab$. 
The dependence on \thetab is evident from the structure of the \Lb spin-density matrix, see Equation~\eqref{eq:spindensity}, arising from the rotation from the initial frame to the \Lstar helicity frame. 
The $\sin\thetab$ dependent terms are accompanied by basis functions
\begin{align}\label{eq:basicfunc_pol}
\begin{split}
f(\angles;l_\text{lep},l_\text{had},m_\text{lep},m_\text{had}) &= 2n^{m_\text{lep}}_{l_\text{lep}}n^{m_\text{had}}_{l_\text{had}}P_{l_\text{lep}}^{|m_\text{lep}|}(\cos\tl) P_{l_\text{had}}^{|m_\text{had}|}(\cos\tp) \\
&\times
\begin{cases} 
\sin ( |m_\text{lep}| \phil + |m_\text{had}| \phip ) & m_\text{lep} \leq 0 \text{ and } m_\text{had} \leq 0 \\
\cos ( |m_\text{lep}| \phil + |m_\text{had}| \phip ) & m_\text{lep} \geq 0 \text{ and } m_\text{had} \geq 0 \end{cases}~,
\end{split}
\end{align}
where $|m_\text{lep} - m_\text{had}| = 1$.
The angular terms proportional to $\sin\thetab$ are given in Table~\ref{tab:orthonormalBasisNew}.
The origin of the numerical factors appearing in Eq.~\eqref{eq:basisfunc_nopol}, Eq.~\eqref{eq:basicfunc_pol} and the $\sqrt{3}$ in front of the $\cos\thetab$ terms is discussed in Sec.~\ref{sec:method-of-moments}.

\begin{table}[htb]
\caption{
Orthogonal basis functions for the angular terms $f_1(\angles)$--$f_{46}(\angles)$ that arise in the unpolarised case, where $P_l^m(\cos\theta)$ are associated Legendre polynomials and $\phi=\phip+\phil$. 
}
\centering
\begin{tabular}{cr|cr}
\toprule
$i$  &  $f_i(\angles)$  &  $i$  & $f_i(\angles)$  \\ 
\midrule
   1 & $\tfrac{1}{\sqrt{3}}P_0^0(\cos\tp)P_0^0(\cos\tl)\hphantom{\cos \phi}$
& 24 & $\tfrac{1}{2} \sqrt{\tfrac{7}{3}} P_3^1(\cos\tp) P_1^1(\cos\tl) \cos \phi\hphantom{2}$ \\
   2 & $P_0^0(\cos\tp)P_1^0(\cos\tl)\hphantom{\cos \phi}$
& 25 & $\tfrac{1}{2} P_4^1(\cos\tp) P_2^1(\cos\tl) \cos \phi\hphantom{2}$ \\
   3 & $\sqrt{\tfrac{5}{3}} P_0^0(\cos\tp)P_2^0(\cos\tl)\hphantom{\cos \phi}$
& 26 & $\tfrac{3}{2 \sqrt{5}} P_4^1(\cos\tp) P_1^1(\cos\tl) \cos \phi\hphantom{2}$ \\
   4 & $P_1^0(\cos\tp)P_0^0(\cos\tl)\hphantom{\cos \phi}$
& 27 & $\tfrac{1}{3} \sqrt{\tfrac{11}{6}} P_5^1(\cos\tp) P_2^1(\cos\tl) \cos \phi\hphantom{2}$ \\
   5 & $\sqrt{3} P_1^0(\cos\tp) P_1^0(\cos\tl)\hphantom{\cos \phi}$
& 28 & $\sqrt{\tfrac{11}{30}} P_5^1(\cos\tp) P_1^1(\cos\tl) \cos \phi\hphantom{2}$ \\
   6 & $\sqrt{5} P_1^0(\cos\tp) P_2^0(\cos\tl)\hphantom{\cos \phi}$
& 29 & $\sqrt{\tfrac{5}{6}} P_1^1(\cos\tp) P_2^1(\cos\tl) \sin \phi\hphantom{2}$ \\
   7 & $\sqrt{\tfrac{5}{3}} P_2^0(\cos\tp)P_0^0(\cos\tl)\hphantom{\cos \phi}$
& 30 & $\sqrt{\tfrac{3}{2}} P_1^1(\cos\tp) P_1^1(\cos\tl) \sin \phi\hphantom{2}$ \\
   8 & $\sqrt{5} P_2^0(\cos\tp) P_1^0(\cos\tl)\hphantom{\cos \phi}$
& 31 & $\tfrac{5}{3 \sqrt{6}} P_2^1(\cos\tp) P_2^1(\cos\tl) \sin \phi\hphantom{2}$ \\
   9 & $\tfrac{5}{\sqrt{3}} P_2^0(\cos\tp) P_2^0(\cos\tl)\hphantom{\cos \phi}$
& 32 & $\sqrt{\tfrac{5}{6}} P_2^1(\cos\tp) P_1^1(\cos\tl) \sin \phi\hphantom{2}$ \\
  10 & $\sqrt{\tfrac{7}{3}} P_3^0(\cos\tp)P_0^0(\cos\tl)\hphantom{\cos \phi}$
& 33 & $\tfrac{1}{6} \sqrt{\tfrac{35}{3}} P_3^1(\cos\tp) P_2^1(\cos\tl) \sin \phi\hphantom{2}$ \\
  11 & $\sqrt{7} P_3^0(\cos\tp) P_1^0(\cos\tl)\hphantom{\cos \phi}$
& 34 & $\tfrac{1}{2} \sqrt{\tfrac{7}{3}} P_3^1(\cos\tp) P_1^1(\cos\tl) \sin \phi\hphantom{2}$ \\
  12 & $\sqrt{\tfrac{35}{3}} P_3^0(\cos\tp) P_2^0(\cos\tl)\hphantom{\cos \phi}$
& 35 & $\tfrac{1}{2} P_4^1(\cos\tp) P_2^1(\cos\tl) \sin \phi\hphantom{2}$ \\
  13 & $\sqrt{3} P_4^0(\cos\tp)P_0^0(\cos\tl)\hphantom{\cos \phi}$
& 36 & $\tfrac{3}{2 \sqrt{5}} P_4^1(\cos\tp) P_1^1(\cos\tl) \sin \phi\hphantom{2}$ \\
  14 & $3 P_4^0(\cos\tp) P_1^0(\cos\tl)\hphantom{\cos \phi}$
& 37 & $\tfrac{1}{3} \sqrt{\tfrac{11}{6}} P_5^1(\cos\tp) P_2^1(\cos\tl) \sin \phi\hphantom{2}$ \\
  15 & $\sqrt{15} P_4^0(\cos\tp) P_2^0(\cos\tl)\hphantom{\cos \phi}$
& 38 & $\sqrt{\tfrac{11}{30}} P_5^1(\cos\tp) P_1^1(\cos\tl) \sin \phi\hphantom{2}$ \\
  16 & $\sqrt{\tfrac{11}{3}} P_5^0(\cos\tp)P_0^0(\cos\tl)\hphantom{\cos \phi}$
& 39 & $\tfrac{5}{12 \sqrt{6}} P_2^2(\cos\tp) P_2^2(\cos\tl) \cos 2\phi$ \\
  17 & $\sqrt{11} P_5^0(\cos\tp) P_1^0(\cos\tl)\hphantom{\cos \phi}$
& 40 & $\tfrac{1}{12} \sqrt{\tfrac{7}{6}} P_3^2(\cos\tp) P_2^2(\cos\tl) \cos 2\phi$ \\
  18 & $\sqrt{\tfrac{55}{3}} P_5^0(\cos\tp) P_2^0(\cos\tl)\hphantom{\cos \phi}$
& 41 & $\tfrac{1}{12 \sqrt{2}} P_4^2(\cos\tp) P_2^2(\cos\tl) \cos 2\phi$ \\
  19 & $\sqrt{\tfrac{5}{6}} P_1^1(\cos\tp) P_2^1(\cos\tl) \cos \phi$
& 42 & $\tfrac{1}{12} \sqrt{\tfrac{11}{42}} P_5^2(\cos\tp) P_2^2(\cos\tl) \cos 2\phi$ \\
  20 & $\sqrt{\tfrac{3}{2}} P_1^1(\cos\tp) P_1^1(\cos\tl) \cos \phi$
& 43 & $\tfrac{5}{12 \sqrt{6}} P_2^2(\cos\tp) P_2^2(\cos\tl) \sin 2\phi$ \\
  21 & $\tfrac{5}{3 \sqrt{6}} P_2^1(\cos\tp) P_2^1(\cos\tl) \cos \phi$
& 44 & $\tfrac{1}{12} \sqrt{\tfrac{7}{6}} P_3^2(\cos\tp) P_2^2(\cos\tl) \sin 2\phi$ \\
  22 & $\sqrt{\tfrac{5}{6}} P_2^1(\cos\tp) P_1^1(\cos\tl) \cos \phi$
& 45 & $\tfrac{1}{12 \sqrt{2}}P_4^2(\cos\tp) P_2^2(\cos\tl) \sin 2\phi$ \\
  23 & $\tfrac{1}{6} \sqrt{\tfrac{35}{3}} P_3^1(\cos\tp) P_2^1(\cos\tl) \cos \phi$
& 46 & $\tfrac{1}{12} \sqrt{\tfrac{11}{42}} P_5^2(\cos\tp) P_2^2(\cos\tl) \sin 2\phi$ \\
\bottomrule
\end{tabular}
\label{tab:legendre:basis}
\end{table}
\begin{table}[htb]
\caption{
Orthogonal basis functions $f_{93}(\Omega)$--$f_{135}(\Omega)$ necessary to describe the angular distribution of polarised \Lb decays with $\cos(|m_\text{had}|\phip+|m_\text{lep}|\phil)$ dependence, where $P_l^m(\cos\theta)$ are associated Legendre polynomials. 
The remaining functions, numbered 136--178 can be obtained by replacing \mbox{$\cos(|m_\text{had}|\phip+|m_\text{lep}|\phil)$} with $\sin(|m_\text{had}|\phip+|m_\text{lep}|\phil)$.
}
\centering
\scalebox{0.9}{
\begin{tabular}{cr|cr}
\toprule
$i$  &  $f_i(\angles)$  &  $i$  & $f_i(\angles)$  \\ 
\midrule
   93 & $\sqrt{\tfrac{3}{2}}   \sin\thetab P_1^1(\cos\tp) P_0^0(\cos\tl) \cos (\phip)$
& 115 & $\sqrt{\tfrac{21}{2}}  \sin\thetab P_3^0(\cos\tp) P_1^1(\cos\tl) \cos (\phil)\hphantom{3+2\phil}$ \\
   94 & $\tfrac{3}{\sqrt{2}}   \sin\thetab P_1^1(\cos\tp) P_1^0(\cos\tl) \cos (\phip)$
& 116 & $\sqrt{\tfrac{15}{2}}  \sin\thetab P_4^0(\cos\tp) P_2^1(\cos\tl) \cos (\phil)\hphantom{3+2\phil}$ \\
   95 & $\sqrt{\tfrac{15}{2}}  \sin\thetab P_1^1(\cos\tp) P_2^0(\cos\tl) \cos (\phip)$
& 117 & $3 \sqrt{\tfrac{3}{2}} \sin\thetab P_4^0(\cos\tp) P_1^1(\cos\tl) \cos (\phil)\hphantom{3+2\phil}$ \\
   96 & $\sqrt{\tfrac{5}{6}}   \sin\thetab P_2^1(\cos\tp) P_0^0(\cos\tl) \cos (\phip)$
& 118 & $\sqrt{\tfrac{55}{6}}  \sin\thetab P_5^0(\cos\tp) P_2^1(\cos\tl) \cos (\phil)\hphantom{3+2\phil}$ \\
   97 & $\sqrt{\tfrac{5}{2}}   \sin\thetab P_2^1(\cos\tp) P_1^0(\cos\tl) \cos (\phip)$
& 119 & $\sqrt{\tfrac{33}{2}}  \sin\thetab P_5^0(\cos\tp) P_1^1(\cos\tl) \cos (\phil)\hphantom{3+2\phil}$ \\
   98 & $\tfrac{5}{\sqrt{6}}   \sin\thetab P_2^1(\cos\tp) P_2^0(\cos\tl) \cos (\phip)$
& 120 & $\tfrac{5}{12}         \sin\thetab P_2^2(\cos\tp) P_2^1(\cos\tl) \cos (2 \phip+\phil)\hphantom{3}$ \\
   99 & $\tfrac{1}{2} \sqrt{\tfrac{7}{3}}  \sin\thetab P_3^1(\cos\tp) P_0^0(\cos\tl) \cos (\phip)$
& 121 & $\tfrac{1}{4} \sqrt{5}            \sin\thetab P_2^2(\cos\tp) P_1^1(\cos\tl) \cos (2 \phip+\phil)\hphantom{3}$ \\
  100 & $\tfrac{1}{2} \sqrt{7}            \sin\thetab P_3^1(\cos\tp) P_1^0(\cos\tl) \cos (\phip)$
& 122 & $\tfrac{1}{12} \sqrt{7}           \sin\thetab P_3^2(\cos\tp) P_2^1(\cos\tl) \cos (2 \phip+\phil)\hphantom{3}$ \\
  101 & $\tfrac{1}{2} \sqrt{\tfrac{35}{3}} \sin\thetab P_3^1(\cos\tp) P_2^0(\cos\tl) \cos (\phip)$
& 123 & $\tfrac{1}{4} \sqrt{\tfrac{7}{5}}  \sin\thetab P_3^2(\cos\tp) P_1^1(\cos\tl) \cos (2 \phip+\phil)\hphantom{3}$ \\
  102 & $\tfrac{3}{2 \sqrt{5}}            \sin\thetab P_4^1(\cos\tp) P_0^0(\cos\tl) \cos (\phip)$
& 124 & $\tfrac{1}{4 \sqrt{3}}            \sin\thetab P_4^2(\cos\tp) P_2^1(\cos\tl) \cos (2 \phip+\phil)\hphantom{3}$ \\
  103 & $\tfrac{3}{2} \sqrt{\tfrac{3}{5}}  \sin\thetab P_4^1(\cos\tp) P_1^0(\cos\tl) \cos (\phip)$
& 125 & $\tfrac{1}{4} \sqrt{\tfrac{3}{5}}  \sin\thetab P_4^2(\cos\tp) P_1^1(\cos\tl) \cos (2 \phip+\phil)\hphantom{3}$ \\
  104 & $\tfrac{3}{2}                     \sin\thetab P_4^1(\cos\tp) P_2^0(\cos\tl) \cos (\phip)$
& 126 & $\tfrac{1}{12} \sqrt{\tfrac{11}{7}} \sin\thetab P_5^2(\cos\tp) P_2^1(\cos\tl) \cos (2 \phip+\phil)\hphantom{3}$ \\
  105 & $\sqrt{\tfrac{11}{30}}             \sin\thetab P_5^1(\cos\tp) P_0^0(\cos\tl) \cos (\phip)$
& 127 & $\tfrac{1}{4} \sqrt{\tfrac{11}{35}} \sin\thetab P_5^2(\cos\tp) P_1^1(\cos\tl) \cos (2 \phip+\phil)\hphantom{3}$ \\
  106 & $\sqrt{\tfrac{11}{10}} \sin\thetab P_5^1(\cos\tp) P_1^0(\cos\tl) \cos (\phip)$
& 128 & $\tfrac{1}{4} \sqrt{5} \sin\thetab P_1^1(\cos\tp) P_2^2(\cos\tl) \cos (\phip+2 \phil)\hphantom{3}$ \\
  107 & $\sqrt{\tfrac{11}{6}}  \sin\thetab P_5^1(\cos\tp) P_2^0(\cos\tl) \cos (\phip)$
& 129 & $\tfrac{5}{12}         \sin\thetab P_2^1(\cos\tp) P_2^2(\cos\tl) \cos (\phip+2 \phil)\hphantom{3}$ \\
  108 & $\sqrt{\tfrac{5}{6}}   \sin\thetab P_0^0(\cos\tp) P_2^1(\cos\tl) \cos (\phil)$
& 130 & $\tfrac{1}{12} \sqrt{\tfrac{35}{2}} \sin\thetab P_3^1(\cos\tp) P_2^2(\cos\tl) \cos (\phip+2 \phil)\hphantom{3}$ \\
  109 & $\sqrt{\tfrac{3}{2}}               \sin\thetab P_0^0(\cos\tp) P_1^1(\cos\tl) \cos (\phil)$
& 131 & $\tfrac{1}{4} \sqrt{\tfrac{3}{2}}   \sin\thetab P_4^1(\cos\tp) P_2^2(\cos\tl) \cos (\phip+2 \phil)\hphantom{3}$ \\
  110 & $\sqrt{\tfrac{5}{2}}               \sin\thetab P_1^0(\cos\tp) P_2^1(\cos\tl) \cos (\phil)$
& 132 & $\tfrac{1}{12} \sqrt{11}           \sin\thetab P_5^1(\cos\tp) P_2^2(\cos\tl) \cos (\phip+2 \phil)\hphantom{3}$ \\
  111 & $\tfrac{3}{\sqrt{2}}               \sin\thetab P_1^0(\cos\tp) P_1^1(\cos\tl) \cos (\phil)$
& 133 & $\tfrac{1}{24} \sqrt{\tfrac{7}{6}}  \sin\thetab P_3^3(\cos\tp) P_2^2(\cos\tl) \cos (3 \phip+2 \phil)$ \\
  112 & $\tfrac{5}{\sqrt{6}}               \sin\thetab P_2^0(\cos\tp) P_2^1(\cos\tl) \cos (\phil)$
& 134 & $\tfrac{1}{8 \sqrt{42}}            \sin\thetab P_4^3(\cos\tp) P_2^2(\cos\tl) \cos (3 \phip+2 \phil)$ \\
  113 & $\sqrt{\tfrac{15}{2}}              \sin\thetab P_2^0(\cos\tp) P_1^1(\cos\tl) \cos (\phil)$
& 135 & $\tfrac{1}{48} \sqrt{\tfrac{11}{42}} \sin\thetab P_5^3(\cos\tp) P_2^2(\cos\tl) \cos (3 \phip+2 \phil)$ \\
  114 & $\sqrt{\tfrac{35}{6}}              \sin\thetab P_3^0(\cos\tp) P_2^1(\cos\tl) \cos (\phil)$
&  &  \\
\bottomrule
\end{tabular}}
\label{tab:orthonormalBasisNew}
\end{table}

The lepton and hadron sides of the decay are fully independent of each other and can generally be considered separately. 
Integrating over all of the angles except for \tp yields 
\begin{align}
\label{eq:thetaB1D}
\begin{split}
    \frac{\deriv^3\Gamma}{\deriv\qsq\,\deriv \mpk\,\deriv\!\cos\tp}=
    &\ \frac{\sqrt{3}}{2}K_1 - \frac{\sqrt{15}}{4}K_7 + 9\frac{\sqrt{3}}{16}K_{13} \\
    &+\left(
    \frac{3}{2}K_4 - 3\frac{\sqrt{21}}{4}K_{10} + 15\frac{\sqrt{33}}{16}K_{16}
    \right)\cos\tp \\
    &+\left(
    3\frac{\sqrt{15}}{4}K_7 - 45\frac{\sqrt{3}}{8}K_{13}
    \right)\cos^2\tp \\
    &+\left(
    5\frac{\sqrt{21}}{4}K_{10} - 35\frac{\sqrt{33}}{8}K_{16}
    \right)\cos^3\tp \\
    &+\frac{105\sqrt{3}}{16}K_{13}\cos^4\tp+\frac{63\sqrt{33}}{16}K_{16}\cos^5\tp \,.
\end{split}
\end{align}
The higher powers of $\cos\tp$ are associated with higher spin combinations. 
Each state individually contributes to powers of $\cos\tp$ up-to $2\jLambda$. 
If interfering resonances have spins \jLambda and $\jLambda^\prime$, their interference can contribute to powers up-to $\jLambda+\jLambda^\prime$. 
The odd powers of $\cos\tp$ result from interference between states with different parities. 
Integrating over all of the angles except \tl instead yields 
\begin{align}\label{eq:angdist_lep}
\frac{\deriv^3\Gamma}{\deriv\qsq\,\deriv \mpk\,\deriv\!\cos\tl}= \frac{\sqrt{3}}{2}K_1 +  \frac{3}{2}K_2\cos\tl + \frac{\sqrt{15}}{4}K_3 ( 3\cos^2\tl - 1 )   \ .
\end{align}
The observable $K_{2}$ generates the lepton-side forward-backward asymmetry that is a feature of \bsll transitions that arises from interference between the vector and axialvector leptonic currents~\cite{Ali:1991is}.

\section{Method of moments}
\label{sec:method-of-moments}

The coefficients $K_i$ can be determined experimentally using the method-of-moments~\cite{Beaujean:2015xea}, with a set of weighting functions, $w_i(\Omega)$, that are orthogonal to the angular terms, \ie 
\begin{align}\label{eq:normalisation}
    \int f_i(\angles)w_j(\angles)\deriv{\angles} = \frac{32\pi^2}{3}\delta_{ij} \,.
\end{align}
The moments can be extracted using an integral over the differential decay rate,
\begin{align}
\int w_i({\angles}) \frac{\deriv^{7}\Gamma}{\deriv\qsq\,\deriv{\mpk}\,\deriv{\angles}}\deriv{\angles} &= \frac{3}{32\pi^2} \int w_i({\angles}) \sum_j K_j(\mpk,\qsq) f_j(\angles)\deriv{\angles} = K_i(\mpk,\qsq) \ .
\end{align}
The moments corresponding to the basis functions proportional to $\cos\thetab$, with indices \mbox{47--92}, can be obtained from the moments for unpolarized \Lb baryons, with indices 1--46, by multiplication with $\tfrac{1}{\sqrt{3}}\pLb$, due to the structure of the spin density matrix, see Eq.~\eqref{eq:spindensity}, and choosing orthogonal basis functions with the normalisation given in Eq.~\eqref{eq:normalisation}.
As a result of choosing $f_1(\angles)=\tfrac{1}{\sqrt{3}}$, the first moment corresponds to the decay rate
\begin{align}
    \frac{\deriv^{2}\Gamma}{\deriv\qsq\,\deriv \mpk} = \sqrt{3}K_1 \ .
\end{align}
It is convenient to define a set of angular observables that are independent of the total rate
\begin{align} \label{eq:observable}
\begin{split} 
\Kobs{i} &=  K_{i} \Big/ \frac{\deriv^{2} \Gamma}{\deriv\qsq\deriv\mpk} =  
\frac{K_{i}}{\sqrt{3} K_{1}}\,. 
\end{split} 
\end{align} 
In a sample with $N_\text{data}$ data points in a given bin of \mpk and \qsq, the average value of \Kobs{i},
\begin{align}
\big\langle\Kobs{i}\big\rangle_\text{bin} = \int_{\text{bin}} K_{i} \deriv\qsq\deriv \mpk \Big/ 
\int_{\text{bin}} \sqrt{3}  K_{1} \deriv\qsq\deriv \mpk\,,
\end{align} 
can be obtained using Mont\'e-Carlo integration as 
\begin{align}
    \big\langle\Kobs{i}\big\rangle^\text{data}_\text{bin} = \frac{1}{N_\text{data}} \sum_{n=1}^{N_\text{data}} w_i(\angles_n)\,.
\end{align}
The method of moments guarantees a Gaussian likelihood for the observables regardless of sample size.
Because of our choice of basis functions, $f_i(\angles)$, the weighting functions are simply \mbox{$w_i(\angles) = f_i(\angles)$}. 
The numerical factors appearing in the basis functions in Sec.~\ref{sec:angular:distribution} are chosen to ensure that Eq.~\eqref{eq:normalisation} is fulfilled.
Beyond this convenient detail, the use of orthogonal functions has the advantage that the correlations between the different observables are minimised in the measurement of the moments.

\section{Explicit expressions for the angular coefficients}
\label{sec:observables}

The angular coefficients, $K_{i}$, involve bilinear combinations of amplitudes that arise from taking the product of ${\cal M}$ and its complex conjugate. In what follows below, we label the indices appearing in ${\cal M}^{\dagger}$ with primes. 
Allowing for even and odd parity as well as spins up to \jFive results in complex and long expressions. 
The structure of the different coefficients is summarised in Table~\ref{tab:K_description}. 
The left-most columns of the table summarise the appearing state combinations, the right-most  columns give details about the structure of the coefficients and the interfering amplitudes.
The fifth column indicates whether the coefficient takes the real or imaginary part of the amplitude product. 
A check mark in the sixth column indicates that the coefficient arises due to vector and axialvector interference of the leptonic currents. 
The last column explains which helicity combinations contribute to the coefficient.
When a basis function is independent of $\phi$, the two amplitudes in a product have the same helicity combination ($\lV=\lV^\prime,\lLambda=\lLambda^\prime$). Basis functions that depend on $\cos\phi$ or $\cos2\phi$  ($\sin\phi$ or $\sin2\phi$) appear with the real (imaginary) part of a bilinear combination of amplitudes. 

\begin{table}[htbp]
    \centering\small
    \caption{
    Amplitude combinations appearing in the coefficient $K_i$. 
    The parity combination and allowed spins indicate which states interfere. 
    Checkmarks in the three columns labelled \textit{single states} indicate whether the coefficient appears in the single resonance case for spin $\jLambda=\jOne,\jThree$, or $\jFive$.
    Some coefficients take the real part (Re) others the imaginary part (Im) of the amplitude products. 
    A checkmark in the column V/A shows that a coefficient arises from vector-axialvector interference. 
    The right-most column indicates the equation defining the observable $K_i$.}
    \label{tab:K_description}
    \scalebox{0.89}{
    \begin{tabular}{c|ccccc|ccc|c}
    \toprule
        \multirow{2}{*}{$i$} & parity & \multirow{2}{*}{$\jLambda+\jLambda^\prime$} & \multicolumn{3}{c|}{single states} & \multirow{2}{*}{Re/Im} & \multirow{2}{*}{V/A} & \multirow{2}{*}{helicity combinations} & \multirow{2}{*}{Eq.} \\
        & combination & & $1/2$ & $3/2$ & $5/2$ & & & & \\
        \midrule
        1   & same     & $\geq 1$ & \checkmark & \checkmark & \checkmark & Re &            & $\jLambda=\jLambda^\prime,(\lLambda,\lV)=(\lLambda,\lV)^\prime$ & \eqref{eq:K1} \\
        2   & same     & $\geq 1$ & \checkmark & \checkmark & \checkmark & Re & \checkmark & $\jLambda=\jLambda^\prime,\lV\not=0$, $(\lLambda,\lV)=(\lLambda,\lV)^\prime$ & \eqref{eq:K2} \\
        3   & same     & $\geq 1$ & \checkmark & \checkmark & \checkmark & Re &            & $\jLambda=\jLambda^\prime,(\lLambda,\lV)=(\lLambda,\lV)^\prime$ & \eqref{eq:K3} \\
        \midrule
        4   & opposite & $\geq 1$ & & & & Re &            & $(\lLambda,\lV)=(\lLambda,\lV)^\prime$ & \eqref{eq:K4} \\
        5   & opposite & $\geq 1$ & & & & Re & \checkmark & $\lV\not=0$, $(\lLambda,\lV)=(\lLambda,\lV)^\prime$ & \eqref{eq:K5} \\
        6   & opposite & $\geq 1$ & & & & Re &            & $(\lLambda,\lV)=(\lLambda,\lV)^\prime$ & \eqref{eq:K6} \\
        7   & same     & $\geq 2$ & & \checkmark & \checkmark & Re &            & $(\lLambda,\lV)=(\lLambda,\lV)^\prime$ & \eqref{eq:K7} \\
        8   & same     & $\geq 2$ & & \checkmark & \checkmark & Re & \checkmark & $\lV\not=0$, $(\lLambda,\lV)=(\lLambda,\lV)^\prime$ & \eqref{eq:K8} \\
        9   & same     & $\geq 2$ & & \checkmark & \checkmark & Re &            & $(\lLambda,\lV)=(\lLambda,\lV)^\prime$ & \eqref{eq:K9} \\
        10  & opposite & $\geq 3$ & & & & Re &            & $(\lLambda,\lV)=(\lLambda,\lV)^\prime$ & \eqref{eq:K10} \\
        11  & opposite & $\geq 3$ & & & & Re & \checkmark & $\lV\not=0$, $(\lLambda,\lV)=(\lLambda,\lV)^\prime$ & \eqref{eq:K11} \\
        12  & opposite & $\geq 3$ & & & & Re &            & $(\lLambda,\lV)=(\lLambda,\lV)^\prime$ & \eqref{eq:K12} \\
        13  & same     & $\geq 4$ & & & \checkmark & Re &            & $(\lLambda,\lV)=(\lLambda,\lV)^\prime$ & \eqref{eq:K13} \\
        14  & same     & $\geq 4$ & & & \checkmark & Re & \checkmark & $\lV\not=0$, $(\lLambda,\lV)=(\lLambda,\lV)^\prime$ & \eqref{eq:K14} \\
        15  & same     & $\geq 4$ & & & \checkmark & Re &            & $(\lLambda,\lV)=(\lLambda,\lV)^\prime$ & \eqref{eq:K15} \\
        16  & opposite & $\geq 5$ & & & & Re &            & $(\lLambda,\lV)=(\lLambda,\lV)^\prime$ & \eqref{eq:K16} \\
        17  & opposite & $\geq 5$ & & & & Re & \checkmark & $\lV\not=0$, $(\lLambda,\lV)=(\lLambda,\lV)^\prime$  & \eqref{eq:K17} \\
        18  & opposite & $\geq 5$ & & & & Re &            & $(\lLambda,\lV)=(\lLambda,\lV)^\prime$ & \eqref{eq:K18} \\
        \midrule
        19  & opposite & $\geq 1$ & & & & Re &            & \multirow{10}{*}{$\lV=0,|\lV^\prime|=1$ (all possible $\lLambda^{(\prime)}$)} & \eqref{eq:K19} \\
        20  & opposite & $\geq 1$ & & & & Re & \checkmark & & \eqref{eq:K20} \\
        21  & same     & $\geq 2$ & & \checkmark & \checkmark & Re &            & & \eqref{eq:K21} \\
        22  & same     & $\geq 2$ & & \checkmark & \checkmark & Re & \checkmark & & \eqref{eq:K22} \\
        23  & opposite & $\geq 3$ & & & & Re &            & & \eqref{eq:K23} \\
        24  & opposite & $\geq 3$ & & & & Re & \checkmark & & \eqref{eq:K24} \\
        25  & same     & $\geq 4$ & & & \checkmark & Re &            & & \eqref{eq:K25} \\
        26  & same     & $\geq 4$ & & & \checkmark & Re & \checkmark & & \eqref{eq:K26} \\
        27  & opposite & $\geq 5$ & & & & Re &            & & \eqref{eq:K27} \\
        28  & opposite & $\geq 5$ & & & & Re & \checkmark & & \eqref{eq:K28} \\
        \midrule
        29  & opposite & $\geq 1$ & & & & Im &            & \multirow{10}{*}{$\lV=0,|\lV^\prime|=1$ (all possible $\lLambda^{(\prime)}$)} & \eqref{eq:K29} \\
        30  & opposite & $\geq 1$ & & & & Im & \checkmark & & \eqref{eq:K30} \\
        31  & same     & $\geq 2$ & & \checkmark & \checkmark & Im &            & & \eqref{eq:K31} \\
        32  & same     & $\geq 2$ & & \checkmark & \checkmark & Im & \checkmark & & \eqref{eq:K32} \\
        33  & opposite & $\geq 3$ & & & & Im &            & & \eqref{eq:K33} \\
        34  & opposite & $\geq 3$ & & & & Im & \checkmark & & \eqref{eq:K34} \\
        35  & same     & $\geq 4$ & & & \checkmark & Im &            & & \eqref{eq:K35} \\
        36  & same     & $\geq 4$ & & & \checkmark & Im & \checkmark & & \eqref{eq:K36} \\
        37  & opposite & $\geq 5$ & & & & Im &            & & \eqref{eq:K37} \\
        38  & opposite & $\geq 5$ & & & & Im & \checkmark & & \eqref{eq:K38} \\
        \midrule
        39  & same     & $\geq 2$ & & \checkmark & \checkmark & Re & & \multirow{8}{*}{$|\lV^{(\prime)}|=1$, $\lLambda=\pm1/2,\lLambda^\prime=\mp3/2$} & \eqref{eq:K39} \\
        40  & opposite & $\geq 3$ & & & & Re & & & \eqref{eq:K40} \\
        41  & same     & $\geq 4$ & & & \checkmark & Re & & & \eqref{eq:K41} \\
        42  & opposite & $\geq 5$ & & & & Re & & & \eqref{eq:K42} \\
        43  & same     & $\geq 2$ & & \checkmark & \checkmark & Im & & & \eqref{eq:K43} \\
        44  & opposite & $\geq 3$ & & & & Im & & & \eqref{eq:K44} \\
        45  & same     & $\geq 4$ & & & \checkmark & Im & & & \eqref{eq:K45} \\
        46  & opposite & $\geq 5$ & & & & Im & & & \eqref{eq:K46} \\
        \bottomrule
    \end{tabular}
    }
\end{table}

In order to have a compact notation for the coefficients, the lepton-side helicity amplitudes, \hLep{}{\jV}{\lLepp}{\lLepn}, are inserted under the assumption that $4 m_\ell^{2} \ll \qsq$. 
The hadron-side helicity amplitudes with negative proton helicity, \hHad{\lLambda}{-1/2}, are replaced using the parity conservation requirement given in Equation~\eqref{eq:hadparity}.
To further simplify the expressions we introduce the symbols
\begin{align}
\begin{split}
    &\Amp{V}{\lLambda}{\lV} = N\sum_{\Lstar} \sum_{i=7^{(\prime)},9^{(\prime)}}\HVL{\lLambda}{\lV}{\Op{i}}\hHad{\lLambda}{1/2} \ , \\
    &\Amp{A}{\lLambda}{\lV} = N\sum_{\Lstar} \sum_{i=10^{(\prime)}}\HVL{\lLambda}{\lV}{\Op{i}}\hHad{\lLambda}{1/2} \ ,
\end{split}
\end{align}
where the sum runs over resonances with the quantum numbers $Q$. 
The reader is reminded that the indices need to satisfy $|\lV-\lLambda|=|\lV^\prime-\lLambda^\prime|=\tfrac{1}{2}$ to conserve helicity.
The normalisation coefficient
\begin{equation}
    N = \sqrt{\frac{N_1^2}{\mLb^22^6(2\pi)^7}\frac{|\vec{k}||\vec{k}_1||\vec{q}_1|}{\sqrt{\qsq}}\ 2\qsq}
\end{equation}
contains the phase-space factors, the normalisation of the weak $\bquark\to\squark$ transition and a common factor of $2\qsq$ stemming from the lepton-side amplitudes, \hLep{}{\jV}{\lLepp}{\lLepn}.

The coefficient $K_1$ is proportional to the total decay rate and equals the sum of all helicity amplitudes squared
\begin{align}\label{eq:K1}
K_1 &= \frac{1}{\sqrt{3}}\sum_{Q}\sum_{\lLambda,\lV}\left( 
\left|\Amp{V}{\lLambda}{\lV}\right|^2 + V\longleftrightarrow A \right)\,.
\end{align}
The coefficient
\begin{align}\label{eq:K2}
K_2 &= 
-\sum_{Q}\sum_{\lambda=\pm1}\lambda\cdot
\text{Re}\left[
\Amp{A*}{\frac{3}{2}\lambda}{\lambda}
\Amp{V}{\frac{3}{2}\lambda}{\lambda}
+
\Amp{A*}{\frac{1}{2}\lambda}{\lambda}
\Amp{V}{\frac{1}{2}\lambda}{\lambda}\right]
\end{align}
generates the lepton-side forward-backward asymmetry, $A_{\rm FB}^{\ell} = \tfrac{3}{2}\Kobs{2}$.
The coefficient $K_{3}$ is the asymmetry in the amplitudes squared between the amplitudes with $|\lV| = 1$ and $\lV = 0$
\begin{align}\label{eq:K3}
    K_3 &= \frac{1}{2\sqrt{15}}
\sum_Q\sum_{\lambda=\pm1} \left( 
 \left|
\Amp{V}{\frac{3}{2}\lambda}{\lambda}\right|^2
+\left|\Amp{V}{\frac{1}{2}\lambda}{\lambda}\right|^2
-2\left|\Amp{V}{\frac{1}{2}\lambda}{0}\right|^2\right) + V\longleftrightarrow A  \ .
\end{align}
A unique feature of $K_1$--$K_{3}$ is that they arise purely due to self-interaction terms and interference between states with the same quantum numbers.
As such, the first three coefficients are non-zero even for single resonances regardless of their spin.

Due to parity conservation in the strong decay of the \Lstar resonance, the $\cos\tp$ distribution must be symmetric for spectra where all states have the same parity. 
Once states with different parities can interfere, a hadron system forward-backward asymmetry is introduced with 
\begin{align} 
A_{\rm FB}^{\proton} = \frac{3}{2} \Kobs{4} - \frac{\sqrt{21}}{8} \Kobs{10} + \frac{\sqrt{33}}{16}\Kobs{16}~.
\end{align} 
One of the three contributing coefficients is
\begin{align}\label{eq:K4}
\begin{split}
K_4 = \frac{1}{105}
\sum_{\lambda=\pm1}
\text{Re}\Big[
&+\lambda\left(
+35 \AmpJP{V*}{\frac{1}{2}\lambda}{0      }{\frac{1}{2}}{+} \AmpJP{V}{\frac{1}{2}\lambda}{0      }{\frac{1}{2}}{-}
+35 \AmpJP{V*}{\frac{1}{2}\lambda}{\lambda}{\frac{1}{2}}{+} \AmpJP{V}{\frac{1}{2}\lambda}{\lambda}{\frac{1}{2}}{-}\right. \\
&\left.+21 \AmpJP{V*}{\frac{3}{2}\lambda}{\lambda}{\frac{3}{2}}{+} \AmpJP{V}{\frac{3}{2}\lambda}{\lambda}{\frac{3}{2}}{-}
+ 7 \AmpJP{V*}{\frac{1}{2}\lambda}{0      }{\frac{3}{2}}{+} \AmpJP{V}{\frac{1}{2}\lambda}{0      }{\frac{3}{2}}{-}
+ 7 \AmpJP{V*}{\frac{1}{2}\lambda}{\lambda}{\frac{3}{2}}{+} \AmpJP{V}{\frac{1}{2}\lambda}{\lambda}{\frac{3}{2}}{-} \right. \\
&\left.+ 3 \AmpJP{V*}{\frac{1}{2}\lambda}{0      }{\frac{5}{2}}{+} \AmpJP{V}{\frac{1}{2}\lambda}{0      }{\frac{5}{2}}{-}
+ 3 \AmpJP{V*}{\frac{1}{2}\lambda}{\lambda}{\frac{5}{2}}{+} \AmpJP{V}{\frac{1}{2}\lambda}{\lambda}{\frac{5}{2}}{-}
+ 9 \AmpJP{V*}{\frac{3}{2}\lambda}{\lambda}{\frac{5}{2}}{+} \AmpJP{V}{\frac{3}{2}\lambda}{\lambda}{\frac{5}{2}}{-}
\right) \\
&+ 84 \AmpJP{V*}{\frac{3}{2}\lambda}{\lambda}{\frac{3}{2}}{+} \AmpJP{V}{\frac{3}{2}\lambda}{\lambda}{\frac{5}{2}}{-}
+ 70 \sqrt{2} \AmpJP{V*}{\frac{1}{2}\lambda}{0      }{\frac{1}{2}}{+} \AmpJP{V}{\frac{1}{2}\lambda}{0      }{\frac{3}{2}}{-}
+ 70 \sqrt{2} \AmpJP{V*}{\frac{1}{2}\lambda}{\lambda}{\frac{1}{2}}{+} \AmpJP{V}{\frac{1}{2}\lambda}{\lambda}{\frac{3}{2}}{-} \\
&+ 42 \sqrt{6} \AmpJP{V*}{\frac{1}{2}\lambda}{0      }{\frac{3}{2}}{+} \AmpJP{V}{\frac{1}{2}\lambda}{0      }{\frac{5}{2}}{-}
+ 42 \sqrt{6} \AmpJP{V*}{\frac{1}{2}\lambda}{\lambda}{\frac{3}{2}}{+} \AmpJP{V}{\frac{1}{2}\lambda}{\lambda}{\frac{5}{2}}{-}
\Big] \\
& + \left(V\longleftrightarrow A\right) + (\pLambda\longrightarrow -\pLambda)\,,
\end{split}
\end{align}
which accompanies the basis function $f_4(\angles) = \cos\tp$. 
Note that interference between \jOne and \jFive states does not contribute to $K_{4}$ and many other observables (see Appendix~\ref{app:observables}).
Another illustrative coefficient is 
\begin{align}\label{eq:K32}
    \begin{split}
    K_{32} =
-\frac{1}{7\sqrt{10}}
\sum_{\lambda=\pm1}\text{Im}\Big[
&+ 4 \sqrt{3} \AmpJP{V*}{\frac{1}{2}\lambda}{0}{\frac{5}{2}}{+} \AmpJP{A}{\frac{3}{2}\lambda}{\lambda}{\frac{5}{2}}{+}
+ 7 \sqrt{2} \AmpJP{V*}{\frac{1}{2}\lambda}{0}{\frac{3}{2}}{+} \AmpJP{A}{\frac{3}{2}\lambda}{\lambda}{\frac{3}{2}}{+} \\
&-\lambda\left(
3 \sqrt{3} \AmpJP{V*}{\frac{1}{2}\lambda}{0}{\frac{5}{2}}{+} \AmpJP{A}{\frac{3}{2}\lambda}{\lambda}{\frac{3}{2}}{+} 
+ \sqrt{2}    \AmpJP{V*}{\frac{1}{2}\lambda}{0}{\frac{3}{2}}{+} \AmpJP{A}{\frac{3}{2}\lambda}{\lambda}{\frac{5}{2}}{+}
\right) \\
&+ 5\lambda \left(\AmpJP{V*}{\frac{1}{2}\lambda}{0}{\frac{3}{2}}{+} \AmpJP{A}{-\frac{1}{2}\lambda}{-\lambda}{\frac{5}{2}}{+}
+\left(\frac{3}{2}\longleftrightarrow\frac{5}{2}\right)\right) \\
&+ 7 \sqrt{2} \left(
\AmpJP{V*}{\frac{1}{2}\lambda}{0}{\frac{1}{2}}{+} \AmpJP{A}{-\frac{1}{2}\lambda}{-\lambda}{\frac{5}{2}}{+}
+ \sqrt{2} \AmpJP{V*}{\frac{1}{2}\lambda}{0}{\frac{1}{2}}{+} \AmpJP{A}{\frac{3}{2}\lambda}{\lambda}{\frac{5}{2}}{+}
-\left(\frac{5}{2}\longleftrightarrow\frac{1}{2}\right)
\right) \\
&\left.+ 7\lambda  \left(
\sqrt{3}\AmpJP{V*}{\frac{1}{2}\lambda}{0}{\frac{1}{2}}{+} \AmpJP{A}{-\frac{1}{2}\lambda}{-\lambda}{\frac{3}{2}}{+}
+ \AmpJP{V*}{\frac{1}{2}\lambda}{0}{\frac{1}{2}}{+} \AmpJP{A}{\frac{3}{2}\lambda}{\lambda}{\frac{3}{2}}{+}
+\left(\frac{3}{2}\longleftrightarrow\frac{1}{2}\right)
\right)\right] \\
+&\left(V\longleftrightarrow A\right)
+\left(\pLambda\longrightarrow-\pLambda\right) \ .
    \end{split}
\end{align}
This coefficient contains terms arising from interference between states with identical quantum numbers and hence exists in the single-state case for $\jLambda\geq\jThree$. 
However, unless the different helicity amplitudes of the \Lstar have independent complex phases, the product of two amplitudes of the same state is always real and $K_{32}$ is zero.
This is the case in na\"ive factorisation. 
Even allowing for large phase differences, the magnitude of $K_{32}$ will remain small for a single state due to the relative suppression of the amplitudes with helicity $|\lLambda|=\tfrac{3}{2}$ compared to amplitudes with $|\lLambda|=\tfrac{1}{2}$. 
In a spectrum with several interfering resonances, the global QCD phase difference between the states can lead to sizeable imaginary terms. 
Moreover,  if there is interference between states with different spins, terms with $|\lLambda^{(\prime)}|=\tfrac{1}{2}$ appear, resulting in large values of $K_{32}$. 
The remaining angular coefficients in the unpolarised case are summarised in Appendix~\ref{app:observables}. 
The additional coefficients appearing in the polarised case are provided in a note book as supplementary material. 

\section{Angular distributions for the decay of unpolarized \texorpdfstring{\boldmath\Lb}{Lb} baryons to individual states}
\label{sec:individual}

When considering only a single strongly decaying spin-\jOne resonance, all but the first three angular coefficients vanish and the angular distribution depends only on $\cos\tl$
\begin{align} 
\frac{32\pi^2}{3} \frac{\deriv^{7}\Gamma}{\deriv\qsq\,\deriv{\mpk}\,\deriv{\angles}} = \frac{1}{\sqrt{3}}K_{1}
-\frac{\sqrt{5}}{2\sqrt{3}}K_3 + K_{2}\cos\tl + \frac{\sqrt{15}}{2} K_3\cos^2\tl ~. 
\end{align}
This reproduces the distribution given in the literature in Ref.~\cite{Dey:2019myq} as well as Refs.~\cite{Boer:2014kda,Blake:2017une} in the case of parity conservation.

For a single spin-\jThree state, in addition to $K_{1,2,3}$, 
\begin{equation}
\begin{aligned}
K_{7} &= +\frac{1}{ \sqrt{15}}
\left(
 \Big|\AmpJP{V}{ \frac{1}{2}}{1}{\frac{3}{2}}{{}}\Big|^2
+\Big|\AmpJP{V}{-\frac{1}{2}}{-1}{\frac{3}{2}}{{}}\Big|^2
+\Big|\AmpJP{V}{ \frac{1}{2}}{0}{\frac{3}{2}}{{}}\Big|^2 \right. \\
&  \qquad\qquad\qquad \left.
+\Big|\AmpJP{V}{-\frac{1}{2}}{0}{\frac{3}{2}}{{}}\Big|^2
-\Big|\AmpJP{V}{ \frac{3}{2}}{1}{\frac{3}{2}}{{}}\Big|^2
-\Big|\AmpJP{V}{-\frac{3}{2}}{-1}{\frac{3}{2}}{{}}\Big|^2
\right) + (V\longleftrightarrow A)  \,, \\
K_{8} &= +\frac{1}{ \sqrt{5}}
\text{Re}\left[
 \AmpJP{A*}{-\frac{1}{2}}{-1}{\frac{3}{2}}{{}} \AmpJP{V}{-\frac{1}{2}}{-1}{\frac{3}{2}}{{}}
-\AmpJP{A*}{-\frac{3}{2}}{-1}{\frac{3}{2}}{{}} \AmpJP{V}{-\frac{3}{2}}{-1}{\frac{3}{2}}{{}}
-\AmpJP{A*}{\frac{1}{2}}{1}{\frac{3}{2}}{{}} \AmpJP{V}{\frac{1}{2}}{1}{\frac{3}{2}}{{}}
+\AmpJP{A*}{\frac{3}{2}}{1}{\frac{3}{2}}{{}} \AmpJP{V}{\frac{3}{2}}{1}{\frac{3}{2}}{{}}
\right]  \,, \\
K_{9} &= +\frac{1}{10\sqrt{3}}
\left(
 \Big|\AmpJP{V}{\frac{1}{2}}{1}{\frac{3}{2}}{{}}\Big|^2
+\Big|\AmpJP{V}{-\frac{1}{2}}{-1}{\frac{3}{2}}{{}}\Big|^2
-2 \Big|\AmpJP{V}{\frac{1}{2}}{0}{\frac{3}{2}}{{}}\Big|^2 \right. \\
& \qquad\qquad\qquad \left.
-2 \Big|\AmpJP{V}{-\frac{1}{2}}{0}{\frac{3}{2}}{{}}\Big|^2
-\Big|\AmpJP{V}{\frac{3}{2}}{1}{\frac{3}{2}}{{}}\Big|^2
-\Big|\AmpJP{V}{-\frac{3}{2}}{-1}{\frac{3}{2}}{{}}\Big|^2
\right) + (V\longleftrightarrow A)  \,, \\
K_{21} &=
-\frac{1}{5} \text{Re}\left[
 \AmpJP{V*}{\frac{1}{2}}{0}{\frac{3}{2}}{{}} \AmpJP{V}{\frac{3}{2}}{1}{\frac{3}{2}}{{}}
+\AmpJP{V*}{-\frac{1}{2}}{0}{\frac{3}{2}}{{}} \AmpJP{V}{-\frac{3}{2}}{-1}{\frac{3}{2}}{{}}
\right]
+ (V\longleftrightarrow A)  \,, \\
K_{22} &= +\frac{1}{\sqrt{5}} \text{Re}\left[
 \AmpJP{V*}{\frac{1}{2}}{0}{\frac{3}{2}}{{}} \AmpJP{A}{\frac{3}{2}}{1}{\frac{3}{2}}{{}}
-\AmpJP{V*}{-\frac{1}{2}}{0}{\frac{3}{2}}{{}} \AmpJP{A}{-\frac{3}{2}}{-1}{\frac{3}{2}}{{}}
\right]
+ (V\longleftrightarrow A)  \,, \\
K_{31} &= +\frac{1}{5} \text{Im}\left[
 \AmpJP{V*}{\frac{1}{2}}{0}{\frac{3}{2}}{{}} \AmpJP{V}{\frac{3}{2}}{1}{\frac{3}{2}}{{}}
-\AmpJP{V*}{-\frac{1}{2}}{0}{\frac{3}{2}}{{}} \AmpJP{V}{-\frac{3}{2}}{-1}{\frac{3}{2}}{{}}
\right]
+ (V\longleftrightarrow A)  \,, \\
K_{32} &=-\frac{1}{\sqrt{5}} \text{Im}\left[
 \AmpJP{V*}{\frac{1}{2}}{0}{\frac{3}{2}}{{}} \AmpJP{A}{\frac{3}{2}}{1}{\frac{3}{2}}{{}}
+\AmpJP{V*}{-\frac{1}{2}}{0}{\frac{3}{2}}{{}} \AmpJP{A}{-\frac{3}{2}}{-1}{\frac{3}{2}}{{}}
\right]
+ (V\longleftrightarrow A)  \,, \\
K_{39} &= -\frac{\sqrt{2}}{5}\text{Re}\left[
\AmpJP{V*}{\frac{1}{2}}{1}{\frac{3}{2}}{{}} \AmpJP{V}{-\frac{3}{2}}{-1}{\frac{3}{2}}{{}}
+\AmpJP{V*}{\frac{3}{2}}{1}{\frac{3}{2}}{{}} \AmpJP{V}{-\frac{1}{2}}{-1}{\frac{3}{2}}{{}}
\right]
+ (V\longleftrightarrow A) \,, \\
K_{43} &= -\frac{\sqrt{2}}{5}\text{Im}\left[
 \AmpJP{V*}{\frac{1}{2}}{1}{\frac{3}{2}}{{}} \AmpJP{V}{-\frac{3}{2}}{-1}{\frac{3}{2}}{{}}
+\AmpJP{V*}{\frac{3}{2}}{1}{\frac{3}{2}}{{}} \AmpJP{V}{-\frac{1}{2}}{-1}{\frac{3}{2}}{{}} \right]
+ (V\longleftrightarrow A) \,,
\end{aligned}
\end{equation}
can be non-zero.
The angular distribution reproduces the result of Ref.~\cite{Descotes-Genon:2019dbw}.
A translation of the coefficients $L_{i}$ used in Ref.~\cite{Descotes-Genon:2019dbw} and the $K_j$ observables used in this paper is given in Appendix~\ref{app:LKtranslation}. 

For a single spin-\jFive particle, the angular distribution receives contributions from $K_{1}$--$K_{3}$, 
\begin{align}
\begin{split}
K_{7} &= \frac{2}{7 \sqrt{15}}\left(
 4\Big|\AmpJP{V}{\frac{1}{2}}{1}{\frac{5}{2}}{{}}\Big|^2
+4\Big|\AmpJP{V}{-\frac{1}{2}}{-1}{\frac{5}{2}}{{}}\Big|^2
+4\Big|\AmpJP{V}{\frac{1}{2}}{0}{\frac{5}{2}}{{}}\Big|^2 \right. \\
&\qquad\qquad\qquad\left.+4\Big|\AmpJP{V}{-\frac{1}{2}}{0}{\frac{5}{2}}{{}}\Big|^2
+ \Big|\AmpJP{V}{\frac{3}{2}}{1}{\frac{5}{2}}{{}}\Big|^2
+ \Big|\AmpJP{V}{-\frac{3}{2}}{-1}{\frac{5}{2}}{{}}\Big|^2
\right) + (V\longleftrightarrow A) \,, \\
K_{8} &= \frac{2}{7 \sqrt{5}}
\text{Re}\left[
  \AmpJP{A*}{-\frac{3}{2}}{-1}{\frac{5}{2}}{{}} \AmpJP{V}{-\frac{3}{2}}{-1}{\frac{5}{2}}{{}}
- \AmpJP{A*}{\frac{3}{2}}{1}{\frac{5}{2}}{{}} \AmpJP{V}{\frac{3}{2}}{1}{\frac{5}{2}}{{}}
+4\AmpJP{A*}{-\frac{1}{2}}{-1}{\frac{5}{2}}{{}} \AmpJP{V}{-\frac{1}{2}}{-1}{\frac{5}{2}}{{}}
-4\AmpJP{A*}{\frac{1}{2}}{1}{\frac{5}{2}}{{}} \AmpJP{V}{\frac{1}{2}}{1}{\frac{5}{2}}{{}}
\right] \,, \\
K_{9} &= \frac{1}{35\sqrt{3}}
\left(
 4\Big|\AmpJP{V}{\frac{1}{2}}{1}{\frac{5}{2}}{{}}\Big|^2
+4\Big|\AmpJP{V}{-\frac{1}{2}}{-1}{\frac{5}{2}}{{}}\Big|^2
-8\Big|\AmpJP{V}{\frac{1}{2}}{0}{\frac{5}{2}}{{}}\Big|^2 \right. \\
&\qquad\qquad\qquad\left.
-8\Big|\AmpJP{V}{-\frac{1}{2}}{0}{\frac{5}{2}}{{}}\Big|^2
+ \Big|\AmpJP{V}{\frac{3}{2}}{1}{\frac{5}{2}}{{}}\Big|^2
+ \Big|\AmpJP{V}{-\frac{3}{2}}{-1}{\frac{5}{2}}{{}}\Big|^2
\right)  + (V\longleftrightarrow A) \,, \\
K_{13} &=
\frac{1}{7\sqrt{3}}\left(
 2 \Big|\AmpJP{V}{\frac{1}{2}}{1}{\frac{5}{2}}{{}}\Big|^2
+2 \Big|\AmpJP{V}{-\frac{1}{2}}{-1}{\frac{5}{2}}{{}}\Big|^2
+2 \Big|\AmpJP{V}{\frac{1}{2}}{0}{\frac{5}{2}}{{}}\Big|^2 \right. \\
&\qquad\qquad\qquad\left.
+2 \Big|\AmpJP{V}{-\frac{1}{2}}{0}{\frac{5}{2}}{{}}\Big|^2
-3 \Big|\AmpJP{V}{\frac{3}{2}}{1}{\frac{5}{2}}{{}}\Big|^2
-3 \Big|\AmpJP{V}{-\frac{3}{2}}{-1}{\frac{5}{2}}{{}}\Big|^2
\right) + (V\longleftrightarrow A) \,, \\
K_{14} &= \frac{1}{7}\text{Re}\left[
-3 \AmpJP{A*}{-\frac{3}{2}}{-1}{\frac{5}{2}}{{}} \AmpJP{V}{-\frac{3}{2}}{-1}{\frac{5}{2}}{{}}
+2 \AmpJP{A*}{-\frac{1}{2}}{-1}{\frac{5}{2}}{{}} \AmpJP{V}{-\frac{1}{2}}{-1}{\frac{5}{2}}{{}}
-2 \AmpJP{A*}{\frac{1}{2}}{1}{\frac{5}{2}}{{}} \AmpJP{V}{\frac{1}{2}}{1}{\frac{5}{2}}{{}}
+3 \AmpJP{A*}{\frac{3}{2}}{1}{\frac{5}{2}}{{}} \AmpJP{V}{\frac{3}{2}}{1}{\frac{5}{2}}{{}}
\right] \,,\\
K_{15} &=
\frac{1}{14\sqrt{15}}\left(
 2 \Big|\AmpJP{V}{\frac{1}{2}}{1}{\frac{5}{2}}{{}}\Big|^2
+2 \Big|\AmpJP{V}{-\frac{1}{2}}{-1}{\frac{5}{2}}{{}}\Big|^2
-4 \Big|\AmpJP{V}{\frac{1}{2}}{0}{\frac{5}{2}}{{}}\Big|^2 \right. \\
&\qquad\qquad\qquad\left.
-4 \Big|\AmpJP{V}{-\frac{1}{2}}{0}{\frac{5}{2}}{{}}\Big|^2
-3 \Big|\AmpJP{V}{\frac{3}{2}}{1}{\frac{5}{2}}{{}}\Big|^2
-3 \Big|\AmpJP{V}{-\frac{3}{2}}{-1}{\frac{5}{2}}{{}}\Big|^2
\right) + (V\longleftrightarrow A) \,, \\
K_{21} &=
-\frac{2\sqrt{6}}{35} \text{Re}\left[
 \AmpJP{V*}{\frac{1}{2}}{0}{\frac{5}{2}}{{}} \AmpJP{V}{\frac{3}{2}}{1}{\frac{5}{2}}{{}}
+\AmpJP{V*}{-\frac{1}{2}}{0}{\frac{5}{2}}{{}} \AmpJP{V}{-\frac{3}{2}}{-1}{\frac{5}{2}}{{}}
\right] + (V\longleftrightarrow A) \,, \\
K_{22} &=\frac{2\sqrt{6}}{7\sqrt{5}} \text{Re}\left[
 \AmpJP{V*}{\frac{1}{2}}{0}{\frac{5}{2}}{{}} \AmpJP{A}{\frac{3}{2}}{1}{\frac{5}{2}}{{}}
-\AmpJP{V*}{-\frac{1}{2}}{0}{\frac{5}{2}}{{}} \AmpJP{A}{-\frac{3}{2}}{-1}{\frac{5}{2}}{{}}
\right] + (V\longleftrightarrow A) \,, \\
K_{31} &= \frac{2\sqrt{6}}{35}
\text{Im}\left[
 \AmpJP{V*}{\frac{1}{2}}{0}{\frac{5}{2}}{{}} \AmpJP{V}{\frac{3}{2}}{1}{\frac{5}{2}}{{}}
-\AmpJP{V*}{-\frac{1}{2}}{0}{\frac{5}{2}}{{}} \AmpJP{V}{-\frac{3}{2}}{-1}{\frac{5}{2}}{{}}
\right] + (V\longleftrightarrow A) \,, \\
K_{32} &=-\frac{2}{7} \sqrt{\frac{6}{5}} \text{Im}\left[
 \AmpJP{V*}{\frac{1}{2}}{0}{\frac{5}{2}}{{}} \AmpJP{A}{\frac{3}{2}}{1}{\frac{5}{2}}{{}}
+\AmpJP{V*}{-\frac{1}{2}}{0}{\frac{5}{2}}{{}} \AmpJP{A}{-\frac{3}{2}}{-1}{\frac{5}{2}}{{}}
\right] + (V\longleftrightarrow A) \,, \\
K_{39} &= -\frac{6\sqrt{3}}{35} \text{Re} \left[
 \AmpJP{V*}{\frac{1}{2}}{1}{\frac{5}{2}}{{}} \AmpJP{V}{-\frac{3}{2}}{-1}{\frac{5}{2}}{{}} 
+\AmpJP{V*}{\frac{3}{2}}{1}{\frac{5}{2}}{{}} \AmpJP{V}{-\frac{1}{2}}{-1}{\frac{5}{2}}{{}} \right]
+ (V\longleftrightarrow A) \,,\\
K_{43} &= -\frac{6\sqrt{3}}{35} \text{Im} \left[
 \AmpJP{V*}{\frac{1}{2}}{1}{\frac{5}{2}}{{}} \AmpJP{V}{-\frac{3}{2}}{-1}{\frac{5}{2}}{{}}
+\AmpJP{V*}{\frac{3}{2}}{1}{\frac{5}{2}}{{}} \AmpJP{V}{-\frac{1}{2}}{-1}{\frac{5}{2}}{{}}\right]
+ (V\longleftrightarrow A) \,,
\end{split}
\end{align}
and $K_{25}$, $K_{26}$, $K_{35}$, $K_{36}$, $K_{41}$, and $K_{45}$. These are trivially related to the other $K_{i}$, 
\begin{equation}\label{eq:dependent5}
\begin{alignedat}{3}
& K_{25} = +\frac{5}{2\sqrt{6}}K_{21} \,,\qquad
& K_{26} = +\frac{5}{2\sqrt{6}}K_{22} \,,\qquad
& K_{35} = +\frac{5}{2\sqrt{6}}K_{31} \,, \\
& K_{36} = +\frac{5}{2\sqrt{6}}K_{32} \,,\qquad 
& K_{41} = \frac{5}{6\sqrt{3}}K_{39}\,,\qquad
& K_{45} = \frac{5}{6\sqrt{3}}K_{43} \,.
\end{alignedat}
\end{equation}
The remaining coefficients vanish.
The relationships in Eq.~\eqref{eq:dependent5} only hold for either a spin-\jFive particle in isolation or for cases where there are only contributions of spin-\jFive states with the same quantum numbers.

In the notation of Refs.~\cite{Descotes-Genon:2019dbw, Boer:2014kda}, the angular distribution of the decay via a spin-\jFive resonance can be expressed as 
\begin{align}\label{eq:singlestate}
\begin{split}
    \frac{8\pi}{3}\frac{\deriv^5\Gamma}{\deriv\mpk\deriv\qsq\deriv\cos\tp\deriv\cos\tl\deriv\phi} = & \cos^2\tp\left(L_{1c}\cos\tl + L_{1cc}\cos^2\tl + L_{1ss}\sin^2\tl\right) \\[-8pt]
    + & \sin^2\tp\left(L_{2c}\cos\tl + L_{2cc}\cos^2\tl + L_{2ss}\sin^2\tl\right) \\
    + & \sin^2\tp\sin^2\tl\left(L_{3ss}\cos^2\phi + L_{4ss}\sin\phi\cos\phi\right) \\
    + & \sin\tp\cos\tp\left(L_{5s}\sin\tl\cos\phi + L_{5sc}\sin\tl\cos\tl\cos\phi\right) \\
    + & \sin\tp\cos\tp\left(L_{6s}\sin\tl\sin\phi + L_{6sc}\sin\tl\cos\tl\sin\phi\right) \\
    + & \cos^4\tp\left(L_{7c}\cos\tl + L_{7cc}\cos^2\tl + L_{7ss}\sin^2\tl\right) \\
    + & \cos^2\tp\sin^2\tp\sin^2\tl\left(L_{8ss}\cos^2\phi + L_{9ss} \sin\phi\cos\phi\right) \\
    + & \cos^3\tp\sin \tp\left(L_{10s} \sin \tl \cos \phi + L_{10sc} \sin \tl\cos\tl \cos \phi\right) \\
    + & \cos^3\tp\sin \tp\left(L_{11s} \sin \tl \sin \phi + L_{11sc} \sin \tl\cos\tl \sin \phi\right) \,.
\end{split}
\end{align}
For both spin-\jThree and spin-\jFive resonances, the structure of the coefficients with the same dependency on \tl, $K_{1,7,13}$, $K_{2,8,14}$, $K_{3,9,15}$, $K_{21,25,31,35}$, $K_{22,26,32,36}$ and $K_{39,41,43,45}$, differ only in the numerical factor appearing in front of the amplitudes with $|\lLambda|=\tfrac{3}{2}$.
Due to the relative suppression of this amplitude compared to amplitudes with $|\lLambda|=\tfrac{1}{2}$, these coefficients in each set are numerically similar to each other and share the same features. 
All of the coefficients depending on the imaginary part of bilinear combinations of amplitudes are zero if there are no relative phases between the different helicity amplitudes as is the case in na\"ive QCD factorisation.
The impact of small corrections to this na\"ive factorisation assumption is considered in Section~\ref{sec:np}.

\section{Predictions for the SM and modified theories}
\label{sec:np}

Predictions for the angular observables for the weakly decaying spin-\jOne \Lnum{1115} resonance and the strongly decaying spin-\jThree \Lnum{1520} resonance have been studied previously. 
In this section we give numerical predictions for the cases that have not been investigated before: the case of the lowest-lying known spin-\jFive resonance, the \Lnum{1820}, and the more general case of an ensemble of different resonances. 
The predictions are made for SM-like values of the Wilson coefficients~\cite{Descotes-Genon:2013vna} and for values of \WC{9} and \WC{10} that are consistent with recent global analyses of mesonic \bsll transitions~\cite{Alguero:2021anc}. The values of the Wilson coefficients used in the two secenarios are given in Table~\ref{tab:wilsoncoeffs} in Appendix~\ref{app:effC9}. 
Four additional non-SM scenarios are considered
\begin{align}
    1)~\WC{9} = -\WC{9}^\text{SM} \qquad\qquad2)~\WC{10} = -\WC{10}^\text{SM} \qquad\qquad3)~\WC{9^\prime} = \WC{9}^\text{SM} \qquad\qquad4)~\WC{10^\prime} = \WC{10}^\text{SM}\,.
\end{align}
These are extreme scenarios, which are used to highlight the sensitivity of the observables to different types of current. 
Each of the scenarios is tested with unpolarized \Lb baryons, \ie with $\pLb=0$. 
The predictions for the different angular observables are obtained numerically by generating large ensembles that are sampled according to the differential decay rate. 
The value of the observable is then determined by evaluating the corresponding moment on the sample as described in Section~\ref{sec:method-of-moments}.
The properties of the \Lstar resonances are taken from Ref.~\cite{PDG2020} and summarised in Table~\ref{tab:resonances}. 
Only established states with the available quark-model form-factor predictions in Ref.~\cite{Mott:2015zma} are considered.
As outlined in Sec.~\ref{sec:helamp:Lstar}, the sub-threshold \Lnum{1405} resonance is described using the Flatt\'e model.

\begin{table}[htb!]
    \centering
    \caption{Resonance parameters used in the predictions presented in this paper. 
    The parameters of the resonances are taken from Ref.~\cite{PDG2020}. 
    The branching fraction of the \Lstar resonance to $p\Km$ is calculated from the centre of the range and scaled according to isospin considerations. 
    The branching fraction of $\Lnum{1405}\to N\overline{K}$ assumes equal partial widths for $\Lnum{1405}\to N\overline{K}$ and $\Lnum{1405}\to \Sigma\pi$.
    }
    \label{tab:resonances}
    \begin{tabular}{c|ccccc}
    \toprule
    resonance & \mLambda~[\gevcc] & \gLambda~[\gevcc] & 2\jLambda & \pLambda & $\mathcal{B}(\Lstar\to N\overline{K})$ \\ 
    \midrule
    \Lnum{1405} & 1.405 & 0.051 & 1 & $-$ & 0.50         \\ 
    \Lnum{1520} & 1.519 & 0.016 & 3 & $-$ & 0.45         \\ 
    \Lnum{1600} & 1.600 & 0.200 & 1 & $+$ & 0.15 -- 0.30 \\ 
    \Lnum{1670} & 1.674 & 0.030 & 1 & $-$ & 0.20 -- 0.30 \\ 
    \Lnum{1690} & 1.690 & 0.070 & 3 & $-$ & 0.20 -- 0.30 \\ 
    \Lnum{1800} & 1.800 & 0.200 & 1 & $-$ & 0.25 -- 0.40 \\ 
    \Lnum{1810} & 1.790 & 0.110 & 1 & $+$ & 0.05 -- 0.35 \\ 
    \Lnum{1820} & 1.820 & 0.080 & 5 & $+$ & 0.55 -- 0.65 \\ 
    \Lnum{1890} & 1.890 & 0.120 & 3 & $+$ & 0.24 -- 0.36 \\ 
    \Lnum{2110} & 2.090 & 0.250 & 5 & $+$ & 0.05 -- 0.25 \\ 
    \bottomrule
    \end{tabular}
\end{table}

\subsection[Predictions for the \texorpdfstring{spin-\jFive \Lnum{1820}}{spin-5/2 L(1820)} resonance]{Predictions for the \texorpdfstring{spin-\boldmath\jFive \Lnum{1820}}{spin-5/2 L(1820)} resonance}

\begin{figure}[htb!]
    \centering
    \includegraphics[width=.45\textwidth]{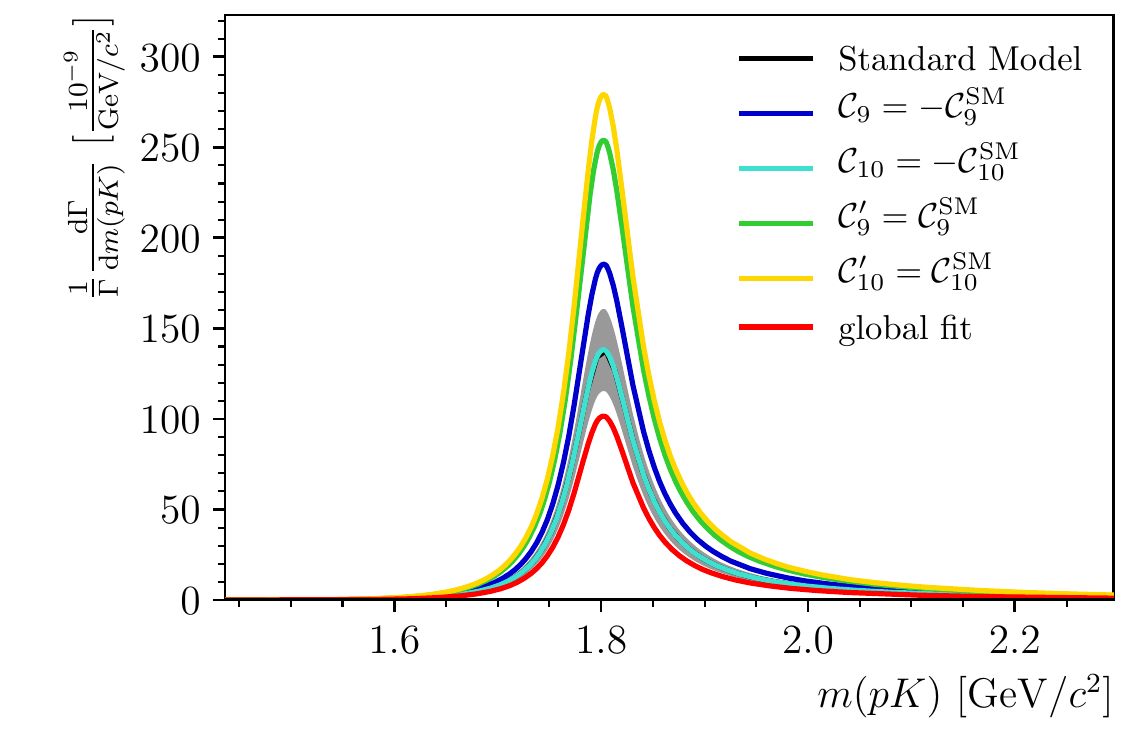}%
    \includegraphics[width=.45\textwidth]{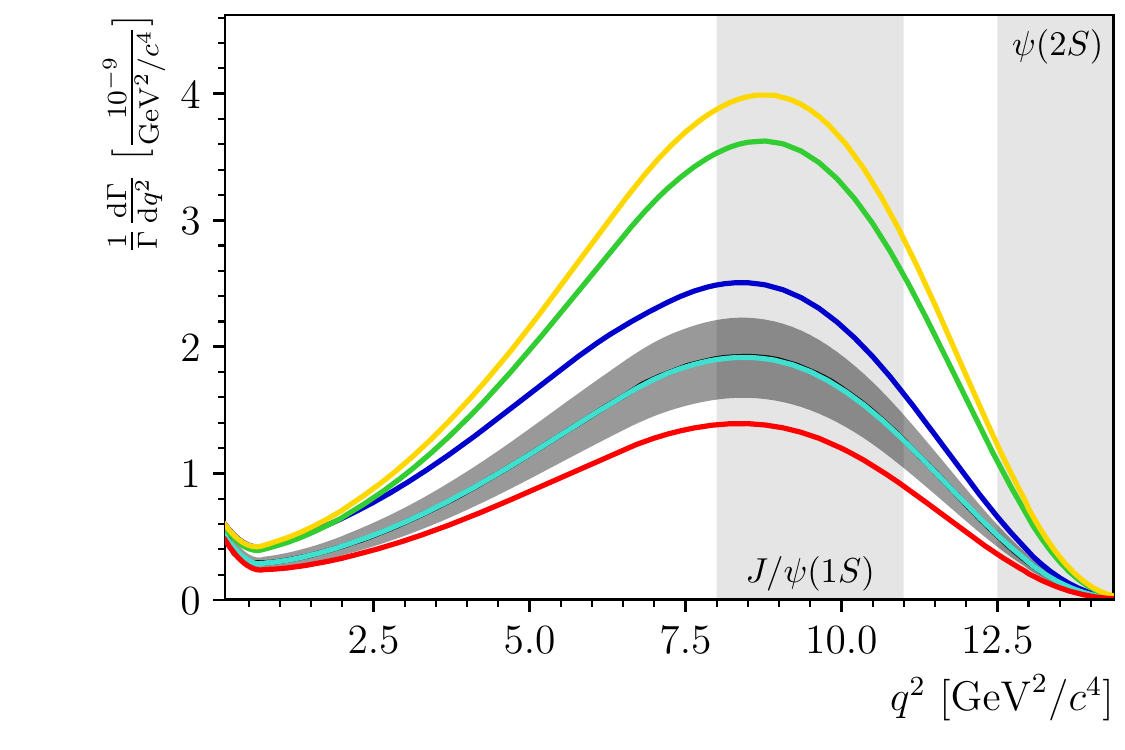}
    \caption{Differential branching fraction in \mpk and \qsq for a single \Lnum{1820} resonance assuming the SM (black line) and different NP scenarios (coloured lines). The SM and $\WC{10} = -\WC{10}^\text{SM}$ scenarios yield identical predictions for the differential branching fraction. The uncertainty on the SM prediction is represented by the gray band.}
    \label{fig:decrate1820}
\end{figure}

Figure~\ref{fig:decrate1820} shows the predicted differential branching fraction for the $\Lb\to\Lnum{1820}\mumu$ decay as a function of \mpk and \qsq for the six different scenarios. 
The SM and $\WC{10} = -\WC{10}^\text{SM}$ scenarios yield identical predictions for the differential branching fraction as its value only depends on $|\WC{10}|^{2}$. 
The gray band in Fig.~\ref{fig:decrate1820} represents an estimate of the theoretical uncertainty on the SM prediction. 
This is determined by varying the magnitude of each form factor, $X_{\Gamma_i}$, according to a normal distribution with a width of 10\%.
Moreover, there can be non-factorisable corrections to the decay amplitudes (which cannot be expressed in terms of local form-factors and Wilson coefficients). 
Such contributions can introduce relative phases between the amplitudes for a single decay. 
This can make observables that depend on the imaginary part of bilinear combinations of amplitudes, like $K_{32}$, non-zero. 
To estimate the uncertainty due to these non-factorisable corrections, each amplitude is varied according to 
\begin{align}
    H\to (1+a)H\,,
\end{align}
where $a$ is uniformly distributed inside a circle of radius 0.1 in the complex plane.
This is similar to the approach used for $\Bz\to\Kstarz\ellell$ decays in Ref.~\cite{Descotes-Genon:2014uoa}.
To propagate these variations to the observables, 200 different SM ensembles are produced and the moments extracted. 
The standard deviation of the resulting moments is taken as the uncertainty on the prediction.

\begin{figure}[htb!]
    \centering
    \includegraphics[width=.33\textwidth]{figs/1820/uncertainties_1820_all_BF_qsq.pdf}%
    \includegraphics[width=.33\textwidth]{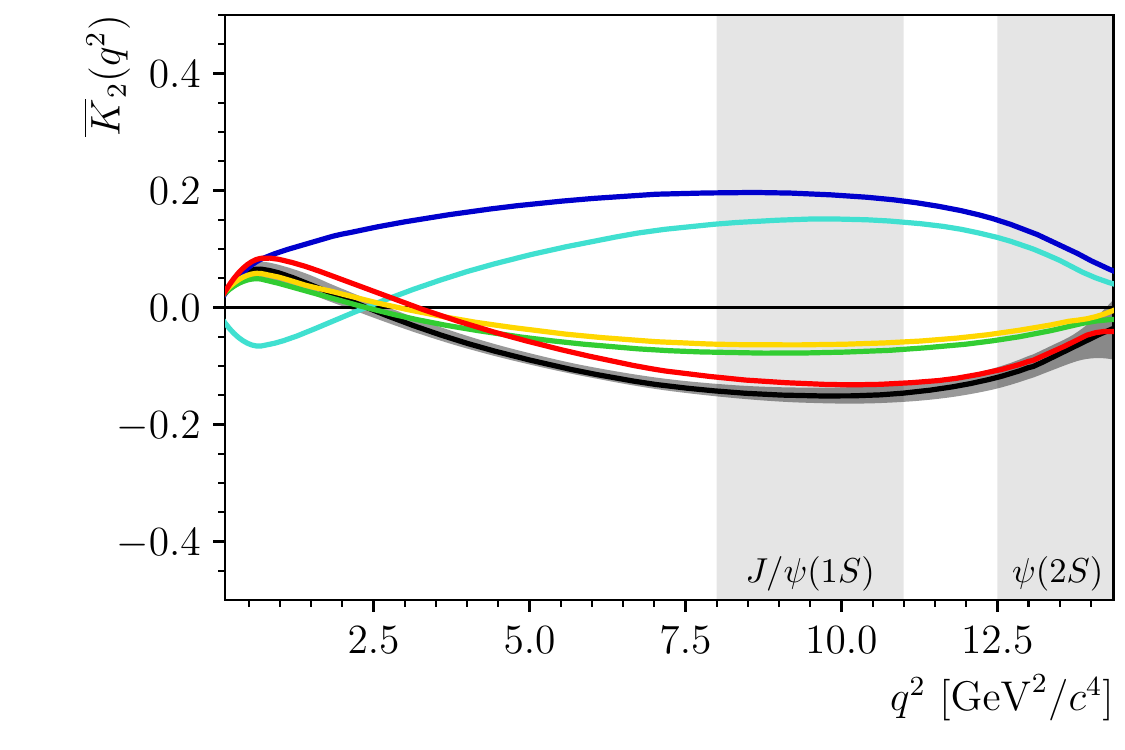}%
    \includegraphics[width=.33\textwidth]{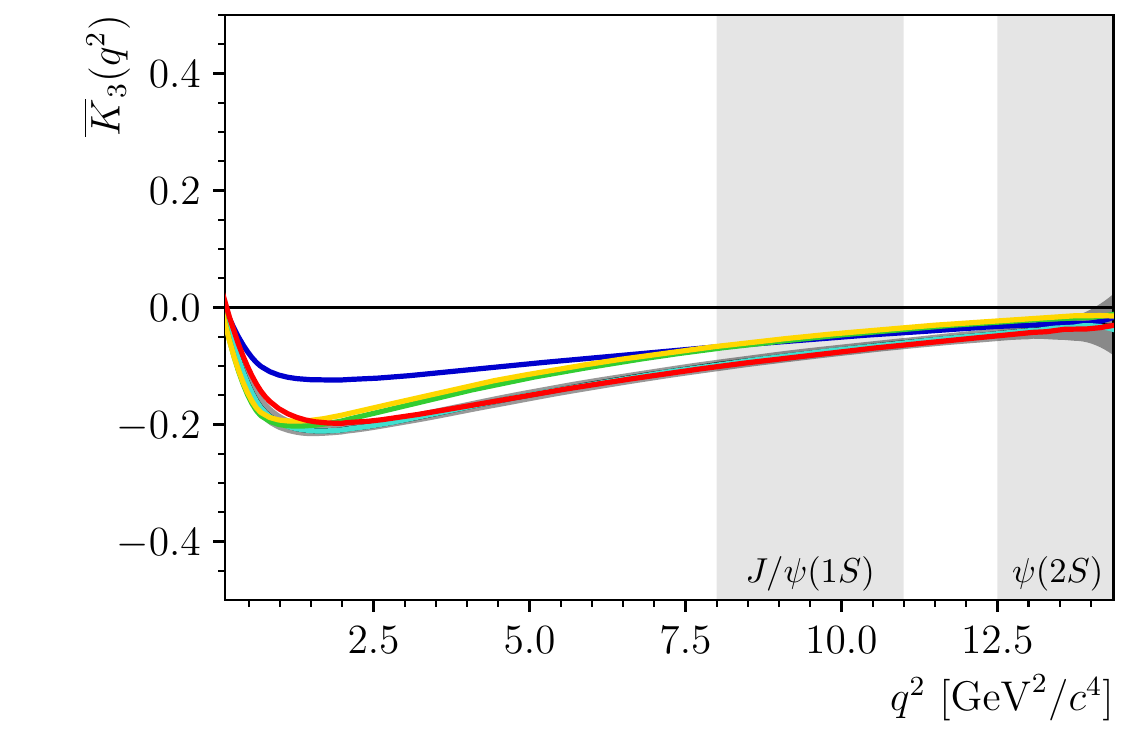}
    \includegraphics[width=.33\textwidth]{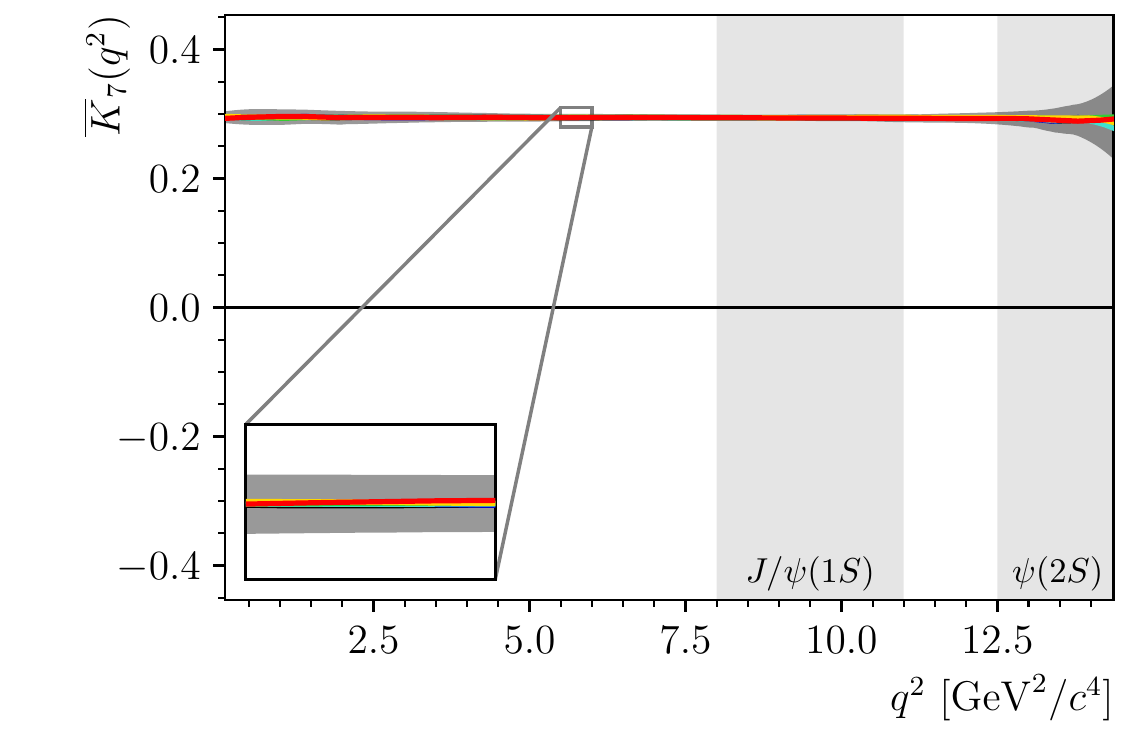}%
    \includegraphics[width=.33\textwidth]{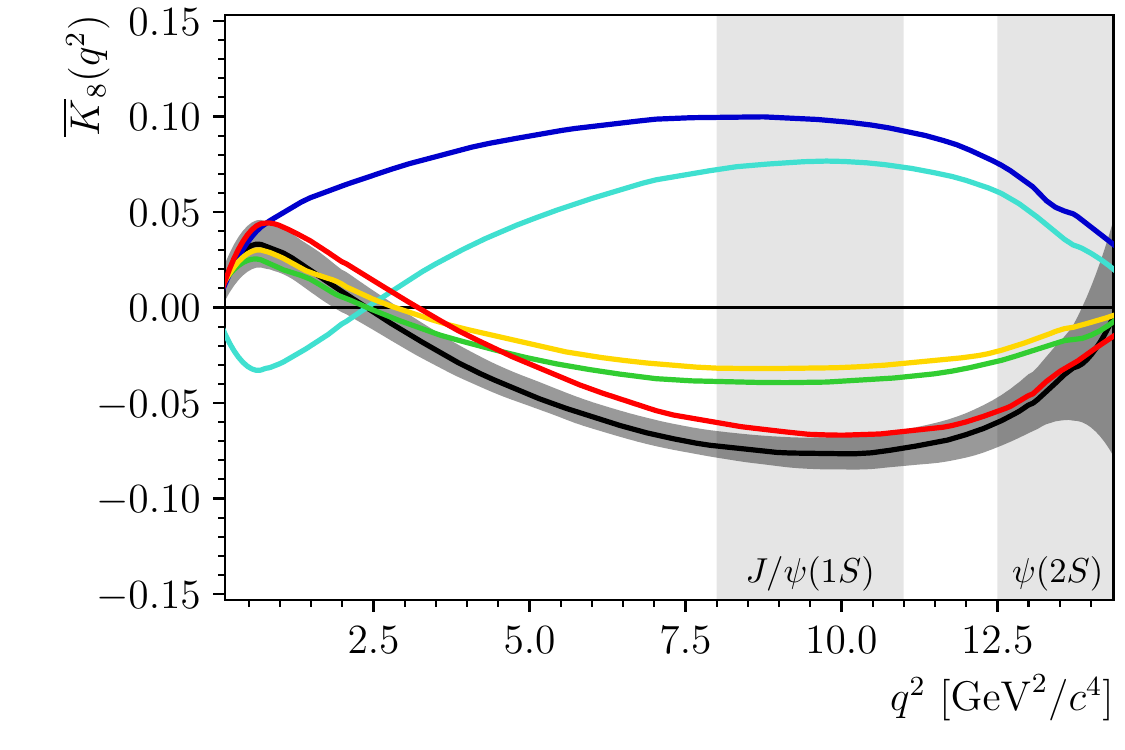}%
    \includegraphics[width=.33\textwidth]{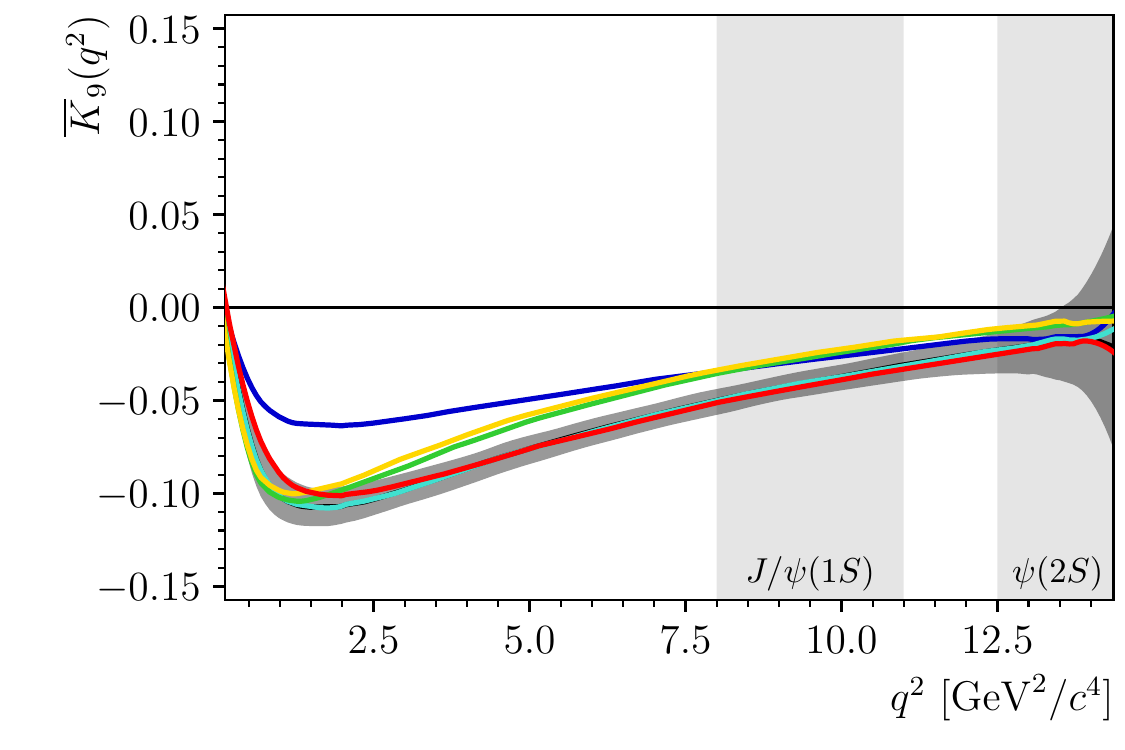}
    \includegraphics[width=.33\textwidth]{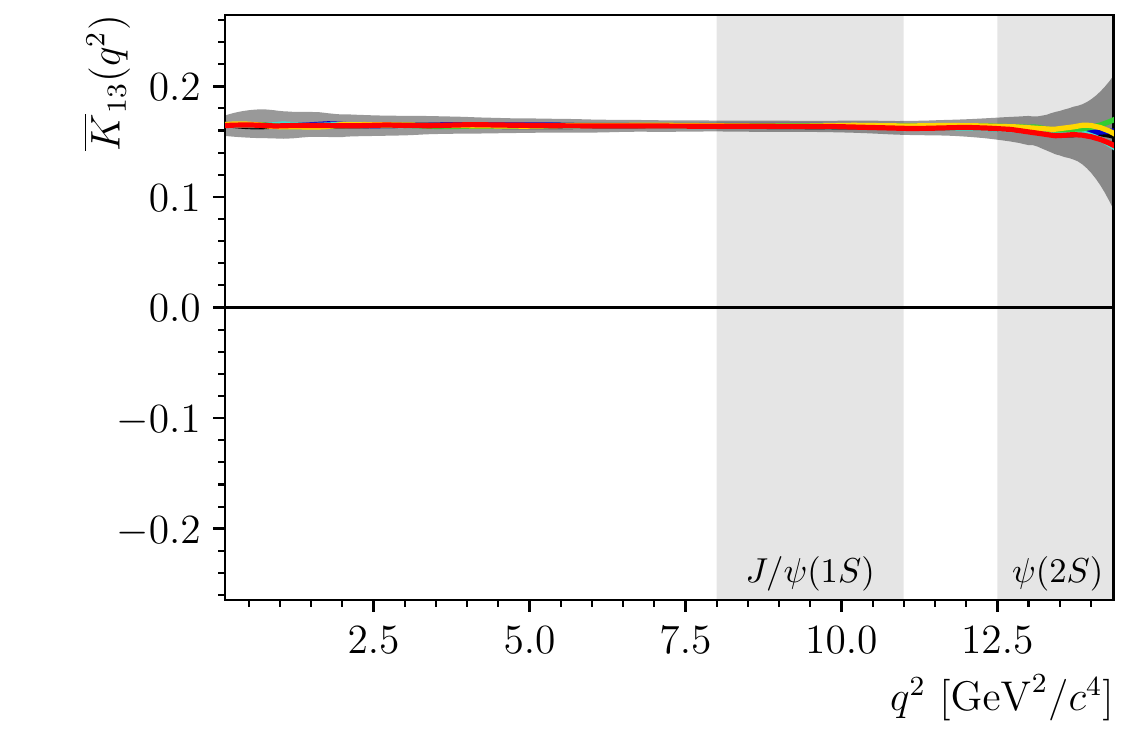}%
    \includegraphics[width=.33\textwidth]{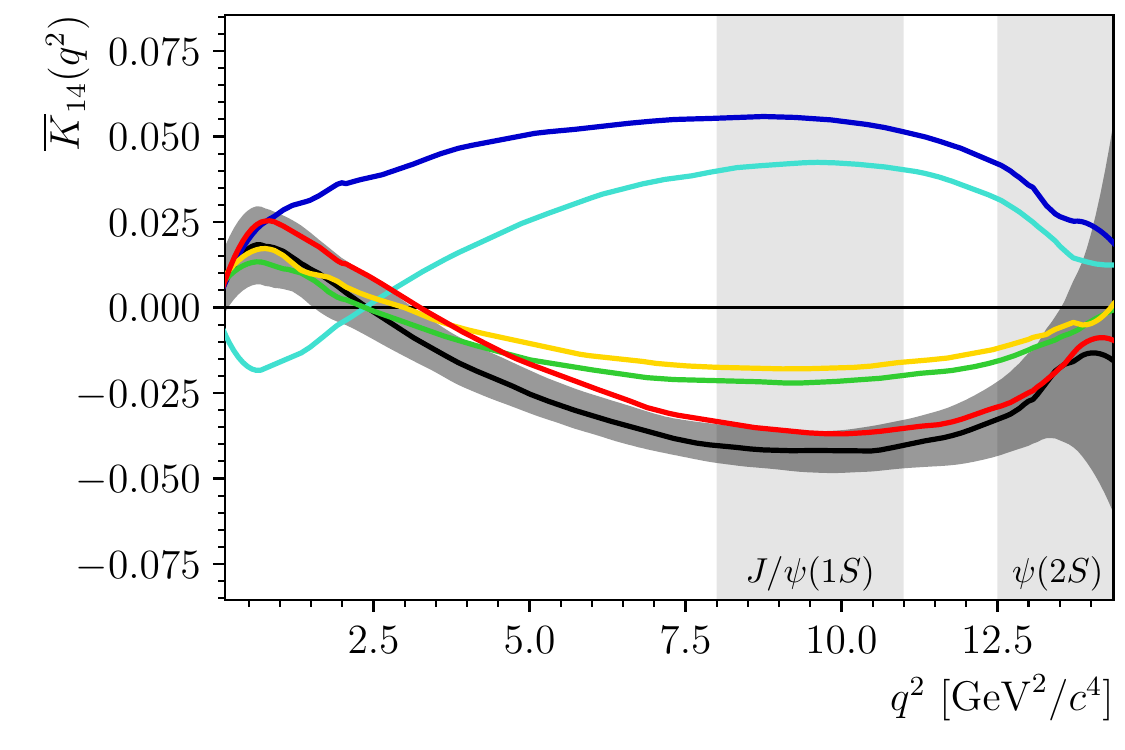}%
    \includegraphics[width=.33\textwidth]{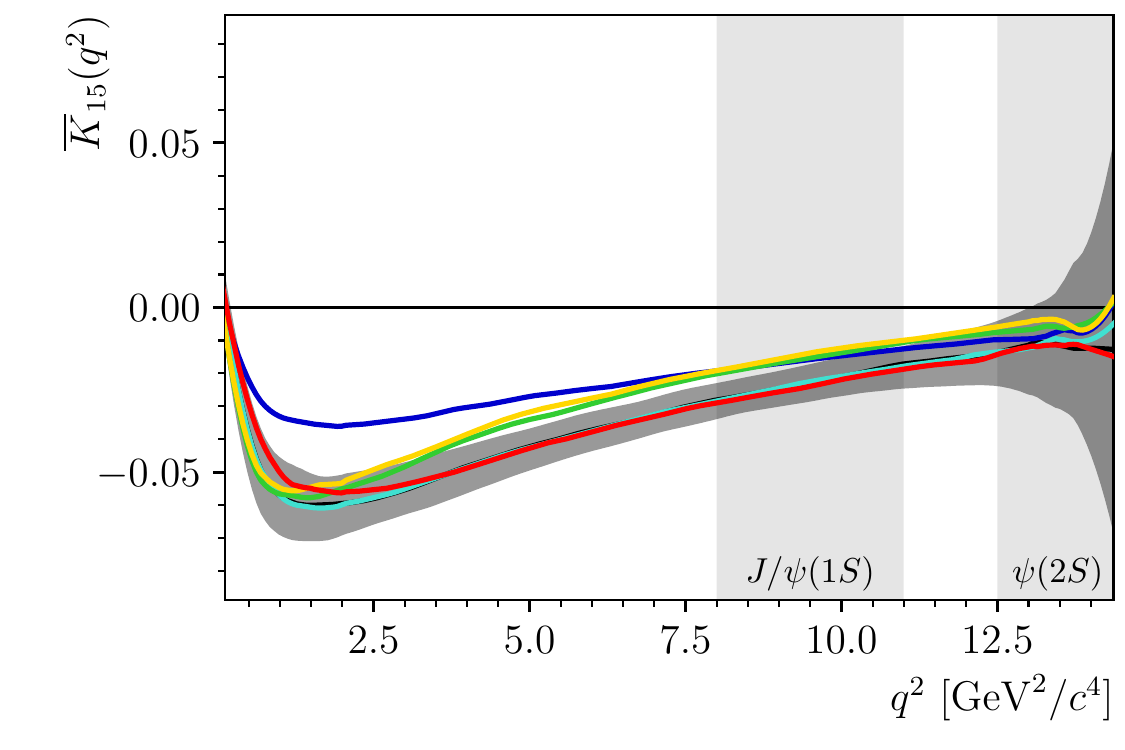}
    \caption{
    Angular observables \Kobs{2-15}, as a function of \qsq for the  spin-\jFive \Lnum{1820} resonance assuming the SM (black line) and different non-SM scenarios (using the same colour code as in Fig.~\ref{fig:decrate1820}). The uncertainty on the SM prediction is represented by the gray band.}
    \label{fig:moments1_allinfo_1820}
\end{figure}

Figure~\ref{fig:moments1_allinfo_1820} shows the angular observables as defined in Equation~\eqref{eq:observable} that are accompanied by basis functions that are independent of $\phi$.
In order to obtain continuous curves for the predictions, the values of the moments are evaluated in fine bins of \qsq and their values are smoothed using Gaussian kernels.
Some residual numerical variation can be seen in the figures when the values of the observables are small, for example in the high \qsq region of \Kobs{14,15}.
The increase in the SM uncertainty band at high \qsq is due to the reduced phase-space, and resulting small sample size, in this region.
The \qsq range in the figures is restricted to the allowed range at the pole mass $\mpk=1.82$\gevcc.
The observables associated with the angular function $P_1^0(\cos\tl)$, \Kobs{2,8,14}, are highly sensitive to modifications of the Wilson coefficients in particular to changes in the left-handed currents.
This is similar to what is seen in the forward-backward asymmetry of other \bsll decays. 
The observables accompanying the basis function $P_2^0(\cos\tl)$, \Kobs{3,9,15},  only differ from the SM for changes in the left-handed vector currents. The differences here are largest for small \qsq, where \WC{9}--\WC{7} interference is important. 
Finally, due to the very similar structures of $K_{1,7,13}$, as discussed Section~\ref{sec:observables}, the values of \Kobs{7,13} are almost identical for the different scenarios considered in this section. 
In the single resonance case, these observables serve as a useful check of the form-factor description.
In general, as the order of the \tp basis function increases ($0,2,4$ for the top, mid, and bottom row of Fig.~\ref{fig:moments1_allinfo_1820}) the magnitude of the corresponding observable decreases. 

\begin{figure}[htb!]
    \centering
    \includegraphics[width=.33\textwidth]{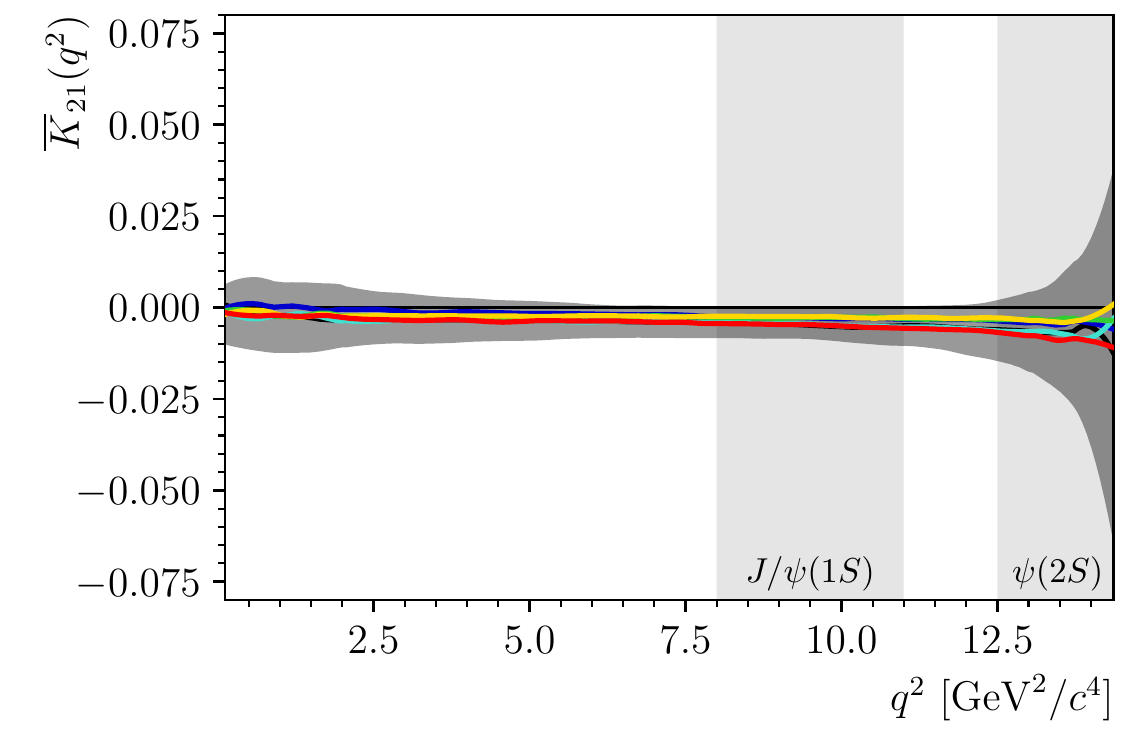}%
    \includegraphics[width=.33\textwidth]{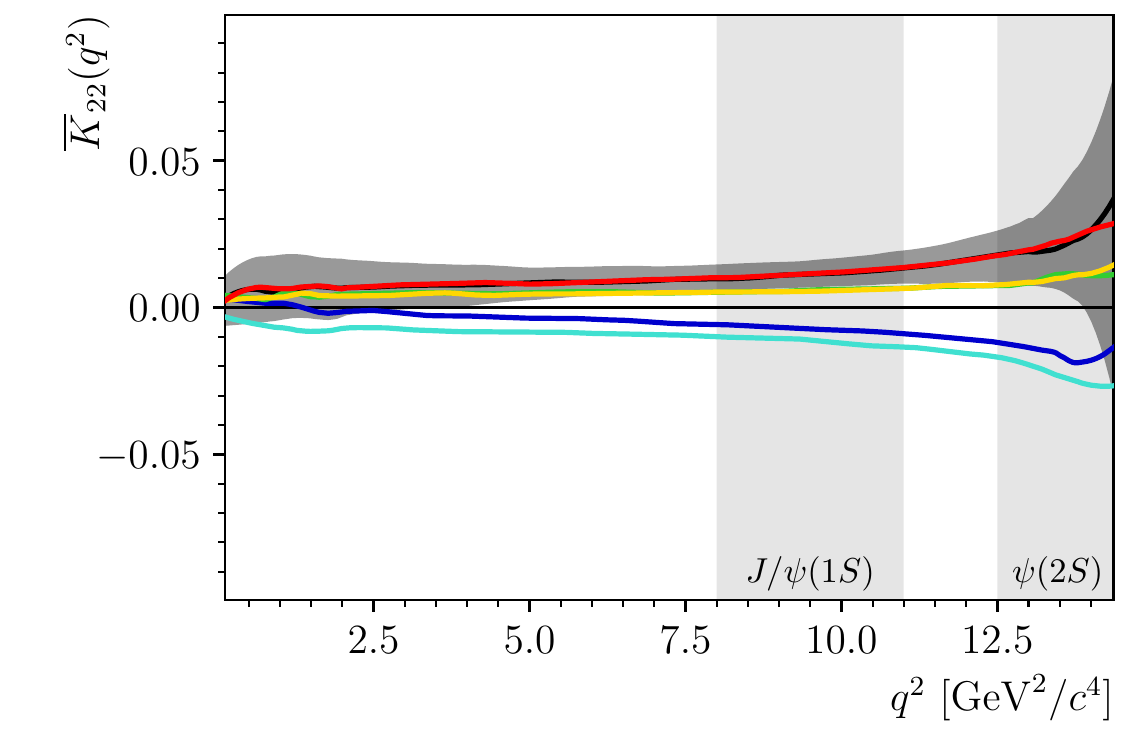}%
    \includegraphics[width=.33\textwidth]{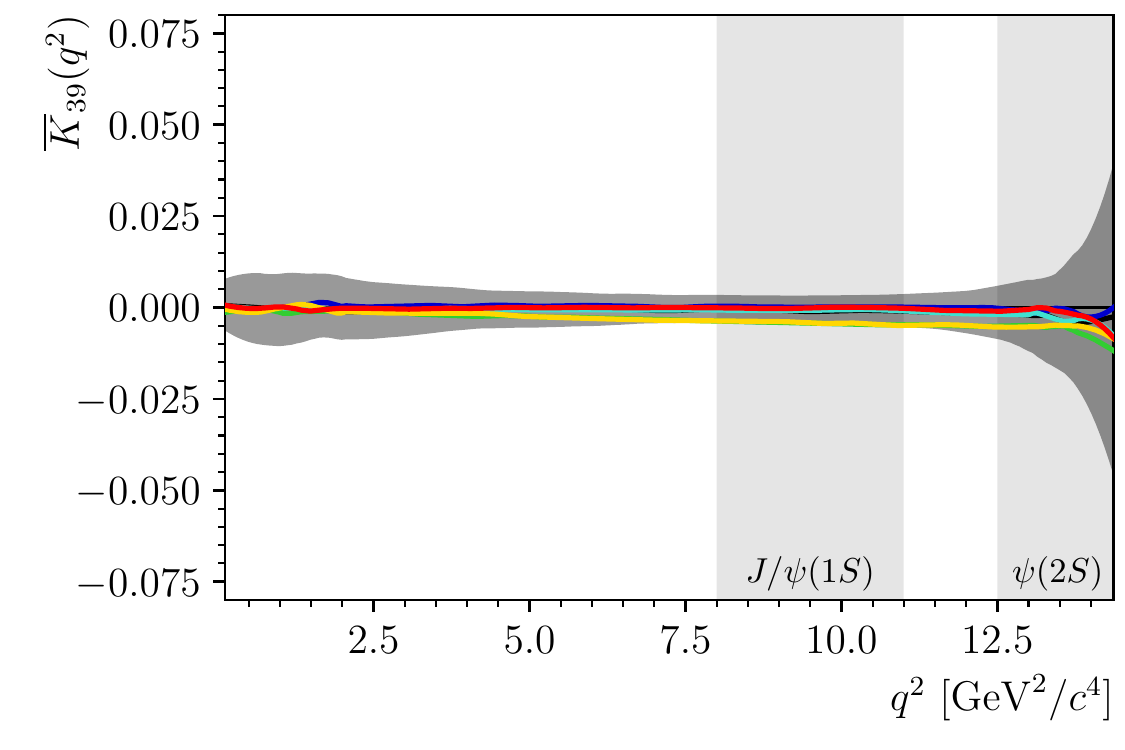}
    \caption{Angular observables \Kobs{21}, \Kobs{22}, and \Kobs{39}, which accompany basis functions with a $\cos\phi$ or $\cos2\phi$ dependence, for the  spin-\jFive \Lnum{1820} resonance as a function of \qsq. 
    The different curves correspond to the same scenarios as labelled in Fig.~\ref{fig:decrate1820}.}
    \label{fig:moments2_allinfo_1820}
\end{figure}

Figure~\ref{fig:moments2_allinfo_1820} shows the angular observables that accompany the basis functions with $\cos\phi$ or $\cos2\phi$ dependency.
Mathematically, these observables depend on the real part of products of different \Lstar helicity amplitudes.
The observables \Kobs{21} and \Kobs{39} are sensitive to the introduction of right-handed currents.
The observable \Kobs{21} is also sensitive to changes in the left-handed vector current.
Even in these extreme scenarios, the changes from the SM prediction only exceed the estimated uncertainty on the predictions for the observable \Kobs{22}.
Like \Kobs{2,8,14}, this observable is sensitive to vector-axialvector interference. 

The remaining observables, \Kobs{31,32,43}, depend on the imaginary part of bilinear combinations of amplitudes and are associated with $\sin\phi$ or $\sin2\phi$ basis functions. These observables are vanishingly small for an individual state because of the small phase-difference between the helicity amplitudes. 
They can be used to test the assumption of QCD factorisation in the SM. 

\subsection{Predictions for an ensemble of different \texorpdfstring{\Lstar}{Lambda} resonances}

Predictions are also obtained for the ensemble of different \Lstar resonances given in Table~\ref{tab:resonances}.
In this case, relative QCD phases between the different states need to be chosen.
By default the phase at the pole of each resonance is set to $\tfrac{\pi}{2}$, \ie $\phase=0$. 
The effect of varying the unknown phase-difference between the different resonances in the range $[-\pi,+\pi]$ is tested and shown in the figures as a separate uncertainty band. 
In this case, 200 different ensembles are produced to estimate the uncertainty. 
Note, unlike the form-factor uncertainties, the uncertainty band corresponds to the full range of the allowed phase-variations. 
For many observables, the variation of the SM predictions due to the phases is larger than the variation between the different scenarios and it will be necessary to understand the phases to interpret the data from experiments. 
Due to the available phase-space the composition of \Lstar resonances changes with \qsq. 
At low \qsq, all of the resonances in Table~\ref{tab:resonances} contribute. 
The composition of the contributing states varies rapidly as \qsq approaches its maximum value as the contribution from heavier states becomes suppressed. 
For example, for $\qsq\gsim16.8\gev^2/c^4$, the ensemble is dominated by the contribution from the \Lnum{1405} resonance. 

\begin{figure}[htb!]
    \centering
    \includegraphics[width=.48\textwidth]{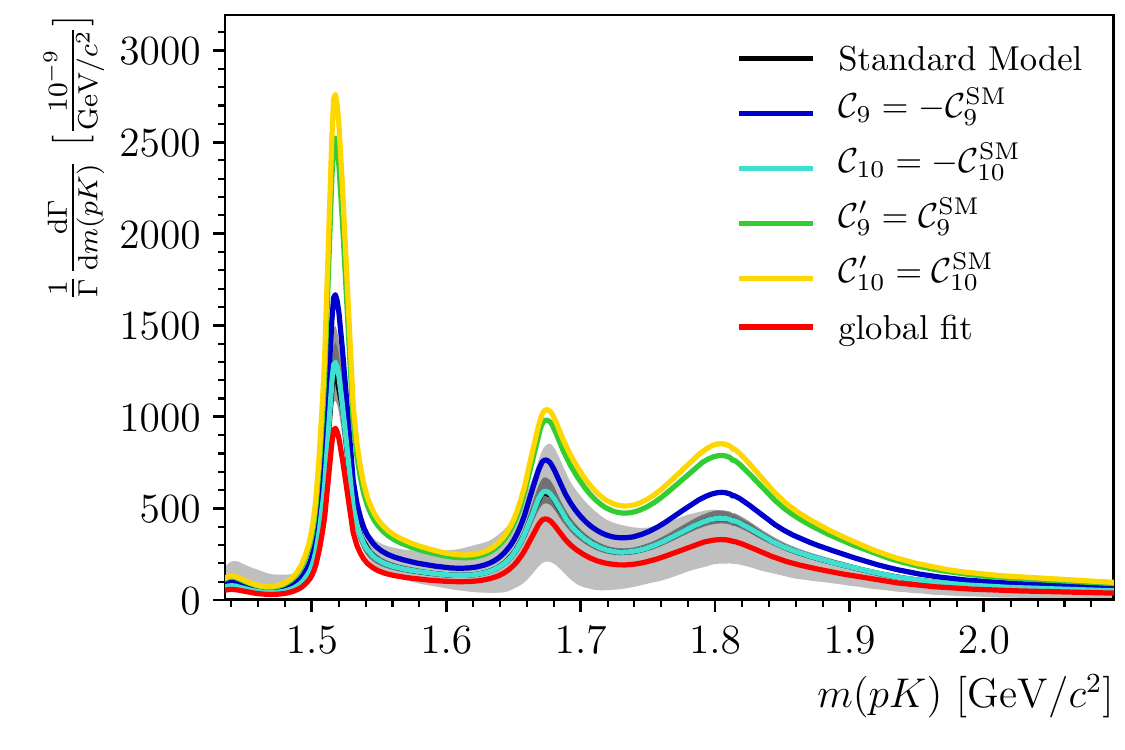}%
    \includegraphics[width=.48\textwidth]{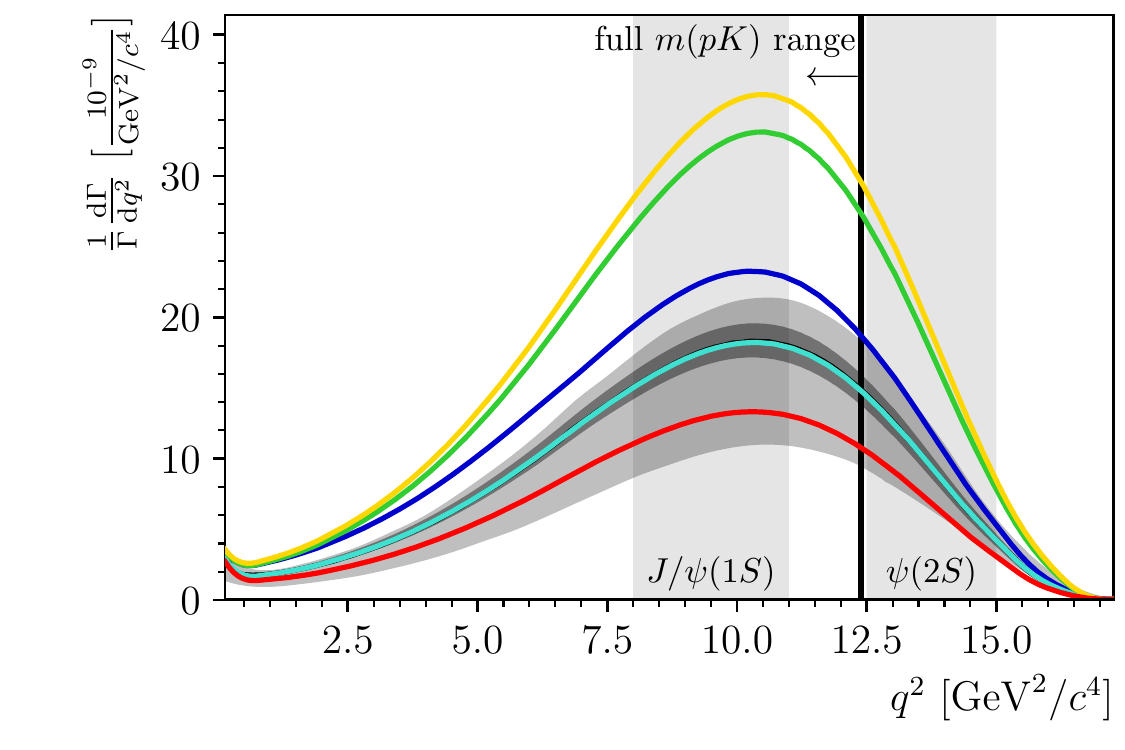}
    \caption{Differential branching fraction as a function of \mpk and \qsq for an ensemble of \Lstar resonances in the SM (black line) and different non-SM scenarios (coloured lines). 
    The possible values given the unknown phases, \phase, is represented by the lighter gray band and the other uncertainties by the darker gray band.
    For $\qsq \gsim 12.4\gev^{2}/c^4$, the available phase-space suppresses the contribution from higher-mass \Lstar resonances. }
    \label{fig:decrate}
\end{figure}

\begin{figure}[htb!]
    \centering
    \includegraphics[width=.48\textwidth]{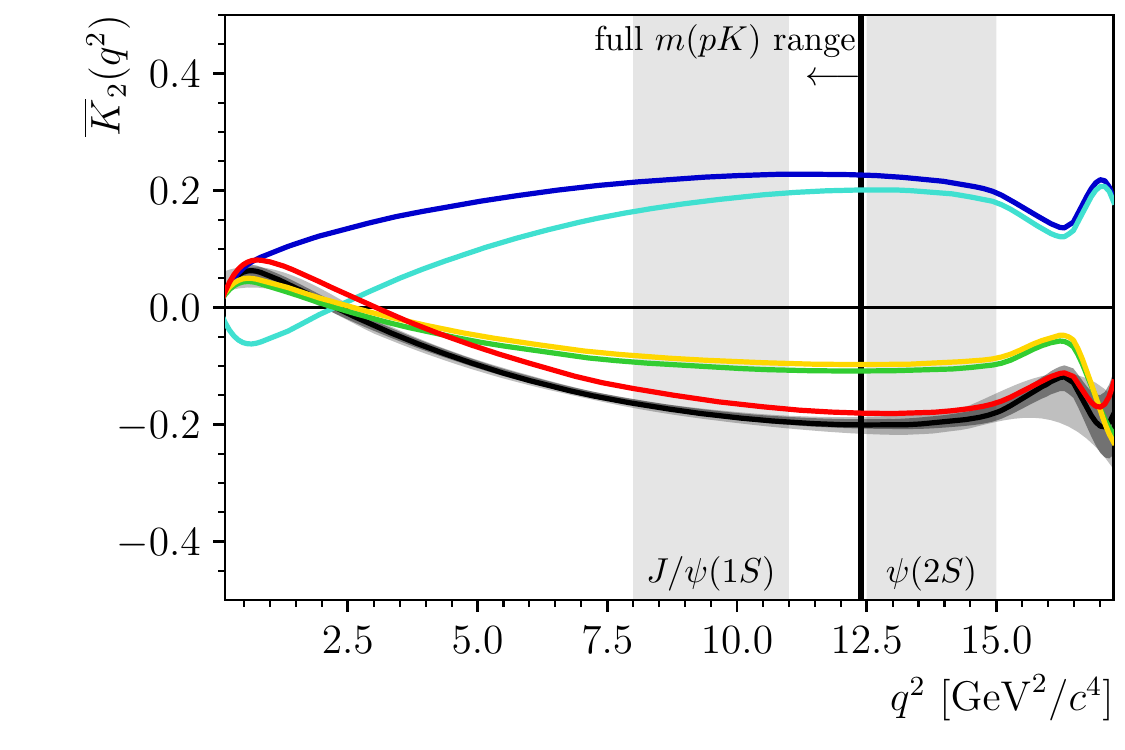}%
    \includegraphics[width=.48\textwidth]{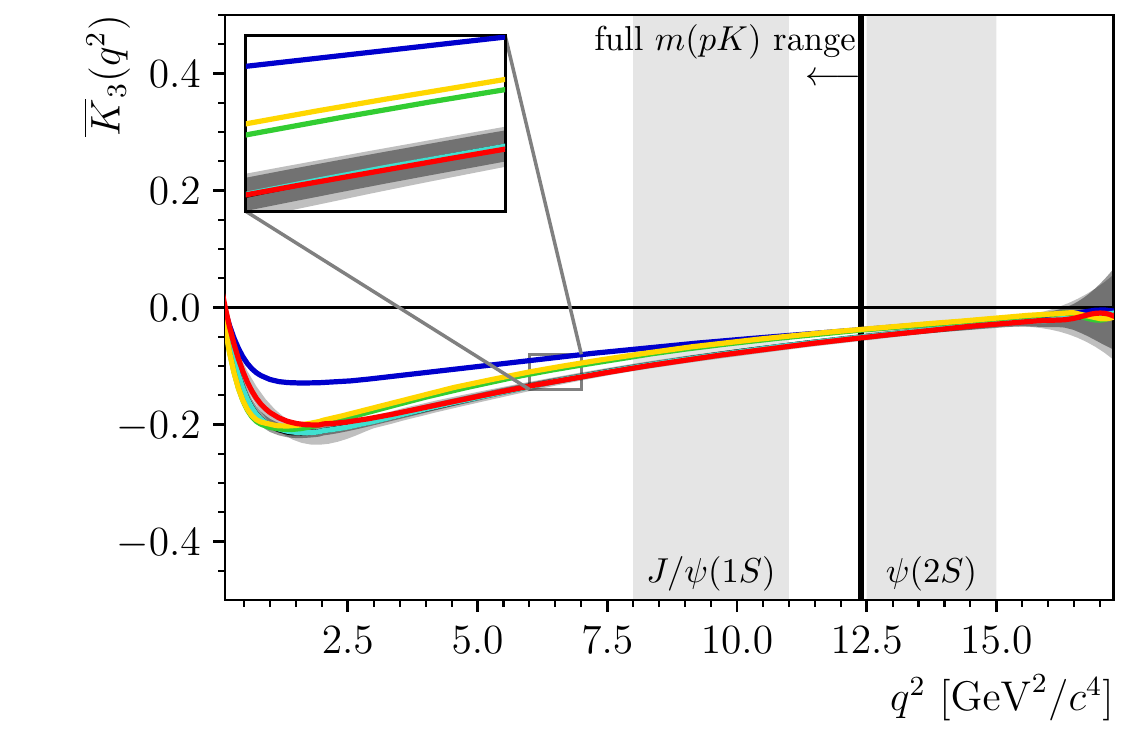}
    \includegraphics[width=.48\textwidth]{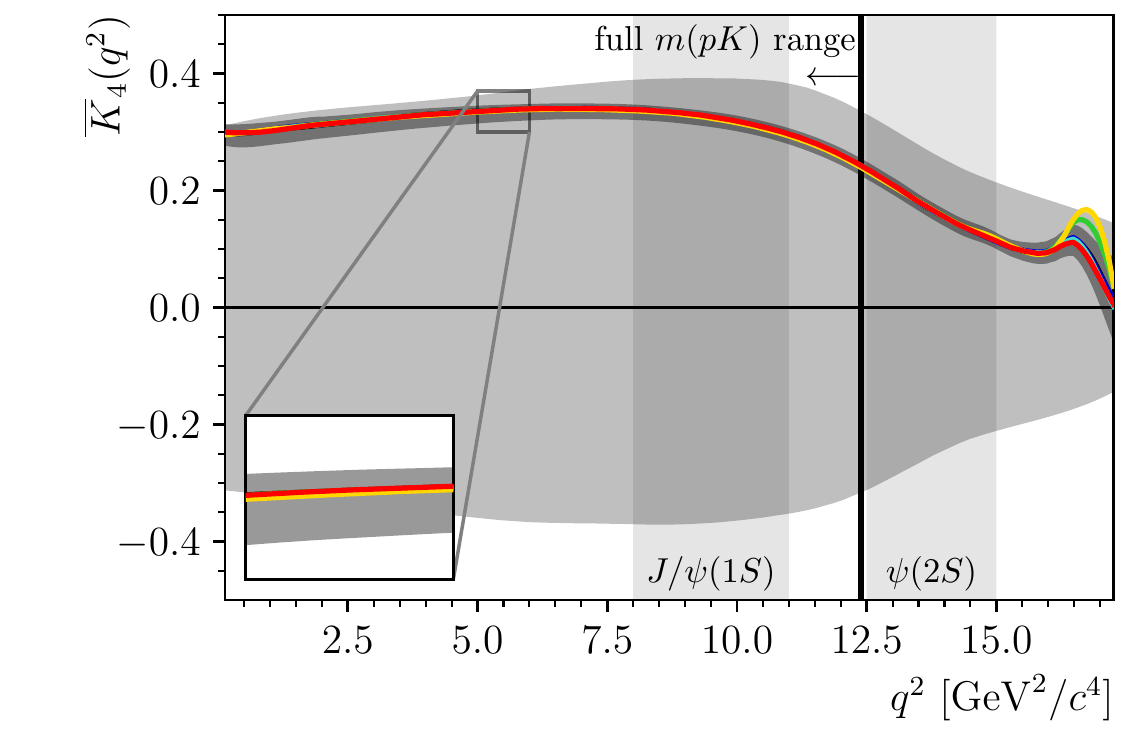}%
    \includegraphics[width=.48\textwidth]{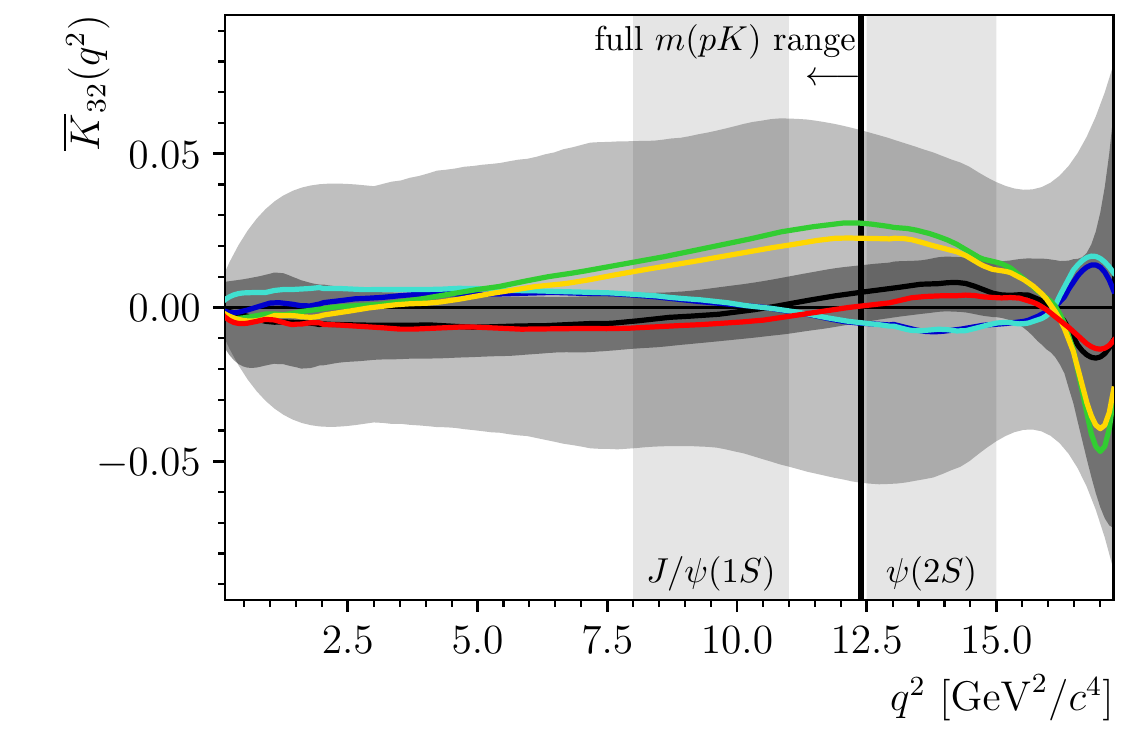}
    \caption{
    Observables \Kobs{2,3,4,32} as a function of \qsq for an ensemble of \Lstar resonances in the SM (black line) and different non-SM scenarios (coloured lines using the same colour code as in Fig.~\ref{fig:decrate}.). 
    The possible values given the unknown phase, \phase, is represented by the lighter gray band and the theory uncertainty by the darker gray band. For $\qsq \gsim 12.4\gev^{2}/c^4$, the available phase-space suppresses the contribution from higher-mass \Lstar resonances. }
    \label{fig:moments_allinfo}
\end{figure}

\begin{figure}[htb!]
    \centering
    \includegraphics[width=.48\textwidth]{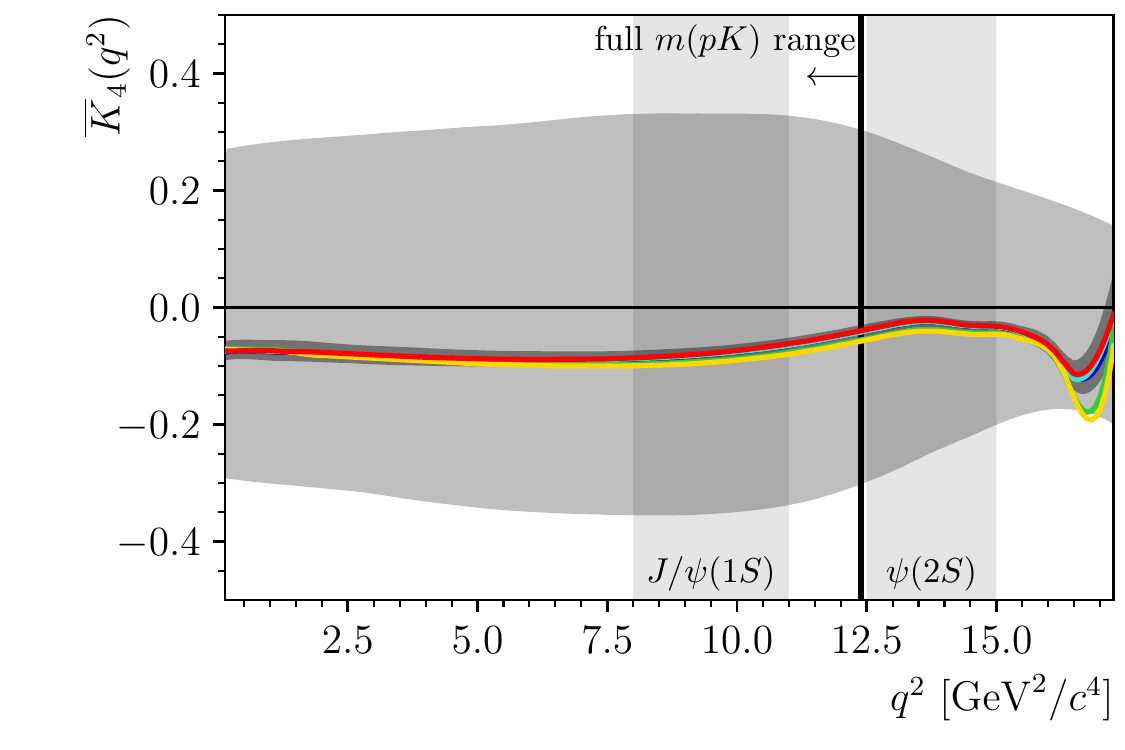}%
    \includegraphics[width=.48\textwidth]{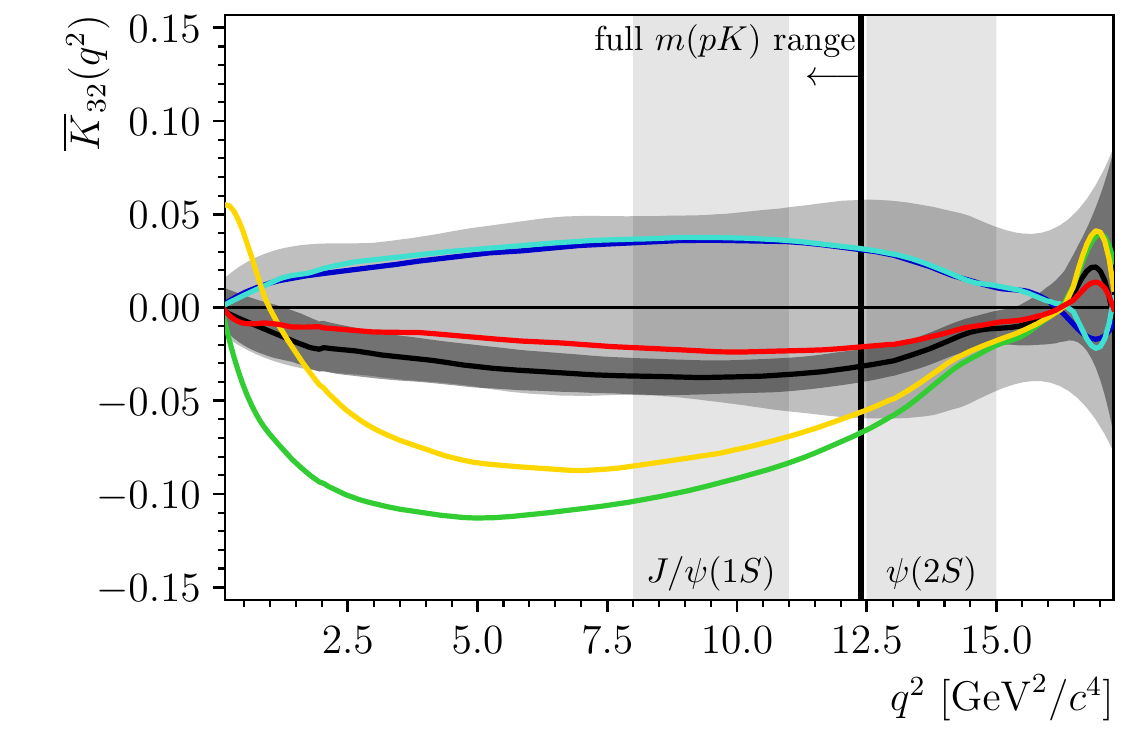}
    \caption{
    Phase-dependent observables \Kobs{4} and \Kobs{32} as a function of \qsq when setting the phases of the spin-\jThree resonances to $\pi$, $\delta_{1520}=\delta_{1690}=\delta_{1890}=\pi$, while keeping all other phases at zero. 
    The lines and bands carry the same meaning as in previous figures. 
    For $\qsq \gsim 12.4\gev^{2}/c^4$, the available phase space suppresses the contribution from higher-mass \Lstar resonances.}
    \label{fig:moments_altphases}
\end{figure}

Figure~\ref{fig:decrate} shows the predicted differential branching fraction for the ensemble of resonances. The angular observables \Kobs{2,3,4,32} are shown in Fig.~\ref{fig:moments_allinfo}. 
Similarly to the single-resonance case considered before, the lepton-forward backward asymmetry (proportional to \Kobs{2}) is very sensitive to changes in the left-handed currents while \Kobs{3} is sensitive to \WC{7}--\WC{9} interference.
Varying the phases between the resonances affects the observables \Kobs{2,3} only slightly, because the overlapping resonances typically have different spin-parity and their interference does not contribute to these observables. 
Conversely, varying the phases completely changes the behaviour of \Kobs{4}.
The observable \Kobs{4} generates a hadron-side forward-backward asymmetry and only arises due to interference between states with different quantum numbers (see Eq.~\ref{eq:K4}) and vanishes for a single state. 
Like \Kobs{4}, all of the other observables that are independent of $\cos\tl$, and arise purely from interference between different resonances, show little dependence on the different non-SM scenarios.
These observables can be used to understand the contribution from the different resonances without being significantly effected by the presence (or not) of new particles that can modify the Wilson coefficients.  
As discusssed in Section~\ref{sec:observables}, the observable \Kobs{32} includes combinations of amplitudes that are present for a single resonance but also terms that only appear due to interference between states with different spins.
If the phases can be measured \Kobs{32} exhibits interesting sensitivity to the different non-SM scenarios. 
Interestingly, different choices of QCD phase give different sensitivities to the different non-SM scenarios. 
This is illustrated in Fig.~\ref{fig:moments_altphases}, which shows the observables \Kobs{4} and \Kobs{32} after changing the phase of all resonances with spin-\jThree to $\pi$, $\delta_{1520}=\delta_{1690}=\delta_{1890}=\pi$, but leaving the others at zero.
With $\phase = 0$, the global-fit values for the Wilson coefficients give rise to observables that are compatible with the SM (Fig.~\ref{fig:moments_allinfo}). 
However, after modifying the phases larger differences are seen in \Kobs{32} between the two scenarios (Fig.~\ref{fig:moments_altphases}). 

\section{Conclusion}

This paper presents a first expression for the angular distribution of \Lbpkll decays comprising a mixture of  \Lstar resonances with spin $\leq\jFive$. 
Considering interference terms gives rise to a complex angular structure and a large number of observables.
The resulting distribution contains 46 (178) angular terms for unpolarised (polarised) \Lb baryons that can be measured.
In this paper, we explore the form of the angular observables and their sensitivity to modifications of the Wilson coefficients. 
A focus is given to observables appearing in the unpolarised case, as the \Lb baryon polarisation at existing experiments is known to be small.  
A particular challenge in interpreting the experimental data on \Lbpkll decays will be the unknown QCD phases between the different resonances. 
Some of the observables explored in this paper only provide useful sensitivity to non-SM scenarios once the phases have been measured.
Others, including the well known lepton forward-backward asymmetry are almost independent of the choice of phase and offer excellent sensitivity to different scenarios. 
There is also a set of observables that arise purely due to interference of different \Lstar resonances. 
These are virtually independent of the values of the Wilson coefficients and can be used to measure the phases and to give valuable input into the validity of form-factor predictions.
The choice of orthogonal basis functions for the angular distribution made in this paper is such that all of the angular observable can be readily extracted from data using a moment analysis using the same set of functions at existing or future experiments.

\section*{Acknowledgements}

We would like to thank Yasmine Amhis, S\'ebastien Descotes-Genon, Biplab Dey, Danny van Dyk, Carla Marin Benito, Mart\'in Novoa-Brunet, Amy Schertz, Javier Virto, Felicia Volle, and Roman Zwicky for their helpful feedback on this paper and discussions on various aspects of this work. 
We would also like to thank the members of the LHCb collaboration for discussions that led to the ideas behind this paper. 
TB and MK would like to acknowledge support from the UK Science and Technology Facilities Council (STFC).
TB would like to acknowledge support from the Royal Society UK. 
\clearpage

\appendix

\section{Effective Wilson coefficients}\label{app:effC9}
In \bsll transitions, tensor-like currents \Op{7}, vector-currents \Op{9} and axialvector currents \Op{10} contribute at leading order.
Their magnitude is given by the corresponding Wilson coefficients, \WC{i}.
However at higher orders, diquark loops represented by four-quark current-current, \Op{1,2}, and QCD penguin, \Op{3-6}, operators become relevant.
Including leading logarithms, the tensor and vector-current coefficients become~\cite{Grinstein:1988me,Beneke:2001at}
\begin{align}
\begin{split}
    \WC{7}^{\rm eff} = \WC{7} &- \frac{1}{9}(4\WC{3} +12\WC{4} -\WC{5} -3\WC{6}) \ , \\
    \WC{9}^{\rm eff}(\qsq) = \WC{9} &+ \frac{2}{9}\left(3\WC{3} + \WC{4 }+ 3\WC{5} + \WC{6}\right) \\
    &-\frac{c_0(\qsq)}{2}\left(\WC{3} + 3\WC{4}\right) \\
    &+ c_1(m_c,\qsq) \left(3\WC{1} + \WC{2} + 3\WC{3} + \WC{4} + 3\WC{5} + \WC{6} \right) \\
    &-\frac{c_1(m_b,\qsq)}{2}\left(4\WC{3} + 4\WC{4} + 3\WC{5} + \WC{6}\right) \ .
\end{split}
\end{align}
The factor $c_0(\qsq)$ appearing with the tensor and axial-tensor QCD penguin operators for light \qqbar pairs, $\Op{3,4}^{(u,d,s)}$, is
\begin{align}
    c_0(\qsq) &= \frac{8}{27} - \frac{4}{9}\log\left(\frac{\qsq}{m_b^2}\right) + \frac{4}{9}i\pi \ .
\end{align}
The situation for the heavy diquark contributions is slightly different above and below the \qqbar threshold and these contribute with a factor
\begin{align}
\begin{split}
    c_1(m_q,\qsq) =-\frac{8}{9}\log\left(\frac{m_q}{m_b}\right) &+ \frac{8}{27} + \frac{16}{9}\frac{m_q^2}{\qsq} \\
    &- \frac{2}{9}(3 - v^2)|v|\times
  \begin{cases}
    2\arctan(|v|^{-1}) &, \qsq<4m_q^2 \\
    \log\left(\frac{1+ |v|}{1 - |v|}\right) - i\pi &, \qsq\geq4m_q^2
  \end{cases}
\end{split}
\end{align}
where $v$ is the velocity of the quark in the dilepton rest frame
\begin{align}
    v^2  = 1 - 4\frac{m_q^2}{\qsq} \ , \quad
    |v| = \sqrt{|v^2|} \ .
\end{align}
The only relevant current-current operators involve the charm quark. 
Note that any long-distance contributions from \qqbar resonances are neglected.
Table~\ref{tab:wilsoncoeffs} provides the SM values for the Wilson coefficients used in this paper.

\begin{table}[htb!]
    \centering
    \caption{Wilson coefficients used in the generator assuming the SM~\cite{Descotes-Genon:2013vna} and a global fit to mesonic \bsll measurements~\cite{Alguero:2021anc}.}
    \label{tab:wilsoncoeffs}
    \begin{tabular}{c|cc}
    \toprule
    & Standard Model & global fit \\
    \midrule
    \WC{1} & $-0.2632$ & \\
    \WC{2} & $\hphantom{-}1.0111$ & \\
    \WC{3} & $-0.0055$ & \\
    \WC{4} & $-0.0806$ & \\
    \WC{5} & $\hphantom{-}0.0004$ & \\
    \WC{6} & $\hphantom{-}0.0009$ & \\
    \midrule
    \WC{7} & $-0.3120$ & $-0.3120$ \\
    \WC{9} & $\hphantom{-}4.0749$ & $\hphantom{-}2.9949$ \\
    \WC{10} & $-4.3085$  & $-4.1585$ \\
    \midrule
    \WC{7^\prime} & $\hphantom{-}0.0000$ & $\hphantom{-}0.0000$ \\
    \WC{9^\prime} & $\hphantom{-}0.0000$ & $\hphantom{-}0.1600$ \\
    \WC{10^\prime} & $\hphantom{-}0.0000$ & $-0.1800$ \\
    \bottomrule
    \end{tabular}
\end{table}

\section{Polarisation vector and Rarita-Schwinger representations}
\label{app:spinors}

In this appendix, we discuss the details of the representation of Dirac spinors, Rarita-Schwinger objects and polarisation vectors.
A spin-\jOne particle with mass $m$ and four-momentum 
\begin{align}
p^\mu=(p^0,|\vec{p}|\cos\phi\sin\theta,|\vec{p}|\sin\phi\sin\theta,|\vec{p}|\cos\theta)
\end{align}
is represented by the following Dirac spinors for positive and negative helicity~\cite{Haber:1994pe}
\begin{align}
    u(p,+\tfrac{1}{2}) = \begin{pmatrix}
   \sqrt{p^0 + m}\cos\frac{\theta}{2} \\
   \sqrt{p^0 + m}\sin\frac{\theta}{2}e^{i\phi} \\
   \sqrt{p^0 - m}\cos\frac{\theta}{2} \\
   \sqrt{p^0 - m}\sin\frac{\theta}{2}e^{i\phi}
    \end{pmatrix} \quad,\quad
    u(p,-\tfrac{1}{2}) = \begin{pmatrix}
   -\sqrt{p^0 + m}\sin\frac{\theta}{2}e^{-i\phi} \\
   \sqrt{p^0 + m}\cos\frac{\theta}{2} \\
   \sqrt{p^0 - m}\sin\frac{\theta}{2}e^{-i\phi} \\
   -\sqrt{p^0 - m}\cos\frac{\theta}{2}
   \end{pmatrix}~.
\end{align}
The spin-\jThree Rarita-Schwinger objects are constructed from the Dirac spinors, $u(k,\pm\tfrac{1}{2})$, and spin-1 polarisation vectors for a massive particle as described in Ref.~\cite{Chung:1971ri}. 
For $\jLambda=\jThree$, 
\begin{align}
    u_\alpha(k,\lLambda) = \sum_{\lambda_\text{i}=-1}^{1}\sum_{\lambda_\text{d}=-1/2}^{1/2} \underbrace{\Bigl< 1\,\lambda_\text{i},\frac{1}{2}\,\lambda_\text{d}\Big|\frac{3}{2}\,\lLambda\Bigr>}_\text{Clebsch-Gordan}\overbrace{e_\alpha(\lambda_\text{i})}^\text{spin-1 vector}\underbrace{u(k,\lambda_\text{d})}_\text{Dirac-spinor} 
\end{align}
and for $\jLambda=\jFive$,
\begin{align}
    u_{\alpha\beta}(k,\lLambda) = \sum_{\lambda_\text{i}=-2}^{2}\sum_{\lambda_\text{d}=-1/2}^{1/2} \underbrace{\Bigl< 2\,\lambda_\text{i},\frac{1}{2}\,\lambda_\text{d}\Big|\frac{5}{2}\,\lLambda\Bigr>}_\text{Clebsch-Gordan}\overbrace{e_{\alpha\beta}(\lambda_\text{i})}^\text{spin-2 tensor}\underbrace{u(k,\lambda_\text{d})}_\text{Dirac-spinor}~,
\end{align}
where \lLambda is the \Lstar helicity and the spin-2 tensor is constructed from polarisation vectors as
\begin{align}
    e^{\alpha\beta}(\lambda_\text{i}) = \sum_{\lambda_1=-1}^1\sum_{\lambda_2=-1}^1
    \underbrace{\Bigl< 1\,\lambda_1,1\,\lambda_2\Big|2\,\lambda_\text{i}\Bigr>}_\text{Clebsch-Gordan}e^\alpha(\lambda_1){e^\beta}(\lambda_2)^\text{T}~.
\end{align}
The Rarita-Schwinger objects satisfy 
\begin{equation}
\begin{alignedat}{3}
    k^\alpha u_{\alpha\beta} &= 0\,,\qquad &&g^{\alpha\beta}u_{\alpha\beta} = 0\,,\qquad&&(\gamma^\mu k_\mu-m)u_{\alpha\beta} = 0 \,, \\
    \gamma^\alpha u_{\alpha\beta} &= 0\,,\qquad&&\hphantom{g^{\alpha\beta}}u_{\alpha\beta} = u_{\beta\alpha}\,. &&
\end{alignedat}
\end{equation}
In the \Lb rest frame, the four-momenta of the \Lb, the \Lstar resonance and the dilepton system are
\begin{align}
    p^\mu = \left( \mLb,\,0,\,0,\,0\right)\,, \qquad
    k^\mu = \left( \mLb-q^0,\, 0,\,0,\,|\vec{k}| \right)\,, \qquad
    q^\mu = \left( q^0,\,0,\,0,\,-|\vec{k}|\right)\,.
\end{align}
The polarisation vectors used to construct the Rarita-Schwinger objects for the \Lstar resonances are given by 
\begin{align}
    && e^\alpha(0) = \frac{1}{\mpk} \left( |\vec{q}\,|,\,0,\,0,\,\mLb-q^0 \right) \,,
    && e^\alpha(\pm1) = \frac{1}{\sqrt{2}}\left( 0,\, \mp1,\,-i,\,0 \right) ~.
    \label{eq:pol:rs}
\end{align}
In the \Lb rest frame, the spin-1 polarisation-vectors representing the vector-boson polarisations are
\begin{align}
\begin{split} 
    \varepsilon^{\mu}(t) &= \frac{1}{\sqrt{\qsq}} \left( q^0,\,0,\,0,\,-|\vec{q}\,| \right)\,, \\
    \varepsilon^{\mu}(0) &= \frac{1}{\sqrt{\qsq}} \left(|\vec{q}\,|,\,0,\,0,\,-q^0 \right)\,, \\
    \varepsilon^{\mu}(\pm) &= \frac{1}{\sqrt{2}} \left(0,\,\pm1,\,-i,\,0\right)\,.
\end{split} 
\end{align}
Note, the relative sign differences w.r.t. Equation~\ref{eq:pol:rs} arise because the dilepton system flies in negative $z$-direction in the \Lb rest frame.
The corresponding spin-1 polarisation objects in the dilepton rest frame (used to calculate the lepton helicity amplitudes) are
\begin{align}
    &&\varepsilon^{\mu}(t) = \left( 1,\,0,\,0,\,0 \right)\,,
    &&\varepsilon^{\mu}(0) = \left(0,\,0,\,0,\,-1\ \right)\,,
    &&\varepsilon^{\mu}(\pm) = \frac{1}{\sqrt{2}} \left( 0,\,\pm1,\,-i,\,0\right)\,.
\end{align}

\section{Definition of the angles}
\label{app:angles}

The main body of this document defines $p$, $k$, $q$, $k_1$ and $q_1$ as the four momentum of the \Lb baryon, the \Lstar baryon, the dilepton system, the proton, and the positively charged lepton, respectively.
Throughout this paper, unless specified otherwise, the four momentum of the \Lb baryon is in the laboratory frame and all other momenta are in their respective parent rest frames.
The decay angles are defined by unit vectors in the direction of the particle three-momentum: $\hat{p}, \hat{k}, \hat{q}, \hat{k}_1, \hat{q}_1$.

The polarization angle \thetab is the angle between the polarisation axis and the \Lstar momentum in the \Lb rest frame
\begin{align}
    \cos\thetab = \hat{n}\cdot\hat{k} \, .
\end{align}
Two common choices for the polarisation axis definition are $\hat{n}_\perp = \hat{p}\times\hat{p}_\text{beam}$ and $\hat{n}_\parallel = \hat{p}$, corresponding to resolving transverse and longitudinal polarisation, respectively. 

The proton (lepton) helicity angle, \tp(\tl), is the angle between the proton ($\ellp$) momentum in the \Lstar (dilepton) rest frame and the \Lstar (dilepton) momentum in the \Lb rest frame, \ie
\begin{align}
    \cos\tp = \hat{k}\cdot\hat{k}_1 \qquad\text{and}\quad\cos\tl = \hat{q}\cdot\hat{q}_1 \, .
\end{align}
The proton and lepton azimuthal angles are
\begin{align}
    \phip = \text{sgn}(k_{1x})\arctan\left(\frac{k_{1y}}{k_{1x}}\right) \qquad\text{and} \qquad\phil = \text{sgn}(q_{1x})\arctan\left(\frac{q_{1y}}{q_{1x}} \right)\, ,
\end{align}
where $k_{1x,1y}$ and $q_{1x,1y}$ are the $x,y$ components of $\vec{k}_1$ and $\vec{q}_1$ given in the coordinate system where the $x$-axis is parallel to $\hat{n}$ and the $z$-axis coincides with $\hat{k}$ and $\hat{q}$ respectively as illustrated using the coordinate systems associated with the \Lstar (dilepton) decay plane in Fig.~\ref{fig:angular:def}.
\section{Translation between lattice and quark-model form factors}\label{app:FFtranslation}

This appendix provides a translation between the form factor parameterisation used in the quark-model prediction by Mott and Roberts in Refs.~\cite{Mott:2011cx,Mott:2015zma} and those used in the lattice predictions. The lattice calculations typically use a helicity based definition of the form factors that follows from Ref.~\cite{Feldmann:2011xf}, whereas the quark-model uses a different expansion as discussed below. 

\subsection[Translation for \texorpdfstring{$\jOne^{+}$ to $\jOne^{+}$}{1/2+ to 1/2+} transitions]{Translation for \texorpdfstring{\boldmath$\jOne^{+}$ to $\jOne^{+}$}{1/2+ to 1/2+} transitions} 

In the quark-model parameterisation of the form factors, the amplitudes for the hadronic transition between a $\jOne^+$ \Lb baryon and a $\jOne^+$ \Lz baryon are expanded in the general parameterisation
\begin{equation}
\begin{aligned}
\bra{\Lstar} \squarkbar \gamma^\mu \bquark \ket{\Lb} &= \bar{u}(k,\lLambda)\left[ F_1 \gamma^{\mu} + F_2 v_p^{\mu} + F_3 v_k^{\mu} \right] & &u(p,\lLb)~, \\
\bra{\Lstar} \squarkbar \gamma^\mu \gamma_5 \bquark \ket{\Lb} &= \bar{u}(k,\lLambda)\left[ G_1 \gamma^{\mu} + G_2 v_p^{\mu} + G_3 v_k^{\mu} \right] &\gamma_5 &u(p,\lLb)~, \\ 
\bra{\Lstar} \squarkbar i\sigma^{\mu\nu} q_{\nu} \bquark \ket{\Lb} &= \bar{u}(k,\lLambda)\left[ F^{\rm T}_1 \gamma^{\mu} + F^{\rm T}_2 v_p^{\mu} + F^{\rm T}_3 v_k^{\mu} \right] & &u(p,\lLb)~,\\ 
\bra{\Lstar} \squarkbar i\sigma^{\mu\nu} \gamma_{5} q_{\nu} \bquark \ket{\Lb} &= \bar{u}(k,\lLambda)\left[ G^{\rm T}_1 \gamma^{\mu} + G^{\rm T}_2 v_p^{\mu} + G^{\rm T}_3 v_k^{\mu} \right] &\gamma_5 &u(p,\lLb)~.
\end{aligned}
\end{equation}
Here, $u(p,\lLb)$ is a Dirac spinor, $F^{(\rm T)}_{1-3}$ and, $G^{(\rm T)}_{1-3}$ are functions of \qsq, and $v_p = p/\mLb$ and $v_k= k/\sqrt{\ksq}$ are the 4-velocities of the \Lb and the \Lz baryons.
In the lattice QCD parameterisation of Ref.~\cite{Detmold:2016pkz}, the vector current is instead expanded in terms of longitudinal and transverse polarisation states as 
\begin{align}
\begin{split}
\bra{\Lstar} \squarkbar \gamma^\mu \bquark \ket{\Lb} = \bar{u}(k,\lLambda)\left[ \right. & f_0 \cdot (\mLb - \mLambda) \frac{q^\mu}{\qsq} + \\ 
& f_{+}\cdot \frac{\mLb+\mLambda}{s_+}\left(p^\mu +  k^\mu - (\mLb^{2} - \mLambda^{2})\frac{q^\mu}{\qsq}\right)+ \\
& \left. f_{\perp} \cdot \left( \gamma^\mu - 2\frac{\mLambda}{s_+} p^\mu - 2\frac{\mLb}{s_+} k^\mu \right) \right] u(p,\lLb)~,
\end{split}
\end{align} 
where $s_{\pm} = (\mLb \pm \mLambda)^2 - \qsq$. 
Comparing terms in the expressions, 
\begin{align}
\begin{split}\label{eq:L0_vec}
F_{1} &= f_{\perp} \,,\\
F_{2} &= f_0 \frac{\mLb(\mLb-\mLambda)}{\qsq} - f_{\perp} \frac{2 \mLambda\mLb}{s_+} + f_+ \frac{\mLb(\mLb + \mLambda)}{s_+}\left( 1 - \frac{(\mLb^2 - \mLambda^{2})}{\qsq}\right) \,, \\ 
F_{3} &= -f_{0}\frac{\mLambda(\mLb-\mLambda)}{\qsq} - f_{\perp} \frac{2\mLambda\mLb}{s_+} + f_{+} \frac{\mLambda(\mLb + \mLambda)}{s_+}\left(1 + \frac{(\mLb^2 - \mLambda^2)}{\qsq} \right)\,.
\end{split}
\end{align}
The axialvector current is 
\begin{align}
\begin{split}
    \bra{\Lstar} \squarkbar \gamma^\mu \gamma_5\bquark \ket{\Lb} = - \bar{u}(k,\lLambda) \gamma_5 \left[ \right. 
    & g_{0} (\mLb + \mLambda) \frac{q^\mu}{\qsq} + \\ 
    & g_{+} \frac{\mLb-\mLambda}{s_-}\left(p^\mu + k^\mu - (\mLb^2 - \mLambda^2)\frac{q^{\mu}}{\qsq}\right) + \\
    & \left. g_{\perp} \left( \gamma^{\mu} + 2 \frac{\mLambda}{s_-}p^{\mu} - 2\frac{\mLb}{s_-} k^\mu \right) \right] u(p,\lLb)~.
\end{split}
\end{align}
Comparing terms in the expressions, 
\begin{align}
\begin{split}\label{eq:L0_axvec}
    G_{1} &= g_{\perp} \,, \\ 
    G_{2} &= -g_0 \frac{\mLb(\mLb + \mLambda)}{\qsq} - g_{\perp} \frac{2 \mLb \mLambda}{s_-} - g_{+} \frac{\mLb(\mLb-\mLambda)}{s_-}\left( 1 - \frac{(\mLb^2 - \mLambda^2)}{\qsq} \right) \,, \\ 
    G_{3} &= g_{0} \frac{\mLambda(\mLb+\mLambda)}{\qsq} + g_{\perp} \frac{2\mLambda\mLb}{s_-} - g_{+} \frac{\mLambda(\mLb - \mLambda)}{s_-}\left( 1 + \frac{(\mLb^2 - \mLambda^{2})}{\qsq}\right)\,.
\end{split}
\end{align}
Note, there is no relative sign between $G_{1}$ and $g_{\perp}$ because $\gamma^\mu\gamma_5 = - \gamma_5\gamma^{\mu}$.  
The tensor current is given by 
\begin{align}
\begin{split}
    \bra{\Lstar} \squarkbar i\sigma^{\mu\nu} q_{\nu} \bquark \ket{\Lb} = - \bar{u}(k,\lLambda) \left[ \right.
    & h_{+} \frac{\qsq}{s_+} \left( p^\mu + k^\mu - (\mLb^{2} - \mLambda^{2})\frac{q^{\mu}}{\qsq} \right) + \\
    & \left. h_{\perp} (\mLb + \mLambda) \left( \gamma^{\mu} - \frac{2\mLambda}{s_+} p^{\mu} - \frac{2\mLb}{s_+} k^\mu  \right)\right] u(p,\lLb)\ ,
\end{split}
\end{align}
where $h$ should not be confused with the hadron helicity amplitudes.
Comparing terms in the two expressions, 
\begin{align}
\begin{split}\label{eq:L0_vectensor}
    F_{1}^{\rm T} &= -h_{\perp}(\mLb + \mLambda) \,, \\ 
    F_{2}^{\rm T} &= -h_{+} \frac{\mLb\qsq}{s_+}\left( 1 - \frac{(\mLb^2 - \mLambda^{2})}{\qsq}\right) + h_{\perp} \frac{2\mLambda\mLb(\mLb+ \mLambda)}{s_+}  \,,\\ 
    F_{3}^{\rm T} &= -h_{+}\frac{\mLambda\qsq}{s_+}\left( 1 + \frac{(\mLb^{2} - \mLambda^{2})}{\qsq} \right) + h_{\perp} \frac{2\mLambda\mLb(\mLb + \mLambda)}{s_+} \,.\\ 
\end{split}
\end{align}
The axialtensor current is given by
\begin{align}
\begin{split}
    \bra{\Lstar} \squarkbar i\sigma^{\mu\nu} q_{\nu}\gamma_{5} \bquark \ket{\Lb} = - \bar{u}(k,\lLambda) \gamma_{5} \left[ \right. 
    & \tilde{h}_{+} \frac{\qsq}{s_-}\left( p^{\mu} + k^\mu - (\mLb^2 - \mLambda^2) \frac{q^{\mu}}{\qsq} \right) + \\
    & \left. \tilde{h}_{\perp} (\mLb - \mLambda) \left( \gamma^{\mu} + \frac{2\mLambda}{s_-} p^{\mu} - \frac{2\mLb}{s_-} k^\mu \right) \right] u(p,\lLb)~,
\end{split}
\end{align}
where $\Tilde{h}$ should not be confused with the hadron helicity amplitudes.
Comparing terms in the two expressions,  
\begin{align}
\begin{split}\label{eq:L0_axtensor}
    G_{1}^{\rm T} &= \tilde{h}_{\perp}(\mLb - \mLambda)\,, \\ 
    G_{2}^{\rm T} &= -\tilde{h}_{+} \frac{\mLb\qsq}{s_-}\left( 1 - \frac{(\mLb^{2} - \mLambda^{2})}{\qsq} \right) - \tilde{h}_{\perp} \frac{2 \mLb \mLambda (\mLb - \mLambda)}{s_-}  \,,\\ 
    G_{3}^{\rm T} &= -\tilde{h}_{+} \frac{\mLambda\qsq}{s_-} \left( 1 + \frac{(\mLb^2 - \mLambda^2)}{\qsq}\right) + \tilde{h}_{\perp} \frac{2 \mLambda \mLb (\mLb - \mLambda)}{s_-} \,.\\
\end{split}
\end{align}

Note that at the kinematic end-points, where either the dilepton or the dihadron system are at rest in the \Lb rest-frame, additional symmetries apply
\begin{equation}
\begin{aligned}
    &f_0(0) = f_+(0) \ , &&g_{\perp}(\qsq_{\rm max}) = g_{+}(\qsq_{\rm max}) \ , \\
    &g_0(0) = g_{+}(0) \ , &&\tilde{h}_{\perp}(\qsq_{\rm max}) = \tilde{h}_{+}(\qsq_{\rm max}) \ .
\end{aligned}
\end{equation}
Furthermore there are inexact relationships resulting from Heavy Quark Effective Theory~\cite{Eichten:1989zv} that imply~\cite{Hiller:2021zth} 
\begin{align}
f_0(\qsq) \approx g_{+}(\qsq) \approx \tilde{h}_+(\qsq) \approx \tilde{h}_{\perp}(\qsq)
\end{align}
and
\begin{align}
g_0(\qsq) \approx f_{+}(\qsq) \approx h_+(\qsq) \approx h_{\perp}(\qsq)
\end{align}
for small \qsq.

\subsection[Translation for \texorpdfstring{$\jOne^{+}$ to $\jThree^{-}$}{1/2+ to 3/2-} transitions]{Translation for \texorpdfstring{\boldmath$\jOne^{+}$ to $\jThree^{-}$}{1/2+ to 3/2-} transitions}

For $\jOne^+$ to $\jThree^-$ transitions two different labelling conventions appear for the helicity based definitions of the form-factors. The form-factors $f_0$ and $f_+$ in Ref.~\cite{Detmold:2016pkz} are referred to as $f_t$ and $f_0$ in Refs.~\cite{Boer:2014kda, Descotes-Genon:2019dbw}, respectively. 
Using the notation of Ref.~\cite{Mott:2011cx}, the vector current is 
\begin{eqnarray}
\bra{ {\Lstar}} \overline{s}\gamma^{\mu}b\ket{\Lb}&=&\overline{u}_\alpha(k,\lLambda)\bigg[v^\alpha\bigg(F_{1}\gamma^{\mu}+F_{2}v^\mu +F_{3}v'^\mu\bigg)+F_4 g^{\alpha\mu}\bigg]u(p,\lLb) \ .
\end{eqnarray}
The expression in Ref.~\cite{Descotes-Genon:2019dbw} is
\begin{equation}
\begin{aligned}
\bra{{\Lstar}} \bar s \gamma^\mu b\ket{\Lb } =\
& \bar u_\alpha(k,\lLambda)\biggl\{p^\alpha\biggl[f_t^V(\qsq) (\mLb-\mLambda)\frac{q^\mu}{\qsq}+\\
& f_0^V(\qsq) \frac{\mLb+\mLambda}{s_+}(p^\mu +k^\mu-\frac{q^\mu}{\qsq}(\mLb^2-m_{{\Lstar}}^2))+\\
& f_\perp^V(\qsq)(\gamma^\mu-2\frac{\mLambda}{s_+}p^\mu -2\frac{\mLb}{s_+}k^\mu)\biggr]+\\
& f_g^V(\qsq) \left[g^{\alpha\mu}+\mLambda\frac{p^\alpha}{s_-} \left(\gamma^\mu - 2 \frac{k^\mu}{\mLambda} +2 \frac{\mLambda p^\mu +\mLb k^\mu}{s_+}\right)\right]\biggr\}u(p,\lLb) \,.
\end{aligned}
\end{equation}
Comparing terms in the two expressions, 
\begin{align}
\begin{split}\label{eq:LStar_vec}
    F_{1} &= \mLb \left( f_\perp^V + f_g^V\frac{\mLambda}{s_-} \right) \,,\\
    F_{2} &= \mLb \Biggl( f_t^V\frac{\mLb(\mLb-\mLambda)}{\qsq} + f_0^V\frac{\mLb(\mLb+\mLambda)}{s_+}\left(1-\frac{\mLb^2-\mLambda^2}{\qsq}\right) \\
    &\qquad\qquad\qquad- f_\perp^V\frac{2\mLb \mLambda}{s_+} + f_g^V\frac{2\mLb \mLambda^2}{s_-s_+} \Biggr)\,, \\
    F_{3} &= \mLb \Biggl( -f_t^V\frac{\mLambda(\mLb-\mLambda)}{\qsq} + f_0^V\frac{\mLambda(\mLb+\mLambda)}{s_+}\left(1+\frac{\mLb^2-\mLambda^2}{\qsq}\right) \\
    &\qquad\qquad\qquad- f_\perp^V\frac{2\mLb \mLambda}{s_+} + f_g^V\frac{2\mLambda}{s_-}\left(\frac{\mLb \mLambda}{s_+}-1\right) \Biggr) \,,\\
    F_{4} &= f_g^V\,.
\end{split}
\end{align}
The axial-vector current in the quark-model prediction is
\begin{eqnarray}
\bra{{\Lstar}} \overline{s}\gamma^{\mu}\gamma_{5}b\ket{\Lb}&=&\overline{u}_\alpha(k,\lLambda)\bigg[v^\alpha\bigg(G_{1}\gamma^{\mu}+G_{2}v^\mu +G_{3}v'^\mu\bigg)+G_4 g^{\alpha\mu}\bigg]\gamma_{5}u(p,\lLb)\,.
\end{eqnarray}
The corresponding expression in Ref.~\cite{Descotes-Genon:2019dbw} is 
\begin{equation}
\begin{aligned}
\bra{{\Lstar}} \bar s \gamma^\mu \gamma ^5 b\ket{\Lb}=& -\bar u_\alpha(k,\lLambda)\gamma^5\biggl\{p^\alpha\biggl[
f_t^A(\qsq) (\mLb+\mLambda)\frac{q^\mu}{\qsq}+\\
    & f_0^A(\qsq) \frac{\mLb-\mLambda}{s_-}(p^\mu +k^\mu-\frac{q^\mu}{\qsq}(\mLb^2-\mLambda^2))+\\
& f_\perp^A(\qsq)(\gamma^\mu+2\frac{\mLambda}{s_-}p^\mu -2\frac{\mLb}{s_-}k^\mu)\biggr]+\\
& f_g^A(\qsq) \left[g^{\alpha\mu}-\mLambda\frac{p^\alpha}{s_+} \left(\gamma^\mu + 2 \frac{k^\mu}{\mLambda} -2 \frac{\mLambda p^\mu -\mLb k^\mu}{s_-}\right)\right]\biggr\}u(p,\lLb)\,.
\end{aligned}
\end{equation}
Comparing terms in the two expressions, 
\begin{align}
\begin{split}\label{eq:LStar_axvec}
    G_{1} &= \mLb \left( f_\perp^A - f_g^A\frac{\mLambda}{s_+} \right)\,, \\
    G_{2} &= \mLb \Biggl( -f_t^A\frac{\mLb(\mLb+\mLambda)}{\qsq} - f_0^A\frac{\mLb(\mLb-\mLambda)}{s_-}\left(1-\frac{\mLb^2-\mLambda^2}{\qsq}\right) \\
    &\qquad\qquad\qquad-f_\perp^A\frac{2\mLambda \mLb}{s_-} - f_g^A\frac{2\mLambda^2\mLb}{s_+s_-} \Biggr) \,,\\
    G_{3} &= \mLb \Biggl( f_t^A\frac{\mLambda(\mLb+\mLambda)}{\qsq} - f_0^A\frac{\mLambda(\mLb-\mLambda)}{s_-}\left(1+\frac{\mLb^2-\mLambda^2}{\qsq}\right)\\
    &\qquad\qquad\qquad+ f_\perp^A\frac{2\mLb \mLambda}{s_-} + f_g^A\frac{2\mLambda}{s_+}\left(1+\frac{\mLb \mLambda}{s_-}\right) \Biggr)\,, \\
    G_{4} &= - f_g^A\,.
\end{split}
\end{align}
The tensor current in the quark-model is 
\begin{equation}
\begin{aligned}
\bra{{\Lstar}}\overline{s}i\sigma^{\mu\nu}q_\nu b\ket{\Lb} &=& \overline{u}_\alpha(k,\lLambda)\bigg[v^\alpha\bigg(F^{T}_1\gamma^\mu+F^{T}_2v^\mu+F^{T}_3v'^\mu\bigg) + F^{T}_4g^{\alpha\mu}\bigg]u(p,\lLb)\,.
\end{aligned}
\end{equation}
The corresponding expression in Ref.~\cite{Descotes-Genon:2019dbw} is
\begin{equation}
\begin{aligned}
\bra{{\Lstar}}\bar si \sigma^{\mu\nu}q_\nu b\ket{\Lb}= - &\bar u_\alpha(k,\lLambda)\biggl\{p^\alpha\biggl[f_0^T(\qsq) \frac{\qsq}{s_+}(p^\mu +k^\mu-\frac{q^\mu}{\qsq}(\mLb^2-\mLambda^2))+\\
& f_\perp^T(\qsq)(\mLb + \mLambda)(\gamma^\mu-2\frac{\mLambda}{s_+}p^\mu -2\frac{\mLb}{s_+}k^\mu)\biggr]+\\
& f_g^T(\qsq)\left[g^{\alpha\mu}+\mLambda\frac{p^\alpha}{s_-} \left(\gamma^\mu - 2 \frac{k^\mu}{\mLambda} +2 \frac{\mLambda p^\mu +\mLb k^\mu}{s_+}\right)\right]\biggr\}u(p,\lLb)\,.
\end{aligned}
\end{equation}
Comparing terms in the two expressions, 
\begin{align}
\begin{split}\label{eq:LStar_vectensor}
    F_{1}^T &= \mLb \left( - f_\perp^T(\mLb+\mLambda) - f_g^T\frac{\mLambda}{s_-} \right)  \,, \\
    F_{2}^T &= \mLb \left(- f_0^T\frac{\mLb\qsq}{s_+}\left( 1-\frac{\mLb^2-\mLambda^2}{\qsq} \right) + f_\perp^T(\mLb+\mLambda)\frac{2\mLambda \mLb}{s_+} - f_g^T\frac{2\mLambda^2 \mLb}{s_-s_+} \right)  \,, \\
    F_{3}^T &= \mLb \Biggl( -f_0^T\frac{\mLambda\qsq}{s_+}\left(1+\frac{\mLb^2-\mLambda^2}{\qsq}\right) \\
    &\qquad\qquad\qquad+ f_\perp^T(\mLb+\mLambda)\frac{2\mLb \mLambda}{s_+} + f_g^T\frac{2\mLambda}{s_-}\left(1-\frac{\mLb \mLambda}{s_+}\right) \Biggr)  \,, \\
    F_{4}^T &= - f_g^T  \,.
\end{split}
\end{align}
The axialtensor current in the quark-model is 
\begin{equation}
\begin{aligned}
\bra{{\Lstar}}\overline{s}i\sigma^{\mu\nu}\gamma_5q_\nu b\ket{\Lb} &=& \overline{u}_\alpha(k,\lLambda)\bigg[v^\alpha\bigg(G^{T}_1\gamma^\mu+G^{T}_2v^\mu+G^{T}_3v'^\mu\bigg)+ G^{T}_4g^{\alpha\mu}\bigg]\gamma_5u(p,\lLb)\,.
\end{aligned}
\end{equation}
The corresponding expression in Ref.~\cite{Descotes-Genon:2019dbw} is
\begin{equation}
\begin{aligned}
\bra{{\Lstar}} \bar si \sigma^{\mu\nu}\gamma^5q_\nu b\ket{\Lb}= -& \bar u_\alpha(k,\lLambda)\gamma^5\biggl\{p^\alpha\biggl[f_0^{T5}(\qsq) \frac{\qsq}{s_-}(p^\mu +k^\mu-\frac{q^\mu}{\qsq}(\mLb^2-\mLambda^2))+\\
& f_\perp^{T5}(\qsq)(\mLb - \mLambda)(\gamma^\mu+2\frac{\mLambda}{s_-}p^\mu -2\frac{\mLb}{s_-}k^\mu)\biggr]+\\
& f_g^{T5}(\qsq)\left[g^{\alpha\mu}-\mLambda\frac{p^\alpha}{s_+} \left(\gamma^\mu + 2 \frac{k^\mu}{\mLambda} -2 \frac{\mLambda p^\mu -\mLb k^\mu}{s_-}\right)\right]\biggr\}u(p,\lLb)\,.
\end{aligned}
\end{equation}
Comparing terms in the two expressions, 
\begin{align}
\begin{split}\label{eq:LStar_axtensor}
    G_{1}^T &= \mLb \left( f_\perp^{T5}(\mLb-\mLambda) - f_g^{T5}\frac{\mLambda}{s_+} \right)  \,,\\
    G_{2}^T &= \mLb \Biggl( - f_0^{T5}\frac{\mLb\qsq}{s_-}\left(1 - \frac{\mLb^2-\mLambda^2}{\qsq}\right) \\
    &\qquad\qquad\qquad- f_\perp^{T5}(\mLb-\mLambda)\frac{2\mLambda \mLb}{s_-} -f_g^{T5}\frac{2\mLambda^2 \mLb}{s_+s_-} \Biggr) \,, \\
    G_{3}^T &= \mLb \Biggl( - f_0^{T5}\frac{\mLambda\qsq}{s_-}\left(1 + \frac{\mLb^2-\mLambda^2}{\qsq}\right) \\
    &\qquad\qquad\qquad+ f_\perp^{T5}(\mLb-\mLambda)\frac{2\mLb \mLambda}{s_-} + f_g^{T5}\frac{2\mLambda}{s_+}\left(1+\frac{\mLb \mLambda}{s_-}\right) \Biggr) \,, \\
    G_{4}^T &= - f_g^{T5}\,.
\end{split}
\end{align}
\section{Blatt-Weisskopf form factors}
\label{app:BW_FF}

The Blatt-Weisskopf form-factor parametrization follows Ref.~\cite{Blatt:1952ije}, \ie 
\begin{align}
\begin{split}
    B_0(z,z_0) &= 1\,, \\
    B_1(z,z_0) &= \sqrt{\frac{1 + z_0}{1 + z}}\,, \\
    B_2(z,z_0) &= \sqrt{\frac{9 + 3z_0 + z_0^2}{9 + 3z  + z^2}}\,, \\
    B_3(z,z_0) &= \sqrt{\frac{225 + 45z_0 + 6z_0^2 + z_0^3}{225 + 45z_{\phantom{0}}  + 6z^2 + z^3}}\,, \\
    B_4(z,z_0) &= \sqrt{\frac{11025 + 1575z_0 + 135z_0^2 + 10z_0^3 + z_0^4}{11025 + 1575z_{\phantom{0}} + 135z^2+ 10z^3 + z^4}}\,, \\
    B_5(z,z_0) &= \sqrt{\frac{893025 + 99225z_0 + 6300z_0^2 + 315z_0^3 + 15z_0^4 + z_0^5}{893025 + 99225z_{\phantom{0}} + 6300z^2 + 315z^3 + 15z^4 + z^5}}\,,
\end{split}
\end{align}
where $z_{(0)}=|\vec{k}_1^{(\Lstar)}|^2r^2$ and $r$ parametrises the size of the parent baryon. A standard value for \Lstar resonances is $r_{\Lstar} = 3.0\gev^{-1}c$. 
The momentum $\vec{k}_1^{(\Lstar)}$ is the proton momentum in the \pk rest frame (at the pole mass).

\section{Full set of angular coefficients}
\label{app:observables}
The full set of angular observables for a mixture of states up-to spin-\jFive are available as a notebook in the supplemental material of this paper.
Expressions for the observables assuming unpolarized \Lb baryons in terms of the amplitudes, ${\cal A}$ defined in Sec.~\ref{sec:observables}, are given below: 

\foreach \x in {5,...,31} {
\begin{align}
\label{eq:K\x}
\begin{split}
\input{moments/K\x\string.tex}
\end{split}
\end{align}
}
\foreach \x in {33,...,46} {
\begin{align}
\label{eq:K\x}
\begin{split}
\input{moments/K\x\string.tex}
\end{split}
\end{align}
}
\section{Translation between the \texorpdfstring{\boldmath$L_i$}{Li} and \texorpdfstring{\boldmath$K_j$}{Kj} basis}
\label{app:LKtranslation}

The angular observables in the $L_i$ basis can be expressed in terms of the $K_j$ observables with 
\begin{align} 
\begin{split} 
L_{1c} 
&= K_{2}+\sqrt{5} K_{8}-\frac{81}{8}K_{14} \,,\quad 
L_{2c} 
= K_{2}-\frac{\sqrt{5}}{2} K_{8} + \frac{9}{8} K_{14} \\
L_{1cc} &=
\frac{1}{8\sqrt{3}}\left(
8K_{1}
+8\sqrt{5} K_{3}
+8\sqrt{5} K_{7}
+40 K_{9}
-81 K_{13}
-81 \sqrt{5} K_{15}
\right) \,, \\
L_{1ss} &=
\frac{1}{16\sqrt{3}}\left(
16 K_{1}
-  8 \sqrt{5} K_{3}
+ 16 \sqrt{5} K_{7}
- 40 K_{9}\right. \\
&\qquad\qquad\qquad\left.-162 K_{13}
+ 81 \sqrt{5} K_{15}
-105 \sqrt{6} K_{41}
\right) \,, \\
L_{2cc} &=
\frac{1}{8\sqrt{3}}\left(
8 K_{1}
+8 \sqrt{5} K_{3}
-4 \sqrt{5} K_{7}
-40 K_{9}
+9 K_{13}
+9 \sqrt{5} K_{15}
\right) \,, \\
L_{2ss} &=
\frac{1}{16\sqrt{3}}\left(
16 K_{1}
- 8 \sqrt{5} K_{3}
- 8 \sqrt{5} K_{7}
+20 K_{9}\right. \\
&\qquad\qquad\qquad\left.
+18 K_{13}
- 9 \sqrt{5} K_{15}
-30 \sqrt{2} K_{39}
+15 \sqrt{6} K_{41}
\right) \,, \\
L_{3ss} &= \frac{5\sqrt{3}}{4\sqrt{2}}\left(2 K_{39}-\sqrt{3} K_{41}\right) \,,\quad
L_{4ss}  = \frac{5\sqrt{3}}{4\sqrt{2}}\left(2 K_{43}-\sqrt{3} K_{45}\right) \,, \\
L_{5s} &= \frac{1}{4} \sqrt{5} \left(2 \sqrt{6} K_{22}-9 K_{26}\right)\,,\quad 
L_{5sc} = \frac{5}{4} \left(2 \sqrt{6} K_{21}-9 K_{25}\right) \,,\\
L_{6s} &= \frac{1}{4} \sqrt{5} \left(2 \sqrt{6} K_{32}-9 K_{36}\right) \,,\quad  
L_{6sc} = \frac{5}{4} \left(2 \sqrt{6} K_{31}-9 K_{35}\right) \,,\\
L_{7cc} &= \frac{35}{8} \sqrt{3} \left(K_{13}+\sqrt{5} K_{15}\right)\,, \\
L_{7ss} &= \frac{35\sqrt{3}}{16}\left(2 K_{13}-\sqrt{5} K_{15}+\sqrt{6} K_{41}\right) \,.
\end{split} 
\end{align}

The remaining observables are directly related, with 
\begin{equation}
\begin{alignedat}{4}
& L_{7c} = \frac{105}{8} K_{14} \,,\quad
& L_{8ss} = \frac{105}{4\sqrt{2}} K_{41} \,,\quad
& L_{9ss} = \frac{105}{4\sqrt{2}} K_{45} \,,\quad & \\
& L_{10s} = \frac{21 \sqrt{5}}{4} K_{26} \,,\quad
& L_{10sc} = \frac{105}{4} K_{25} \,,\quad
& L_{11s} = \frac{21 \sqrt{5}}{4} K_{36} \,,\quad
& L_{11sc} = \frac{105}{4} K_{35}\,.
\end{alignedat}
\end{equation}
Note that for the spin-\jThree case, the coefficients $L_i$ with an index larger than six are zero.
On top, only the $K_j$ indicated in Table~\ref{tab:K_description} are non-zero for a spin-\jThree state.
\clearpage

\addcontentsline{toc}{section}{References}
\bibliographystyle{LHCb}
\bibliography{main.bib,standard.bib}

\end{document}